\documentclass[11pt,a4paper]{article}
\pdfoutput=1
\usepackage{jheppub}

\usepackage{multirow, graphicx,amssymb,url,mathrsfs,amsmath}
\usepackage{wrapfig,boxedminipage,setspace,subfigure,epsfig}
\usepackage{amsxtra,amstext,latexsym,dsfont,amsfonts}
\usepackage{color,eucal}
\usepackage[dvipsnames]{xcolor}
\usepackage{float}
\usepackage{slashed,comment}
\usepackage{kotex}
\usepackage{tikz}
\usetikzlibrary{calc,patterns,angles,quotes}
\usetikzlibrary{decorations.pathreplacing,decorations.markings,snakes}
\usepackage{tabularx, array}
\usepackage{mdframed,mathtools}
\newcolumntype{L}[1]{>{\raggedright\arraybackslash}p{#1}}
\newcolumntype{C}[1]{>{\centering\arraybackslash}p{#1}}
\newcolumntype{R}[1]{>{\raggedleft\arraybackslash}p{#1}}

\DeclareMathOperator{\csch}{csch}
\DeclareMathOperator{\sech}{sech}

\newcommand{\nn}{\nonumber}

\newcommand{\dd}{\mathrm{d}}

\newcommand{\abs}[1]{\left| #1 \right|}
\newcommand{\bra}[1]{\mbox{$\langle #1 |$}}
\newcommand{\ket}[1]{\mbox{$| #1 \rangle$}}

\newcommand{\be}{\begin{equation}}
\newcommand{\ee}{\end{equation}}
\newcommand{\bea}{\begin{eqnarray}}
\newcommand{\eea}{\end{eqnarray}}

\title{\huge Krylov complexity of density matrix operators}

\author[a,b]{Pawel Caputa,}
\author[c,d]{Hyun-Sik Jeong,}
\author[a]{Sinong Liu,}
\author[c]{Juan F. Pedraza}
\author[c,d]{and Le-Chen Qu}

\emailAdd{pawel.caputa@fuw.edu.pl}
\emailAdd{hyunsik.jeong@csic.es}
\emailAdd{sinong.liu@fuw.edu.pl}
\emailAdd{j.pedraza@csic.es}
\emailAdd{lechen.qu@ift.csic.es}

\preprint{YITP-24-21, IFT-UAM/CSIC-24-25}

\affiliation[a]{Faculty of Physics, University of Warsaw, Pasteura 5, 02-093 Warsaw, Poland}
\affiliation[b]{Yukawa Institute for Theoretical Physics, Kyoto University, Kitashirakawa Oiwakecho, Sakyo-ku, Kyoto 606-8502, Japan}
\affiliation[c]{Instituto de F\'isica Te\'orica UAM/CSIC, Calle Nicol\'as Cabrera 13-15, 28049 Madrid, Spain}
\affiliation[d]{Departamento de F\'isica Te\'orica, Universidad Aut{\'o}noma de Madrid, 28049 Madrid, Spain}

\abstract{Quantifying complexity in quantum systems has witnessed a surge of interest in recent years, with Krylov-based measures such as Krylov complexity ($C_K$) and Spread complexity ($C_S$) gaining prominence.  In this study, we investigate their interplay by considering the complexity of states represented by \textit{density matrix operators}. After setting up the problem, we analyze a handful of analytical and numerical examples spanning generic two-dimensional Hilbert spaces, qubit states, quantum harmonic oscillators, and random matrix theories, uncovering insightful relationships. For generic pure states, our analysis reveals two key findings: (I) a correspondence between moment-generating functions (of Lanczos coefficients) and survival amplitudes, and (II) an early-time equivalence between $C_K$ and $2C_S$. Furthermore, for maximally entangled pure states, we find that the moment-generating function of $C_K$ becomes the Spectral Form Factor and, at late-times, $C_K$ is simply related to $NC_S$ for $N\geq2$ within the $N$-dimensional Hilbert space. Notably, we confirm that $C_K = 2C_S$ holds across all times when $N=2$. Through the lens of random matrix theories, we also discuss deviations between complexities at intermediate times and highlight subtleties in the averaging approach at the level of the survival amplitude.
}
\begin{document}
\maketitle

%
\section{Introduction}\label{}
Understanding complexity is inherently intuitive yet challenging to precisely quantify. It involves gauging the difficulty of tasks within the constraints of limited resources. For instance, information theorists \cite{shannon1949synthesis} often adopt the circuit model, where a task can be accomplished by using a set of elementary gates and their minimal number that does the job is the estimate for complexity. In recent years, quantifying complexity in quantum systems has emerged as a focal point, spanning disciplines from quantum field theories \cite{Dowling:aa,Jefferson:2017sdb,Caputa:2017yrh,Chapman:2017rqy,Magan:2018nmu,Caputa:2018kdj} to quantum gravity \cite{Stanford:2014jda,Brown:2015bva,Brown:2015lvg,Couch:2016exn,Belin:2021bga}. This exploration has given rise to various physically-motivated measures of quantum complexity, which are now being tested as innovative tools for unraveling the mysteries of the quantum realm.

One of the most promising recent definitions in the realm of quantum complexity is Krylov complexity, developed to quantify \emph{operator} size and growth in quantum many-body systems and quantum field theories \cite{Parker:2018yvk} (see also \cite{Roberts:2018mnp}). Essentially, the idea is to map the growth of the Heisenberg operator into a one-dimensional chain, representing the minimal orthonormal basis capturing operator evolution. As time progresses, operator growth manifests as motion along this chain, with the average position defining the Krylov complexity (i.e. the operator size). Recently, this idea has been extended to quantum \emph{states} undergoing unitary evolution and spreading through the Hilbert space \cite{Balasubramanian:2022tpr}. Here, Spread complexity quantifies the average position of the state's spread along a similar one-dimensional chain. These two physical definitions have provided significant insights into quantum systems, such as e.g. sensitivity to quantum chaos \cite{Rabinovici:2022beu,Balasubramanian:2022tpr,Balasubramanian:2023kwd}, topological phase transitions \cite{Caputa:2022eye,Caputa:2022yju}, and also play important roles in random matrix theories \cite{Balasubramanian:2022dnj}, quantum gravity and the AdS/CFT correspondence \cite{Lin:2022rbf,Rabinovici:2023yex}.
See also~\cite{Barbon:2019wsy,Avdoshkin:2019trj,Dymarsky:2019elm,Magan:2020iac,Rabinovici:2020ryf,Cao:2020zls,Jian:2020qpp,Kim:2021okd,Rabinovici:2021qqt,Dymarsky:2021bjq,Caputa:2021sib,Patramanis:2021lkx,Caputa:2021ori,Trigueros:2021rwj,Fan:2022xaa,Heveling:2022hth,Bhattacharjee:2022vlt,Muck:2022xfc,He:2022ryk,Hornedal:2022pkc,Alishahiha:2022anw,Camargo:2022rnt,Baek:2022pkt,Bhattacharya:2022gbz,Liu:2022god,Bhattacharjee:2022lzy,Afrasiar:2022efk,Avdoshkin:2022xuw,Erdmenger:2023wjg,Hashimoto:2023swv,Camargo:2023eev,Bhattacharya:2023yec,Vasli:2023syq,Patramanis:2023cwz,Chattopadhyay:2023fob,Iizuka:2023pov,Huh:2023jxt,Pal:2023yik,Bhattacharya:2023zqt,Li:2024kfm,Anegawa:2024wov,Beetar:2023mfn,Loc:2024oen,adhikari2023krylov,adhikari2022cosmological} and the references therein for some of the recent development in Krylov and Spread complexity.

One of the interesting open questions within the the Krylov-based complexity program is the exploration of whether and how the two above definitions differ and how they encode the information about the system's dynamics, such as quantum chaos or thermalization. To make progress in this direction, in this work, we investigate the Krylov complexity of the density matrix operator's growth, corresponding to a pure state, denoted as $\rho(t)=\ket{\psi(t)}\bra{\psi(t)}$ and compare it with the Spread complexity of $\ket{\psi(t)}=\exp(-iHt)\ket{\psi(0)}$.\footnote{In fact this could have been an alternative way to \cite{Balasubramanian:2022tpr} to define quantum Krylov complexity for states.} In this context, once we find the Krylov basis and Lanczos coefficients for $\ket{\psi(t)}$ and compute the time evolution of the Spread complexity, it is natural to ask how various features (e.g. the growth, the time evolution and various timescales) get repackaged into the Krylov complexity of $\rho(t)$. As we will see, even though the Krylov complexity may appear more natural for pure states (e.g. regarding the return amplitude) the Spread complexity is often amenable to an elegant, analytical treatment. More generally, in this paper we explore similarities and differences between these two approaches for pure states with a hope to pave the way towards more efficient and sensitive definitions of quantum complexity.\footnote{While our primary focus lies on scenarios involving \emph{pure} states, we also extend our discussion to encompass the Krylov complexity associated with density matrices of \emph{mixed} states within the framework of two-dimensional Hilbert spaces.}

One intriguing observation emerges when we examine the time evolution of the thermofield double state $\ket{\psi(t)}=\exp\left(-i(H_L+H_R)t\right)\ket{TFD}$. Notably, previous work \cite{Balasubramanian:2022tpr} highlighted that the Lanczos coefficients in this context are encoded in the return amplitude given by the amplitude of the spectral form factor (see also related \cite{Papadodimas:2015xma,delCampo:2017bzr}). In our investigation, we find a similar pattern when considering the evolution of $\rho(t)$. Here, the return amplitude aligns with the spectral form factor itself. This observation carries significant implications, particularly in the context of averaging and ensembles of theories, such as Random Matrices or 2D quantum gravity \cite{Saad:2018bqo,Saad:2019lba,Chen:2023hra}. Namely, it has been recently realised that averages over such return amplitudes in theories dual to gravity probe contributions from gravitational wormholes in the bulk. Following these ideas in the context of complexity of a pure $\rho(t)$ we can then hope to make a link between the Krylov basis and wormholes. However, we point out a few subtleties that stress the fact that the order of taking an average can sometimes lead to incorrect conclusions.

This paper is structured as follows: Section \ref{SECTION2} provides essential background information on Krylov complexity for operators. In Section \ref{SECTION3}, we analyze several analytical models, computing the Spread and Krylov complexity for pure states. Section \ref{SECTION5} expands our examination to examples within random matrix theories. Towards the conclusion of this section, in \ref{SECTION6}, we present a simple toy model to highlight the subtleties surrounding the order of averaging in computing Krylov complexity. Section \ref{SECTION4} discusses the repackaging of Lanczos coefficients and Krylov basis data for $\ket{\psi(t)}$ into those of $\rho(t)=\ket{\psi(t)}\bra{\psi(t)}$. Finally, Section \ref{SECTION7} presents our main conclusions and discusses some interesting open questions. Additional details on Spread complexity are provided in Appendix \ref{appena}.

\section{Preliminaries}\label{SECTION2}
In this section, we review the formalism of the operator growth within the context of the Krylov complexity~\cite{Parker:2018yvk}. Specifically, tailored to our objectives, we present the formalism pertaining to the \textit{density matrix} case.

\subsection{Operator growth and Lanczos algorithm}\label{}

\paragraph{Density matrix in quantum mechanics.}
In quantum mechanics, a system is described by a state vector, typically denoted as $\ket{\psi}$, which belongs to a Hilbert space. Nevertheless, in many practical situations, a quantum systems may not be in a pure state but rather in a mixed state. A mixed state occurs when a system is in a statistical ensemble of different pure states with certain probabilities.

The notion of the density matrix, denoted as $\rho$, proves to be a valuable asset in quantum mechanics, offering a means to represent the quantum state of a system -- be it pure or mixed. It incorporates information regarding the probabilities associated with various outcomes during measurements.
To illustrate, the density matrix takes the form of the outer product of the state vector:
\begin{align}\label{DOEX}
\begin{split}
\rho = \ket{\psi} \bra{\psi} \,, \qquad  \rho =  \sum_{i} p_{i} \ket{\psi_{i}} \bra{\psi_{i}} \,, 
\end{split}
\end{align}
where the former represents a pure state, and the latter characterizes a mixed state. Note that the latter is constructed as a summation of the outer products of individual pure states, each weighted by its respective probability $p_i$. Density matrices emerge as indispensable tools in branches of quantum mechanics dealing with mixed states, such as quantum statistical mechanics, open quantum systems, and quantum information.

\paragraph{Liouville-von Neumann equation.}
The Liouville-von Neumann equation describes the time evolution of a quantum systems in terms of its density matrix, i.e., it describes how the density matrix evolves with time under the influence of the Hamiltonian $H$. It is a fundamental equation in quantum mechanics and quantum information theory.\footnote{An illustrative application of this equation in quantum mechanics pertains to its role in deriving the optical Bloch equations, elucidating the behavior of atoms engaged in interactions with light. Additionally, the (reduced) density matrix can serve as a widely adopted approach within the von Neumann entropy for comprehending and quantifying quantum entanglement through the evaluation of entanglement entropy.}

The Liouville-von Neumann equation -- e.g., for the mixed state; the latter in \eqref{DOEX} -- asserts that
\begin{align}\label{von}
\begin{split}
\partial_t \rho(t) \,=\,& \sum_{i} \,p_{i}\, \left(\partial_t\ket{\psi_{i}}\right)\,\bra{\psi_{i}} \,+\, \sum_{i} \,p_{i}\, \ket{\psi_{i}}\,\left(\partial_t\bra{\psi_{i}}\right) \\
\,=\,& -\frac{i}{\hbar} \sum_{i} \,p_{i}\, H \ket{\psi_{i}} \,\bra{\psi_{i}} \,+\, \frac{i}{\hbar} \sum_{i} \,p_{i}\,\ket{\psi_{i}}\, \bra{\psi_{i}} H \\
\,=\,& -\frac{i}{\hbar} \left[H, \rho(t)\right] \,, 
\end{split}
\end{align}
where the Schrödinger equation $\partial_t \ket{\psi_{i}} = -\frac{i}{\hbar} H \ket{\psi_{i}}$ is employed in the second equality. For simplicity, we adopt $\hbar=1$ hereafter.

Several observations are in order. Firstly, the Liouville-von Neumann equation \eqref{von} is evidently applicable to pure states as well. Secondly, though \eqref{von} bears resemblance to the Heisenberg equation, a crucial sign distinction exists; caution is warranted to prevent misconstruing this similarity. The Heisenberg equation operates within the Heisenberg picture, where states remain time-independent and operators evolve over time. In contrast, \eqref{von} adheres to the Schrödinger picture, where states (within in the density matrix) exhibit the time-dependence, and the Hamiltonian $H$ remains unaltered over time. Lastly, \eqref{von} constitutes the quantum mechanical analogue of the classical Liouville equation, which, in classical mechanics, gives rise to a significant result known as Liouville's theorem.

\paragraph{Krylov basis and Lanczos algorithm.}
The Liouville-von Neumann equation \eqref{von} can be formally solved by
\begin{align}\label{KBEOM1}
\begin{split}
\rho(t) = e^{-i H t} \,\rho_0\, e^{i H t} = \sum_{n=0}^{\infty} \frac{(-i t)^n}{n !} \mathcal{L}^{n} \rho_0 \,, \qquad \mathcal{L}=[H, \cdot] \,,
\end{split}
\end{align}
where $\rho_0:=\rho(0)$ and $\mathcal{L}$ is the Liouvillian superoperator.
Note that the expression \eqref{KBEOM1} indicates that the time evolution of the density matrix, $\rho(t)$, can be articulated as a linear combination involving a sequence of operators 
\begin{align}\label{KBEOM2}
\begin{split}
\rho_0 \,,\,\, \mathcal{L} \rho_0 \,,\,\, \mathcal{L}^2 \rho_0 \,,\, \cdots \,.
\end{split}
\end{align}
In other words, the initial density matrix $\rho_0$ undergoes the time evolution within the operator (sub)space spanned by \eqref{KBEOM2}, which is called the Krylov subspace. The common lore is that a more ``chaotic" $H$ (within $\mathcal{L}$) may result in a more complex $\rho(t)$ compared to a scenario where $H$ is less ``chaotic".

The Krylov subspace can admit an orthonormal basis denoted as $|\rho_n)$, commonly referred to as the \textit{Krylov basis}. This basis can be systematically generated through the application of the Gram-Schmidt orthogonalization procedure on the subspace: this procedure is known as the \textit{Lanczos algorithm}. To carry out the Lanczos algorithm, an essential requirement is the determination of the inner product between operators $\rho_1$ and $\rho_2$ within the Krylov space. In this work, we will pick an inner product, denoted as $\left(\rho_1 | \rho_2\right)$, defined as \cite{Hashimoto:2023swv}
\begin{align}\label{inner}
\begin{split}
\left(\rho_1 | \rho_2\right) := \text{Tr}\left[\rho_1^{\dagger} \rho_2 \right].
\end{split}
\end{align}
An alternative definitions for the inner product can be adopted (e.g. the Wightman inner product), as discussed in \cite{Caputa:2021sib}. 

Utilizing the inner product \eqref{inner}, one can construct the Krylov basis $|\rho_n)$ through the Lanczos algorithm:
\begin{enumerate}
\item{Define $|A_0):=|\rho_0)$ and normalize it with $b_0 : = \sqrt{(A_0|A_0)}$ to obtain the orthonormal basis element $|\rho_0) : = b_0^{-1} |A_0)$.}
\item{Define $|A_1):=\mathcal{L}|\rho_0)$ and normalize it with $b_1 : = \sqrt{(A_1|A_1)}$ to obtain the orthonormal basis element $|\rho_1) : = b_1^{-1} |A_1)$.}
\item{For $n\geq2$, generate the successive element as $|A_n):=\mathcal{L}|\rho_{n-1}) - b_{n-1}|\rho_{n-2})$. Normalize it by $b_n : = \sqrt{(A_n|A_n)}$ and define the $n$-th basis element as $|\rho_n) : = b_n^{-1} |A_n)$.}
\item{Terminate the algorithm when $b_n$ becomes zero.}
\end{enumerate} 
In this way, the Lanczos algorithm provides the Krylov basis $\left\{ |\rho_n) \right\}$ and the associated normalization coefficients $\left\{ b_n \right\}$ commonly referred to as the \textit{Lanczos coefficients}.

\paragraph{Krylov complexity of density matrix.}
Upon obtaining the Krylov basis using the Lanczos algorithm, it becomes feasible to represent the time-dependent density matrix, $\rho(t)$, in terms of the Krylov basis
\begin{align}\label{owf}
\begin{split}
|\rho(t)) = \sum_n i^n \varphi_n(t) |\rho_n) \,.
\end{split}
\end{align}
Here, the transition amplitudes, denoted as $\varphi_n(t)$, adheres to the normalization condition $\sum |\varphi_n(t)|^2 =1$.\footnote{It is noteworthy that $|\rho(t))$ can be interpreted as the Gelfand-Naimark-Segal (GNS) state.}
Then, by substituting \eqref{owf} into the Liouville-von Neumann equation \eqref{von},\footnote{Strictly speaking, we need to plug the operator version of \eqref{owf} into \eqref{von}.} one can find the recursion relation for the transition amplitude as 
\begin{align}\label{dse}
\begin{split}
\partial_t \varphi_n(t) = b_{n+1} \, \varphi_{n+1}(t) - b_n \, \varphi_{n-1}(t) \,,
\end{split}
\end{align}
subject to the initial condition $\varphi_n(0) = \delta_{n 0}$ by definition.
Note that the solution of the equation \eqref{dse}, $\varphi_n(t)$, can be achieved once $b_n$ is determined through the Lanczos algorithm. The transition amplitude $\varphi_n(t)$ constitutes a fundamental component in constructing the Krylov complexity \eqref{Kcom}, as defined subsequently.

Furthermore, it is insightful to regard the Lanczos coefficients as hopping amplitudes facilitating the traversal of the initial operator along the ``Krylov chain". In addition, the functions $\varphi_n(t)$ can be thought of as wave packets moving along this chain~\cite{Parker:2018yvk}. This conceptualization motivates a definition of the \textit{Krylov complexity} ($C_{K}$)
\begin{align}\label{Kcom}
\begin{split}
C_{K} := \sum_n n\left|\varphi_n(t)\right|^2 \,,
\end{split}
\end{align}
which is the average position on the chain i.e., quantifies the average position in the Krylov basis.\footnote{In a similar vein, Krylov entropy can be defined as $S_{K} := -\sum_n\left|\varphi_n(t)\right|^2 \log \left|\varphi_n(t)\right|^2$. It gauges the level of randomness within the distribution~\cite{Barbon:2019wsy}. In \cite{Fan:2022xaa}, the relationship between $S_{K}$ and $C_{K}$ is reported as $S_K = -C_K \log C_K + C_K$, valid in the early time regime.}

\subsection{More on Krylov complexity of density matrix}

\subsubsection{Lanczos coefficients by moment method}
Since we are primarily concerned with $\varphi_n(t)$, it is instructive to introduce an alternative method for computing the Lanczos coefficients, known as the moment method~\cite{Parker:2018yvk}. The procedural steps for this calculation are outlined as follows.
To begin with, the moments $\mu_{2n}$ are defined through the moment-generating function $G(t)$ as
\begin{equation}\label{OCF}
\mu_{2 n} := \left.\frac{d^{2 n}}{d t^{2 n}} {G}(it)\right|_{t=0}, \qquad G(t) = C(t) := (\rho(t) | \rho_0) \,,
\end{equation}
where $G(t)$ is chosen as the autocorrelation function $C(t)$ for the Krylov complexity.
Subsequently, by leveraging the determinant of the Hankel matrix $M_{ij}=\mu_{i+j}$ constructed from these moments, a relation between the Lanczos coefficients $b_n$ and the moment $\mu_{2n}$ is established:
\begin{equation}\label{OCF1}
b_1^{2 n} \cdots b_n^2 = \text{det}\left(\mu_{i+j}\right)_{0 \leq i, j \leq n} \,.
\end{equation}
Alternatively, this relationship can be expressed through a recursion relation:
\begin{equation}\label{OCF2}
\begin{aligned}
M_{2 \ell}^{(j)} & =\frac{M_{2 \ell}^{(j-1)}}{b_{j-1}^2}-\frac{M_{2 \ell-2}^{(j-2)}}{b_{j-2}^2} \quad \text { with } \quad \ell=j, \ldots, n, \\
M_{2 \ell}^{(0)} & =\mu_{2 \ell}, \quad b_{-1}:=1,\quad b_0:=\text{Tr}\left[\rho_0^{\dagger} \rho_0 \right], \quad M_{2 \ell}^{(-1)}=0, \\
b_n & =\sqrt{M_{2 n}^{(n)}}.
\end{aligned}
\end{equation}
%

\subsubsection{Density matrix of the thermofield double state}
In the preceding subsection, we presented the formalism for the Lanczos algorithm and outlined the computation of Krylov complexity for the density matrix of a generic (pure or mixed) state $\ket{\psi}$.

The primary focus of this manuscript is to investigate the Krylov complexity of the density matrix in the context of the thermofield double (TFD) state $\ket{\psi_{\beta}}$, given by
\begin{equation}\label{TFDeq}
\ket{\psi_{\beta}} := \frac{1}{\sqrt{Z_\beta}} \sum_n e^{-\frac{\beta E_n}{2}}|n, n\rangle \,, \qquad Z_\beta=\sum_n e^{-{\beta E_n}} \,,
\end{equation}
where $E_n$ denotes the eigenvalues, $\ket{n}$ represents the eigenstates, and $\beta=1/T$ denotes the inverse temperature.
Using the TFD state \eqref{TFDeq}, we can construct the density matrix of the TFD, denoted as $\rho_\beta$:
\begin{equation}\label{TFDDO0}
\rho_\beta=\left|\psi_\beta\right\rangle \left\langle\psi_\beta \right|=
\frac{1}{{Z_\beta}}\sum_{n,m}e^{-\frac{\beta (E_n+E_m)}{2}}|n, n\rangle\langle m,m|,
\end{equation}
and its time evolution is given by \eqref{KBEOM1}:
\begin{equation}\label{TFDDO}
\rho_\beta(t)=e^{-i \frac{H_L+H_R}{2} t} \,\,\rho_\beta\,\, e^{i \frac{H_L+H_R}{2} t}=
\frac{1}{{Z_\beta}}\sum_{n,m}e^{-i(E_n-E_m)t}e^{-\frac{\beta (E_n+E_m)}{2}}|n, n\rangle\langle m,m| \,.
\end{equation}
Recall that the TFD state evolves by $(H_L + H_R)/2$, where $H_{L, R}$ act independently on the left and right copies of the Hamiltonian.\\

The motivation for considering this particular pure state, the TFD state, arises from the exploration of Spread complexity~\cite{Balasubramanian:2022tpr}. In that study, the authors used the TFD state to identify new features in the evolution of the Spread complexity that are universal characteristics of chaotic systems, such as the ramp-peak-slope and plateau. In the subsequent sections of this study, we test and compare these features within the Krylov complexity of the density matrix using the TFD state. Furthermore, we will explore the connection between the Krylov complexity of the density matrix \eqref{TFDDO} and Spread complexity of our specific \textit{pure} state, the TFD state. We will also discuss the case of the \textit{mixed} state within certain analytic models.

\paragraph{Autocorrelation function of the TFD state.}
Before concluding this section, it is instructive to discuss the autocorrelation function \eqref{OCF} for the TFD state, a key quantity for the analysis of Krylov complexity, especially for evaluating the Lanczos coefficients.

For the TFD case, utilizing \eqref{inner} together with \eqref{TFDDO}, the autocorrelation function is constructed as follows:
\begin{align}\label{autocorr}
\begin{split}
C(t) &= (\rho_\beta(t) | \rho_\beta) \\
&= \operatorname{Tr}\left[\frac{1}{{Z_\beta}^2}\sum_{n,m,p,q}e^{-i(E_n-E_m)t}e^{-\frac{\beta (E_n+E_m+E_p+E_q)}{2}}|n, n\rangle\langle m,m||p, p\rangle\langle q,q|\right]\\
&=\operatorname{Tr}\left[\frac{1}{{Z_\beta}^2}\sum_{n,m,q}e^{-i(E_n-E_m)t}e^{-\frac{\beta (E_n+2E_m+E_q)}{2}}|n, n\rangle\langle q,q|\right]\\
&=\frac{1}{{Z_\beta}^2}\sum_{n,m}e^{-i(E_n-E_m)t}e^{-\beta (E_n+E_m)} \,.
\end{split}
\end{align}
The final result implies that the autocorrelation function of the density matrix of the TFD state is the spectral form factor (SFF) of the TFD:
\begin{equation}\label{SFFRE}
\text{SFF} := \frac{\left|Z_{\beta-i t}\right|^2}{\left|Z_\beta\right|^2} = \frac{1}{{Z_\beta}^2}\sum_{n,m}e^{-i(E_n-E_m)t}e^{-\beta (E_n+E_m)} \,,
\end{equation}
where the partition function is defined in \eqref{TFDeq}. In other words, the Lanczos coefficient of the density matrix of the TFD state can be exclusively determined by the energy spectrum through the SFF.

Furthermore, the relationship \eqref{autocorr} between the autocorrelation function and the SFF can also be illuminated through the survival amplitude $S(t)$, introduced in the study of the Spread complexity of the TFD state~\cite{Balasubramanian:2022tpr}. Expressing \eqref{autocorr} in terms of $S(t)$, we have:
\begin{align}\label{autocorr22}
\begin{split}
C(t) \,=\, (\rho_{\beta}(t) | \rho_{\beta}) \,=\, \operatorname{Tr}\left[ \ket{\psi_{\beta}(t)} \bra{\psi_{\beta}(t)}\, \ket{\psi_{\beta}} \bra{\psi_{\beta}}\right] \,=\, \langle\psi_{\beta}|\psi_{\beta}(t)\rangle\langle\psi_{\beta}(t)|\psi_{\beta}\rangle \,=\,  |S(t)|^2,
\end{split}
\end{align}
where \eqref{DOEX} and \eqref{inner} are employed in the second equality. In the last equality, $S(t):=\langle\psi_{\beta}(t)|\psi_{\beta}\rangle$ is referred to as the survival amplitude (or return amplitude)~\cite{Balasubramanian:2022tpr,cotler2017black,saad2018semiclassical}, and the authors have established its relationship with the SFF as:
\begin{align}\label{SSFFRE}
\begin{split}
|S(t)| = \sqrt{\text{SFF}} \,.
\end{split}
\end{align}
Consequently, the autocorrelation function \eqref{autocorr22} simplifies to $C(t) = \text{SFF}$ as depicted in \eqref{autocorr}-\eqref{SFFRE}. It is worth noting that the autocorrelation function is not equivalent to the SFF for general operators and inner products. Our observation reveals that when we select the density matrix associated with the TFD state as the initial operator and opt for the inner product as defined in Eq.~\eqref{inner}, the resulting autocorrelation function precisely aligns with the SFF. 

It is also worth emphasizing that the moment-generating function of the Krylov complexity (associated with the density matrix of the \textit{generic pure} state), and that of the Spread complexity~\cite{Balasubramanian:2022tpr} can be succinctly summarized as
\begin{align}\label{GKS}
\begin{split}
G(t) =  
\begin{cases}
      \, |S(t)|^2  & \text{for Krylov complexity \,,}  \\
      \, S(t)  & \text{for Spread complexity \,,}
\end{cases} 
\end{split}
\end{align}
where one can find the further relation with the SFF \eqref{SSFFRE} for the TFD state.
This result \eqref{GKS} may highlight a certain similarity between the case of Krylov complexity and that of Spread complexity, representing one of the main objectives in this manuscript.\footnote{Regarding \eqref{GKS}, we can further elaborate on $G(t)$ within the context of the ``phase" of $S(t)$. Note that the phase information is integrated out for the case of the Krylov complexity, i.e., $|S|^2$, unlike the Spread complexity.} 

\paragraph{The Krylov and Spread complexity at early time.}
One intriguing implication arising from \eqref{GKS} becomes evident when examining the Lanczos coefficients in both Krylov and Spread complexity. To elucidate this, consider the expression
\begin{align}\label{NER11}
\begin{split}
S(t) = S_{\text{Re}}(t) + i \, S_{\text{Im}}(t) \,, 
\end{split}
\end{align}
where $S(0)=1$ by definition.

For Krylov complexity, $G(t)=|S(t)|^2$, utilizing the moment method \eqref{OCF}-\eqref{OCF2} enables the evaluation of the first Lanczos coefficient $b_1$:
\begin{align}\label{NER22}
\begin{split}
b_{\text{ Krylov}}^2 \,:=\, b_1^2 \,=\, - 2 S_{\text{Re}}{'}(0)^2  - 2 S_{\text{Re}}{''}(0) - 2 S_{\text{Im}}{'}(0)^2 \,.
\end{split}
\end{align}
Similarly, for Spread complexity (refer to the appendix \ref{appena} for a formalism review) with $G(t)=S(t)$, the Lanczos coefficient is given by
\begin{align}\label{NER33}
\begin{split}
b_{\text{ Spread}}^2 \,:=\, b_1^2 \,=\, S_{\text{Re}}{'}(0)^2   - S_{\text{Re}}{''}(0) - S_{\text{Im}}{'}(0)^2    \,,
\end{split}
\end{align}
where the imaginary part is omitted by definition.

Then, as the objective function $|S(t)|^2$ is an even function of time by definition, it follows that
\begin{align}\label{NER44}
\begin{split}
 \partial_t  \, |S(t)|^2 \, \big|_{t=0} = 2 S_{\text{Re}}{'}(0) \,\quad\rightarrow\quad S_{\text{Re}}{'}(0) = 0 \,.
\end{split}
\end{align}
Comparing \eqref{NER22} with \eqref{NER33} using \eqref{NER44}, one deduces that
\begin{align}\label{NER55}
\begin{split}
b_{\text{ Krylov}}^2 \,=\, 2 \, b_{\text{ Spread}}^2   \,.
\end{split}
\end{align}

Note that analytical studies \cite{Fan:2022xaa,Huh:2023jxt} have shown that both $C_K$ and $C_S$ exhibit quadratic behavior in the early time regime\footnote{
In this context, we are specifically referring to the early time limit, which should not be confused with the early time scale \cite{rabinovici2022krylov}. The early time limit, primarily influenced by $b_1$, corresponds to the regime as $t \rightarrow 0$, whereas the early time scale, predominantly dictated by $b_n \approx \alpha n$, extends over the interval $0 \lesssim t \lesssim \log S$ for finite chaotic systems with $S$ degrees of freedom.}:
\begin{align}\label{NER66}
\begin{split}
C_K \,\approx\, b_{\text{ Krylov}}^2 \, t^2 \,, \qquad C_S \,\approx\, b_{\text{ Spread}}^2 \,t^2   \,.
\end{split}
\end{align}
As such, combining \eqref{NER66} with \eqref{NER55}, the conclusion emerges that:\\  

$\bullet$\,\, In the \textit{early} time regime, the Krylov complexity associated with the density matrix of the \textit{general pure} state is always twice the Spread complexity: \, $C_K \,=\, 2 C_S$ \, ($t\ll1$).\\

In the subsequent sections, we will demonstrate that $C_K \,=\, 2 C_S$ can hold even beyond the early time regime, specifically when examining the maximally entangled state within the context of a two-dimensional Hilbert space.

\section{Analytically solvable models}\label{SECTION3}
In this section, we employ the methodology introduced in the preceding section to investigate the Krylov complexity associated with the density matrix of the pure/mixed state. Our primary focus is not only on developing the further understanding of the Krylov complexity, but also on conducting a comparative analysis with the Spread complexity of the TFD, given in previous literature.

For this purpose, we focus on cases where analytical examination is viable. Specifically, we consider scenarios involving: 
\begin{enumerate}
\item{General two-dimensional Hilbert space,}
\item{Qubit state (within modular Hamiltonian),}
\item{Quantum harmonic oscillators.}
\end{enumerate}
It is noteworthy that the first two models pertain to the case of a \textit{two}-dimensional Hilbert space, while the third one pertains to an \textit{infinite}-dimensional Hilbert space. Our investigations of these three models are geared towards achieving comparability with the Spread complexity outlined in \cite{Balasubramanian:2022tpr, Caputa:2023vyr}.

The analysis of the first model will be expanded to \textit{higher} (yet finite)-dimensional Hilbert spaces in the subsequent section, employing random matrix theories, in particular.

\subsection{Two-dimensional Hilbert space}
Let us first consider an arbitrary two-dimensional system with a Hamiltonian in the energy basis represented as
\begin{equation}\label{ASM1}
H=\left(\begin{array}{cc}
E_1 & 0 \\
0 & E_2
\end{array}\right) \,, \qquad E_2 > E_1 \,,
\end{equation}
accompanied by a generic (pure or mixed) density matrix by
\begin{equation}\label{DM1}
\rho_0=\left(\begin{array}{cc}
A_{11} & A_{12} + i B_{12} \\
 A_{12}- i B_{12} & A_{22}
\end{array}\right) \,,
\end{equation}
that is Hermitian. Additionally, the normalization condition for probabilities ($\text{Tr}[\rho_0]=1$) and $\text{Det}[\rho_0]\geq 0$ result in 
\begin{align}\label{CONTDH}
\begin{split}
A_{11}+A_{22} = 1 \,, \qquad A_{12}^2+B_{12}^2 \,\leq\, A_{11} A_{22} \,.
\end{split}
\end{align}
Recall that the density matrix is a Hermitian and positive semi-definite operator, implying that all of its eigenvalues are real and non-negative. Consequently, the product of its eigenvalues, i.e., the determinant of it ($\text{Det}[\rho_0]$), is non-negative.

Also recall that to compute the Krylov complexity given by \eqref{Kcom}, one needs to find the transition amplitude $\varphi_n(t)$ by solving the recursion equation \eqref{dse} where the computation of the Lanczos coefficients $b_n$ is required in advance. In this study, the moment method \eqref{OCF}-\eqref{OCF2} is employed to compute $b_n$, where the initial step involves the computation of the auto-correlation function $C(t)$.

\paragraph{Krylov complexity of the \textit{generic} state.}
For the case of a (general) density matrix \eqref{DM1}, the auto-correlation function can be read as
\begin{align}\label{ACFR1}
\begin{split}
C(t) \,=\, (\rho(t) | \rho_0) \,=\,  \operatorname{Tr}[e^{-i H t} \rho_0 e^{i H t} \, \rho_0] \,=\, A_{11}^2+A_{22}^2+2(A_{12}^2+B_{12}^2)\cos [\Delta E \, t] \,,
\end{split}
\end{align}
where $\Delta E := E_2-E_1$. Here we use \eqref{inner} in the second equality, and \eqref{ASM1}-\eqref{DM1} are applied in the last equality.

Subsequently, employing the moment method \eqref{OCF}-\eqref{OCF2} for the given \eqref{ACFR1}, the non-vanishing Lanczos coefficients can be determined as
\begin{equation}\label{LC1}
\begin{aligned}
b_1 = \sqrt{\frac{2\left(A_{12}^2+B_{12}^2\right)(\Delta E)^2}{A_{11}^2+A_{22}^2+2(A_{12}^2+B_{12}^2)}} \,, \qquad 
b_2 = \sqrt{\frac{\left(A_{11}^2+A_{22}^2\right)(\Delta E)^2}{A_{11}^2+A_{22}^2+2(A_{12}^2+B_{12}^2)}} \,.
\end{aligned}
\end{equation}
Finally, by solving the recursion equation \eqref{dse} with \eqref{LC1}, the Krylov complexity of the generic operator \eqref{DM1} can be achieved as
\begin{align}\label{K2d}
\begin{split}
C_K = \frac{2\left(A_{12}^2+B_{12}^2\right)}{A_{11}^2+A_{22}^2+2(A_{12}^2+B_{12}^2)}\bigg\{\sin^2 [\Delta E \, t] + \frac{8\left(A_{11}^2+A_{22}^2\right)}{A_{11}^2+A_{22}^2+2(A_{12}^2+B_{12}^2)}\sin^4 \left[\frac{\Delta E}{2}t\right] \bigg\} \,.
\end{split}
\end{align}
In what follows, we will investigate \eqref{K2d} for both pure and mixed states, respectively.\\

\paragraph{Krylov complexity of the \textit{pure} state.}
Next, let us examine the case of the arbitrary pure state given by
\begin{equation}\label{PSS}
|\psi\rangle = \cos\theta |E_1\rangle + \sin\theta \, e^{i \phi} |E_2\rangle \,,
\end{equation}
which leads to the density matrix $\rho_0 = \ket{\psi}\bra{\psi}$ following \eqref{DOEX} with
\begin{equation}\label{PSCON}
A_{11} = \cos^2\theta \,,\quad A_{22}=\sin^2\theta \,, \quad
A_{12} = \cos\theta \sin\theta \cos \phi \,, \quad
B_{12} = -\cos\theta \sin\theta \sin \phi \,.
\end{equation}
Then, plugging this into \eqref{K2d}, the Krylov complexity is obtained as
\begin{equation}\label{KWITHP}
C_K = 2 \cos^2(\theta)\sin^2(\theta) \bigg\{\sin^2 [\Delta E \, t] + 2(3+\cos(4\theta))\sin^4 \left[\frac{\Delta E}{2}t\right] \bigg\} \,.
\end{equation}
It is worth noting that the phase factor $\phi$ in \eqref{PSS} does not contribute to the Krylov complexity since it cancels out in $A_{12}^2+B_{12}^2$.

One particularly interesting pure state arises from the mapping
\begin{equation}\label{MAPP2}
\cos(\theta)=\frac{e^{-\frac{\beta E_1}{2}}}{\sqrt{Z_\beta}} \,, \qquad \sin(\theta)= \sqrt{1-\cos^2(\theta)} = \frac{e^{-\frac{\beta E_2}{2}}}{\sqrt{Z_\beta}}  \,, \qquad \phi = 0 \,,
\end{equation}
where $Z_\beta = e^{-{\beta E_1}} + e^{-{\beta E_2}}$. This mapping yields the form of the TFD \eqref{TFDeq} as
\begin{equation}
|\psi\rangle = \frac{e^{-\frac{\beta E_1}{2}}}{\sqrt{Z_\beta}}  |E_1\rangle + \frac{e^{-\frac{\beta E_2}{2}}}{\sqrt{Z_\beta}} |E_2\rangle  \,.
\end{equation}
The corresponding the Krylov complexity from \eqref{KWITHP} is expressed as
\begin{equation}\label{TFDCKTD1}
(\text{TFD}): \quad C_K = \frac{1}{2}\sech^2 \left[\frac{\beta\Delta E}{2}\right] \, \sin^2 \left[\Delta E \,t\right]  \,+\,  2\cosh[\beta \Delta E] \sech^4 \left[\frac{\beta\Delta E}{2}\right] \, \sin^4 \left[\frac{\Delta E}{2}t\right]\,.
\end{equation}
Shortly, we will demonstrate that this Krylov complexity can be directly associated with the Spread complexity of the TFD.\\

\begin{figure}[]
 \centering
     {\includegraphics[width=12.1cm]{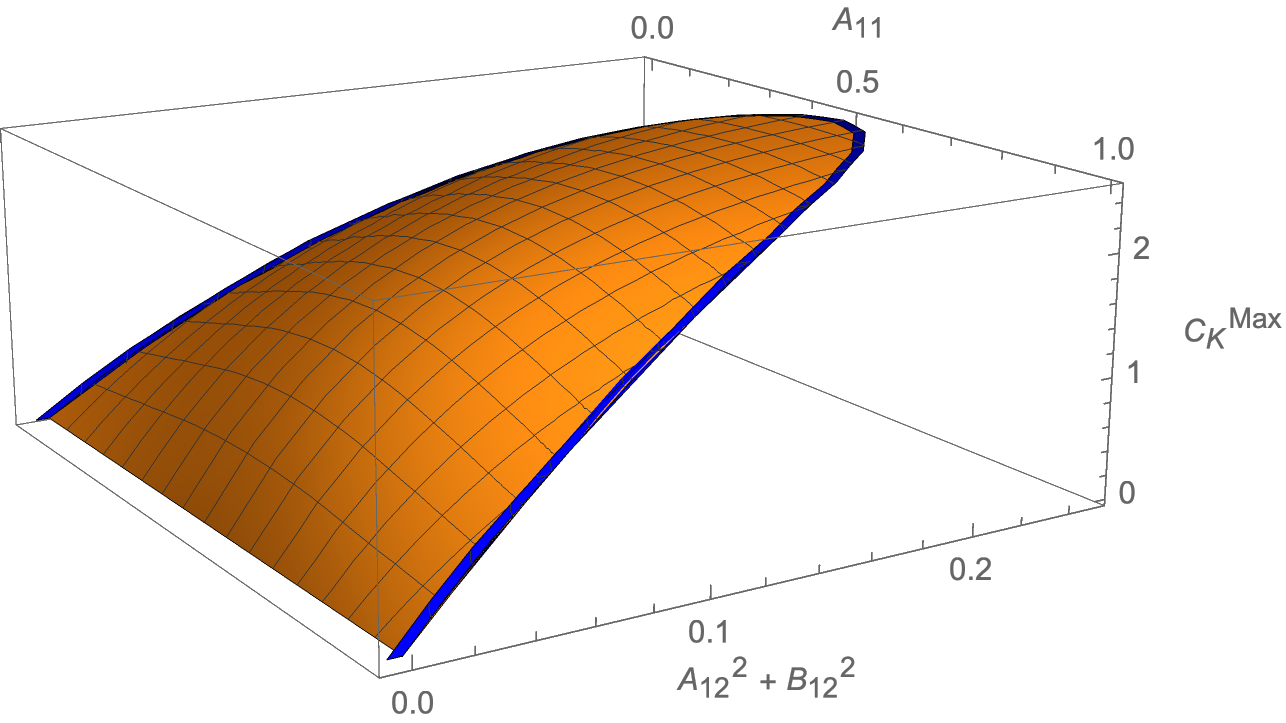} \label{}}
\caption{$C_K^{\text{Max}}$ in \eqref{CKMAX} where $A_{22} = 1-A_{11}$. The yellow surface represents the case for the mixed state, while the blue line corresponds to $C_K^{\text{Max}}$ at $A_{12}^2+B_{12}^2 = A_{11} A_{22}$, indicating a pure state.}\label{MVFIG}
\end{figure}

\paragraph{Krylov complexity of the \textit{mixed} state.}
To investigate the scenario of a mixed state, we examine the maximum value of the Krylov complexity in \eqref{K2d}
\begin{align}\label{CKMAX}
\begin{split}
C_K^{\text{Max}} = \frac{2\left(A_{12}^2+B_{12}^2\right)}{A_{11}^2+A_{22}^2+2(A_{12}^2+B_{12}^2)}\bigg\{1 + \frac{8\left(A_{11}^2+A_{22}^2\right)}{A_{11}^2+A_{22}^2+2(A_{12}^2+B_{12}^2)} \bigg\} \,.
\end{split}
\end{align}
Then, the maximization condition for $C_K^{\text{Max}}$,  i.e., the maximum of $C_K^{\text{Max}}$, is obtained when
\begin{align}\label{}
\begin{split}
A_{12}^2+B_{12}^2  = \frac{14}{9} \left(A_{11}^2+A_{22}^2\right) \,.
\end{split}
\end{align}
However, considering the second condition in \eqref{CONTDH}, it is evident that
\begin{align}\label{}
\begin{split}
A_{12}^2+B_{12}^2 \,\leq\, A_{11} A_{22} \,<\, \frac{14}{9} \left(A_{11}^2+A_{22}^2\right)  \,.
\end{split}
\end{align}
Since $C_K^{\text{Max}}$ in \eqref{CKMAX} is a monotonically increasing function in the region of $A_{12}^2+B_{12}^2 \,\leq\, A_{11} A_{22}$, it is observed that $C_K^{\text{Max}}$ reaches its maximum value when $A_{12}^2+B_{12}^2 \,=\, A_{11} A_{22}$, which is supported by numerical verification in Fig. \ref{MVFIG}, detailed below.

It is imperative to note that $C_K^{\text{Max}}$ in \eqref{CKMAX} is a function of two quantities: 
\begin{align}\label{}
\begin{split}
A_{12}^2+B_{12}^2 \,,\qquad A_{11}^2+A_{22}^2 \,.
\end{split}
\end{align}
However, utilizing the first condition in \eqref{CONTDH}, $A_{22} = 1-A_{11}$, we can express $C_K^{\text{Max}}$ in terms of $A_{12}^2+B_{12}^2$ and $A_{11}$.
In Fig. \ref{MVFIG}, we make a plot of $C_K^{\text{Max}}$ in terms of such variables where the blue line denotes $C_K^{\text{Max}}$ at the upper bound of $A_{12}^2+B_{12}^2$, which is $A_{11} A_{22}$.

This analysis demonstrates that $C_K^{\text{Max}}$, ``at fixed $A_{11}$", attains its highest value when $A_{12}^2+B_{12}^2=A_{11} A_{22}$.\footnote{By examining \eqref{PSCON}, it becomes evident that $A_{12}^2+B_{12}^2=A_{11} A_{22}$ implies a pure state.} As this upper bound implies that the density matrix corresponds to a pure state (the blue line in the figure), we can conclude that the Krylov complexity of the density matrix for a pure state is consistently greater than that of a mixed state (the yellow surface in the figure).\footnote{This observation aligns with the study of circuit complexity~\cite{Caceres:2019pgf}, where it was noted that the complexity of a mixed state has the lowest value, optimized over all possible purifications of a given mixed state.}

One remark is in order. The examination of $C_K^{\text{Max}}$ can be conventionally carried out with a fixed value of $A_{11}$ (rather than $A_{12}^2 + B_{12}^2$), as it is directly linked to fixing the energy eigenvalue of the system. To illustrate, the (average) energy of the system can be determined as follows
\begin{equation}\label{KCMSE1}
\langle E \rangle = \operatorname{Tr}[\rho_0 H] = E_1A_{11} + E_2A_{22}.
\end{equation}
Here, \eqref{ASM1}-\eqref{DM1} have been employed. Additionally, by utilizing \eqref{KCMSE1} along with the first condition in \eqref{CONTDH}, $A_{11} + A_{22} = 1$, we can derive the following expressions
\begin{equation}
\begin{aligned}
A_{11} = \frac{E_2 - \langle E \rangle}{E_2 - E_1} \,, \qquad
A_{22} = \frac{\langle E \rangle - E_1}{E_2 - E_1} \,.
\end{aligned}
\end{equation}
Consequently, it is appropriate to discuss $C_K^{\text{Max}}$ as a function of $A_{12}^2 + B_{12}^2$ when the energy eigenvalues ($E_1, E_2$) and $\langle E \rangle$ of the system are specified, i.e., under the condition of fixed $A_{11}$ (or $A_{22}=1-A_{11}$).\\

\paragraph{Spread complexity of the \textit{pure} state.}
We now investigate one of the pivotal objectives of this manuscript, namely, to establish a connection between Krylov complexity and Spread complexity.
For this purpose, we compute the Spread complexity of the state $\ket{\psi}$ in \eqref{PSS}, allowing for a comparison with the Krylov complexity of the pure state \eqref{KWITHP}.\footnote{It is important to recall that, by definition, Spread complexity can only be evaluated for pure states~\cite{Balasubramanian:2022tpr}.}

Similar to the Krylov complexity, the computation of Lanczos coefficients is performed using the moment method \eqref{GKS}, where the survival amplitude is given by
\begin{equation}
S(t)=\langle\psi(t) | \psi\rangle=\left\langle\psi\left|e^{i H t}\right| \psi\right\rangle=e^{i t E_1}\cos^2\theta+e^{i t E_2}\sin^2\theta \,.
\end{equation}
The non-vanishing Lanczos coefficients are determined as follows
\begin{align}
\begin{split}
a_0=E_1\cos^2\theta+E_2\sin^2\theta \,, \qquad b_1={\cos\theta \, \sin\theta \, \Delta E} \,, \qquad
a_1=\frac{1}{2}\left(E_1+E_2+ \Delta E \cos (2\theta) \right) \,.
\end{split}
\end{align}
Finally, the Spread complexity is obtained
\begin{equation}\label{eqgss}
C_{S} = \sin^2 (2 \theta) \, \sin^2 \left[\frac{\Delta E}{2}t\right] \,.
\end{equation}
A brief overview of the Spread complexity computation~\cite{Balasubramanian:2022tpr} is provided in the appendix \ref{appena}.

Upon inspection, it is observed that both the Krylov complexity of the pure state \eqref{KWITHP}, denoted as $C_K$, and the Spread complexity \eqref{eqgss}, denoted as $C_S$, exhibit oscillatory behavior over time. However, their quantitative characteristics differ.\\

Importantly, the Spread complexity \eqref{eqgss}, when mapped into the TFD using \eqref{MAPP2}, yields more intriguing results:
\begin{equation}\label{SCRMTN2}
(\text{TFD}): \quad  C_{S} = \sech^2 \left[\frac{\beta\Delta E}{2}\right] \, \sin^2 \left[\frac{\Delta E}{2}t\right] \,,
\end{equation}
which can be compared with \eqref{TFDCKTD1}. Notably, when $\beta=0$, the Krylov complexity associated with the density matrix of the TFD is twice the Spread complexity of the TFD, i.e.,
\begin{equation}\label{MCR}
(\text{TFD} \,\,\,\&\,\,\, \beta=0): \quad  C_{K} \,=\, 2C_{S} \,.
\end{equation}
From these observations, we propose the main conclusion for the case of the \textit{two}-dimensional Hilbert space:\\

$\bullet$\,\, For the \textit{maximally entangled} pure state, Krylov complexity becomes exactly twice Spread complexity: \, $C_K \,=\, 2 C_S$ \, (for any $t\geq0$).\\

In addition, beyond the maximally entangled state (such as the TFD at finite $\beta$), Krylov complexity can exhibit qualitatively similar behavior to Spread complexity.
These assertions are the primary highlights of this manuscript, and we provide additional evidence through various examples in the subsequent sections.

\subsection{Qubit state: modular Hamiltonian}
Subsequently, to provide additional evidence for the aforementioned proposal, we study another analytical example within the context of qubit states. In particular we consider two entangled qubits, whose state can be expressed in the Schmidt form as follows
\begin{equation}\label{QSS}
\ket{\psi} = \sum_{j} \sqrt{\lambda_j} \, \ket{j j} \,, \qquad \ket{j j} := \ket{j}_A \ket{j}_{A^c} \,,
\end{equation}
where $\ket{j}$ are the basis vectors. This pure state resides in the Hilbert space $H_{A} \otimes H_{A^{c}}$, where $A$ represents a qubit subsystem and $A^{c}$ its complement. 
In addition, it is noteworthy that one can define the reduced operator $\rho_A$ for the subsystem as
\begin{equation}\label{DMR}
\rho_A = \text{Tr}_{A^c} \big(\ket{\psi}\bra{\psi}\big) =: e^{-H_A} \,,
\end{equation}
where $H_A$ is referred to as the modular Hamiltonian.

Importantly, the state $\ket{\psi}$ in \eqref{QSS} is invariant under the time evolution with the total (modular) Hamiltonian $H_A \otimes 1_{A_c} - 1_{A} \otimes H_{A_c}$~\cite{Haag_1996}. Thus, to investigate the non-trivial time evolution from $\ket{\psi}$, it is necessary to consider the evolution with $H_A \otimes 1_{A_c}$, expressed as:
\begin{equation}\label{QSS2}
\ket{\psi(t)} = e^{-i t \, H_A \otimes 1_{A_c}} \ket{\psi} \,,
\end{equation}
where $\ket{\psi}$ is in \eqref{QSS}.

\paragraph{Qubit state with \textit{two} energy levels.}
Let us commence by examining the Krylov complexity for a qubit state with \textit{two} energy levels, expressed as:
\begin{equation}\label{QSS3}
\ket{\psi} = \sqrt{p} \, \ket{00} + \sqrt{1-p} \, \ket{11} \,, \qquad \rho_A = p \, |0\rangle_{A} {}_{A}\langle0|+(1-p) \, |1\rangle_{A} {}_{A}\langle1| \,.
\end{equation}
Here, $\rho_A$ is obtained by utilizing \eqref{DMR}, and the corresponding $H_A$ can be derived. It is pertinent to note that the modular Spread complexity analysis for the same state has been presented in \cite{Caputa:2023vyr}.

Next, determining the survival amplitude as
\begin{equation}
S(t) = \langle\psi(t) | \psi\rangle=p^{(1-i t)}+(1-p)^{(1-i t)} \,,
\end{equation}
where \eqref{QSS2} is used with the given $\rho_A$ in \eqref{QSS3}, one can employ the moment method \eqref{GKS} to compute the non-vanishing Lanczos coefficients as
\begin{equation}
\begin{aligned}
b_1 = \sqrt{2p(1-p)(\log(1-p)-\log p)^2} \,, \qquad 
b_2 = \sqrt{\left(1-2 p+2 p^2\right)(\log (1-p)-\log p)^2} \,.
\end{aligned}
\end{equation}

Then, solving the recursion equation \eqref{dse} with $b_n$, the Krylov complexity can be expressed as
\begin{equation}\label{Kqubit}
C_K = 2p(1-p) \left\{\sin^2\left[ {t} \log \frac{1-p}{p}\right] + 8\left(1-2 p+2 p^2\right) \sin^4\left[ \frac{t}{2} \log \frac{1-p}{p}\right]\right\} \,,
\end{equation}
which can be compared with \eqref{KWITHP}. This can be contrasted with the Spread complexity of the two energy level state \eqref{QSS3} from \cite{Caputa:2023vyr}, given by:
\begin{equation}\label{Kqubit2}
C_S = 4 p(1-p) \sin^2\left[\frac{t}{2} \log \frac{1-p}{p}\right] \,,
\end{equation}
which can also be comparable with \eqref{eqgss}. A qualitative similarity between $C_K$ and $C_S$ can be observed.

In order to verify our relation, we first need to identify the maximally entangled state from the entanglement entropy:
\begin{equation}
S_{\text{EE}} := -\operatorname{Tr}\left(\rho_A \log \rho_A\right)=-p\log p-(1-p)\log (1-p)  \,,
\end{equation}
where $\rho_A$ is given by \eqref{QSS3}. Notably, $p=1/2$ corresponds to the maximally entangled state. 

Therefore, expanding \eqref{Kqubit}-\eqref{Kqubit2} near $p=1/2$ yields
\begin{align}\label{}
\begin{split}
\frac{C_K}{C_S} \,=\, 2 + 32 \, t^2 \left(p - \frac{1}{2}\right)^4 + \mathcal{O}\left(p - \frac{1}{2}\right)^6\,,
\end{split}
\end{align}
which aligns with the observation below \eqref{MCR}: $C_K = 2C_S$ for the maximally entangled state in the two-dimensional Hilbert space.

\paragraph{Qudit state with \textit{three} energy levels.}
Similarly to the two-energy-level state \eqref{QSS3}, we explore the qudit state with \textit{three} energy levels:
\begin{equation}\label{QSS4}
\ket{\psi} = \sqrt{p} \, |00\rangle + \sqrt{q} \, |11\rangle+\sqrt{1-p-q} \, |22\rangle \,.
\end{equation}
The corresponding density matrix is obtained as
\begin{equation}\label{QSS333}
\begin{aligned}
    \rho_A =p \, |0\rangle_{A} {}_{A}\langle0| \,+\, q \, |1\rangle_{A} {}_{A}\langle1| \,+\, (1-p-q) \, |2\rangle_{A} {}_{A}\langle2|   \,,
\end{aligned}
\end{equation}
along with the survival amplitude
\begin{equation}
S(t)=p^{(1-i t)}+q^{(1-i t)}+(1-p-q)^{(1-i t)} \,.
\end{equation}

We find that, for the case of $q=p$, the Lanczos coefficients can be computed analytically
\begin{equation}
\begin{aligned}
b_1 = \sqrt{4p(1-2p)(\log( 1-2p)-\log p)^2} \,, \qquad
b_2= \sqrt{\left(1-4 p+8 p^2\right)(\log (1-2p)-\log p)^2}
\end{aligned}
\end{equation}
Finally, the resulting Krylov complexity is then given by
\begin{equation}\label{Kqubit3}
C_K = 4p(1-2p) \left\{\sin^2\left[ {t} \log \frac{1-2p}{p}\right] + 8\left(1-4 p+8 p^2\right) \sin^4\left[ \frac{t}{2} \log \frac{1-2p}{p}\right]\right\} \,.
\end{equation}
Additionally, using the method in the appendix, we compute the Spread complexity for \eqref{QSS4}: 
\begin{equation}\label{Kqubit4}
C_S = 8 p(1-2p) \sin ^2\left[ \frac{t}{2} \,{\log \frac{1-2p}{p}} \right] \,.
\end{equation}

Again, to identify the maximally entangled state, we calculate the entanglement entropy
\begin{equation}
S_{\text{EE}} =  -\operatorname{Tr}\left(\rho_A \log \rho_A\right)=-2p\log p-(1-2p)\log (1-2p)   \,,
\end{equation}
where $\rho_A$ is given by \eqref{QSS333} and one can find that $p=1/3$ is the maximally entangled state.

Therefore, expanding \eqref{Kqubit3}-\eqref{Kqubit4} near $p=1/3$ as
\begin{align}\label{}
\begin{split}
\frac{C_K}{C_S} \,=\, 2 + \frac{9}{2} \, t^2 \left(p - \frac{1}{3}\right)^2 + \mathcal{O}\left(p - \frac{1}{3}\right)^3\,,
\end{split}
\end{align}
we can confirm that $C_K = 2C_S$, also holds for qudit states with three energy levels \eqref{QSS4}.

\subsection{Quantum harmonic oscillator}
To conclude this section, we explore the analytical relationship between $C_K$ and $C_S$ for the \textit{infinite}-dimensional Hilbert space.

For this purpose, we focus on the Krylov complexity associated with the density matrix of the TFD for the case of the quantum harmonic oscillator with frequency $\omega$, which has been examined by circuit complexity \cite{ali2020chaos,qu2022chaos,qu2022chaos1}, Krylov complexity \cite{Baek:2022pkt} and spread complexity \cite{Balasubramanian:2022tpr,Huh:2023jxt}. The spectrum and its partition function are
\begin{equation}\label{HOBASIC}
E_n = \omega n \,, \qquad Z_{\beta}=\frac{e^{\beta  \omega}}{e^{\beta \omega}-1} \,,
\end{equation}
where \eqref{TFDeq} is employed for the partition function. Here, we use the same notation for the integer $n$ as used in the study of the Spread complexity in \cite{Balasubramanian:2022tpr}.

Since we are interested in the Krylov complexity from the TFD, the autocorrelation function $C(t)$, computed through \eqref{GKS}, relies on the survival amplitude $S(t)$:
\begin{equation}
S(t) = \frac{1-e^{-\beta \omega}}{1-e^{-(\beta-i t) \omega}} \,.
\end{equation}
This expression is derived from \eqref{SSFFRE}, utilizing \eqref{SFFRE}, and \eqref{HOBASIC}.

\begin{figure}[]
 \centering
     \subfigure[$b_n$ vs. $n$]
     {\includegraphics[width=6.6cm]{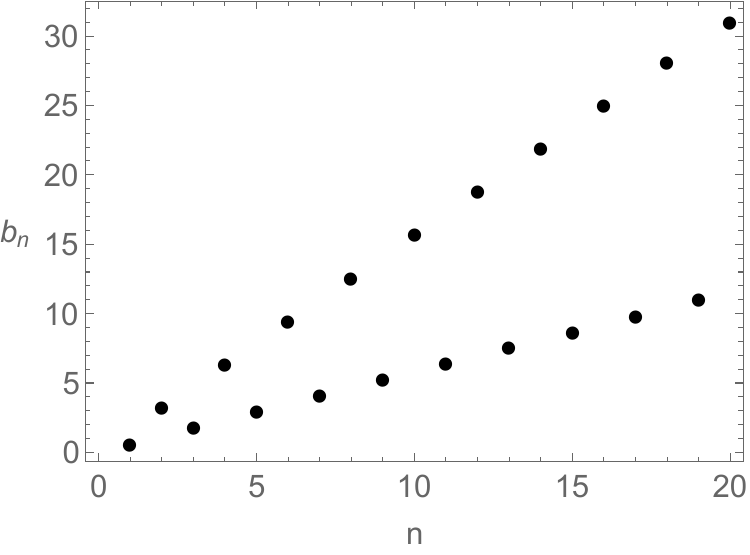} \label{}}
\quad
     \subfigure[$\left(C_K,\, C_S\right)$ vs. $t$]
     {\includegraphics[width=6.6cm]{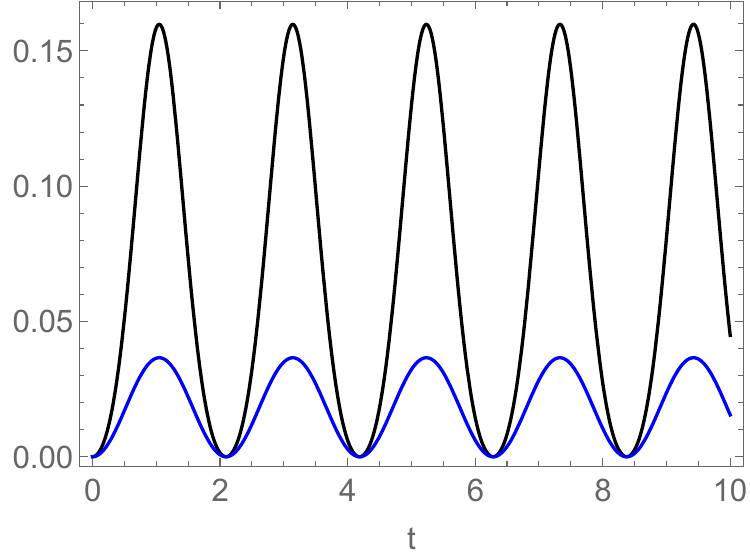} \label{}}
\caption{The numerical results for quantum harmonic oscillators with $(\beta,\,\omega)=(\pi/2,\,3)$. In the right panel, $C_K$ is depicted as the black line, while $C_S$ is with the blue line.}\label{HOFIG1}
\end{figure}

Then, utilizing the moment method, we determine the non-vanishing Lanczos coefficients $b_n$ for the quantum harmonic oscillator as
\begin{equation}\label{LZCHO}
\begin{aligned}
b_1 = \sqrt{\frac{1}{2} \omega^2 \csch^2 \frac{\beta \omega}{2}}\,, \qquad
b_2 = \sqrt{\frac{1}{2} \omega^2 \csch^2 \frac{\beta \omega}{2}(4+\cosh\beta\omega)} \,, \qquad
b_3 \,,\,\, \cdots \,.
\end{aligned}
\end{equation}
Nevertheless, we numerically find that $b_n$ oscillates in small $n$ regime and remains finite even with increasing values of $n$. 
As an illustration, we present a representative plot with $(\beta,\,\omega)=(\pi/2,\,3)$ in the left panel in Fig. \ref{HOFIG1}.
Note that the non-vanishing $b_n$ in larger $n$ is also observed for the case of the Spread complexity in~\cite{Balasubramanian:2022tpr}.

Finally, by solving the recursion equation \eqref{dse} using the numerically obtained $b_n$, we determine the Krylov complexity of the quantum harmonic oscillator. The results are presented as the black line in the right panel of Fig. \ref{HOFIG1}. Notably, the qualitative behavior of $C_K$ is akin to the Spread complexity described in \cite{Balasubramanian:2022tpr}, given by
\begin{equation}\label{HORESULT}
C_S = \frac{\sin^2 (\omega t /2)}{\sinh^2 (\beta \omega /2)} \,.
\end{equation}
This Spread complexity is depicted as the blue line in the right panel of Fig. \ref{HOFIG1}.\\

\paragraph{Inverted quantum harmonic oscillator.}
One might contemplate whether an exact duality between $C_K$ and $C_S$, as proposed in \eqref{MCR}, can be identified even for the infinite-dimensional Hilbert space.
However, unlike the two-dimensional Hilbert space example, setting $\beta$ to zero for the harmonic oscillators is not feasible, as indicated in \eqref{HOBASIC}. Moreover, the generic ``analytic" expressions for both $b_n$ and $C_K$ are not readily available in this case.

\begin{figure}[]
 \centering
     {\includegraphics[width=7.1cm]{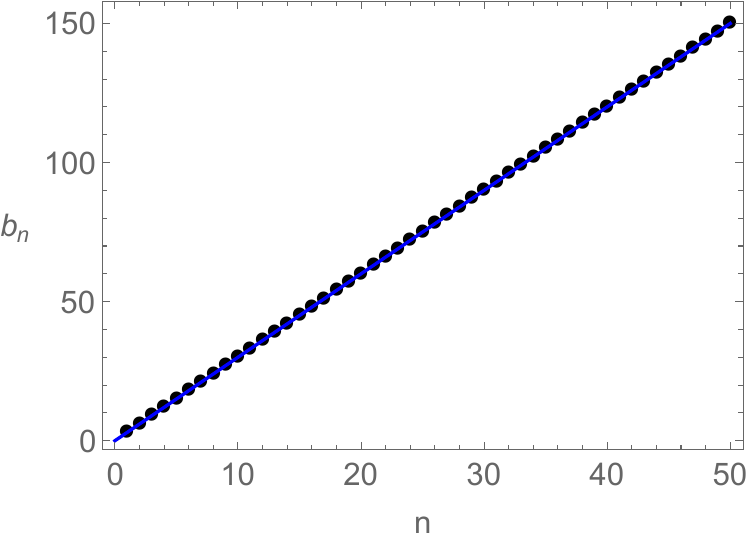} \label{}}
\caption{Lanczos coefficients for the inverted harmonic oscillator with $\Omega=3$. The black dots represent the numerical data and the blue solid line is the analytic result \eqref{IHOANARE}.}\label{LCHOFIG}
\end{figure}

Nevertheless, some progress can been made in the case of the \textit{inverted} harmonic oscillator as well, where $\omega \rightarrow-i \Omega$. Specifically, we have found that the Lanczos coefficients given in \eqref{LZCHO} can be expressed as:
\begin{equation}\label{LZCHO2}
\begin{aligned}
b_1 = \sqrt{\frac{1}{2} \Omega^2 \csc^2 \frac{\beta \Omega}{2}}\,, \qquad
b_2 = \sqrt{\frac{1}{2} \Omega^2 \csc^2 \frac{\beta \Omega}{2}(4+\cos\beta\Omega)} \,, \qquad
b_3 \,,\,\, \cdots \,.
\end{aligned}
\end{equation}
Especially, when considering $\cos\beta\Omega=0$, which establishes the relationship between $\beta$ and the frequency as
\begin{equation}\label{IMPCON}
\Omega = \left(\frac{1}{2} + \ell \right)\frac{\pi}{\beta} \,, \qquad \ell = 0\,,1\,,2\,,\cdots \,,
\end{equation}
the Lanczos coefficients can be found analytically
\begin{equation}\label{IHOANARE}
b_n = \Omega \, n \,,
\end{equation}
which is verified in Fig. \ref{LCHOFIG}.
Given this analytic Lanczos coefficients, we solve the equation \eqref{dse} and obtain the analytic expression for $\varphi_n(t)$
\begin{equation}
\varphi_n(t) = \sqrt{\frac{\Gamma(1+n)}{n ! }} \frac{\tanh ^n(\Omega t)}{\cosh(\Omega t)} \,.
\end{equation}
This provides the Krylov complexity of the inverted harmonic oscillator as
\begin{equation}\label{HOMR1}
C_K = \sinh ^2(\Omega t) \,.
\end{equation}
Having the analytic expression for the Krylov complexity, finally we can quantitatively compare it with the Spread complexity for the inverted harmonic oscillator:
\begin{equation}\label{HOMR2}
C_S = \frac{\sinh^2 (\Omega t /2)}{\sin^2 (\beta \Omega /2)} \,,
\end{equation}
which is obtained from \eqref{HORESULT} after substituting $\omega \rightarrow-i \Omega$.

Comparing \eqref{HOMR1} with \eqref{HOMR2} alongside \eqref{IMPCON}, we derive 
\begin{equation}\label{HORESFINAL}
C_K = \sinh ^2(\Omega t) \,, \qquad
C_S = 2 \sinh^2 (\Omega t /2) \,,
\end{equation}
and confirm that $C_K = 2C_{S}$ in \eqref{NER66} is valid in the early time regime.\\

Recall that for quantum harmonic oscillators, setting $\beta=0$ for the TFD is not well-defined (divergent). Hence we cannot claim that $C_K = 2C_{S}$ for the maximally entangled state in the quantum harmonic oscillator. 
Nevertheless, we also observe that the derived complexities of the inverted harmonic oscillator, \eqref{HORESFINAL}, in the late time regime ($t\gg1$) scale as
\begin{equation}\label{}
C_K \approx \frac{1}{4}e^{2 \Omega t} \,, \qquad
C_S \approx \frac{1}{2}e^{\Omega t} \,,
\end{equation}
leading to the relation $C_{K} \,=\, C_{S}^2$ that will be worth exploring in other examples in the future.
   
\section{Random matrix theories}\label{SECTION5}

\subsection{Average Krylov complexity}
In this section, we continue exploring the relationship between $C_K$ and $C_S$ within the context of a \textit{higher} (yet finite)-dimensional Hilbert space. In particular, we consider \textit{random matrix theories} (RMT) for this purpose. It is worth noting that the analysis of $C_S$ for RMT can be found in \cite{Balasubramanian:2022tpr} (and \cite{Balasubramanian:2022dnj,Balasubramanian:2023kwd,Tang:2023ocr}), allowing us to draw comparisons between our findings in $C_K$ and $C_S$ presented in that literature.

RMT serves not only as a suitable model for our purpose, but also holds significance in physics, particularly in identifying potentially universal features within general chaotic systems. A foundational conjecture suggests that the detailed structure of the spectrum of a quantum chaotic Hamiltonian can be effectively approximated by the statistical behaviors of random matrices~\cite{Dyson_1962,Bohigas:1984aa}: see also~\cite{Guhr:1997ve,MEHTA1960395,GAUDIN1961447,Dyson_1962III} for the review.

Within the scope of this manuscript, we consider all three universality classes: Gaussian Unitary Ensemble (GUE), Gaussian Orthogonal Ensemble (GOE), and Gaussian Symplectic Ensemble (GSE). These terms denote specific ensembles of random matrices. Each ensemble aligns with a distinct symmetry class of random matrices with Gaussian measures, respectively, as
\begin{equation}\label{CLASSRMT}
\frac{1}{Z_{\mathrm{GUE}}} e^{-\frac{N}{2 \mathcal{E}^2} \operatorname{Tr}\left(H^2\right)} \,, \qquad 
\frac{1}{Z_{\mathrm{GOE}}} e^{-\frac{N}{4 \mathcal{E}^2} \operatorname{Tr}\left(H^2\right)} \,, \qquad
\frac{1}{Z_{\operatorname{GSE}}} e^{-\frac{N}{\mathcal{E}^2}\operatorname{Tr}\left(H^2\right)} \,,
\end{equation}
where $H$ is the $N \times N$ matrix, $\mathcal{E}$ is the scaling factor which can be chosen $\mathcal{E}=1$, $Z_{\mathrm{GUE \,\,or\,\, GOE \,\,or\,\, GSE}}$ is a normalization constant. Note that here $N$ denotes the dimension of the Hilbert space.\\

\paragraph{Average Krylov complexity.}
It is worth to note that in RMT, the Hamiltonian can be evaluated by the random values within the classes in \eqref{CLASSRMT}. Consequently, the resulting $C_K$ from RMT is subject to this inherent ``randomness". Thus, to mitigate this randomness and uncover the essential structure in $C_K$, this paper introduces the average Krylov complexity, denoted as
\begin{equation}\label{MKC}
\begin{aligned}
\bar{C}_K := \int_{-\infty}^{\infty} \dd E_i \,\, \big[ \rho\left(E_i\right) \, C_K\left(E_i\right) \big] \,,
\end{aligned}
\end{equation}
where the probability distribution of eigenvalues for a $N \times N$ Gaussian matrix, $\rho\left(E_i\right)$, is given by
\begin{equation}\label{MKC2}
\rho\left(E_i\right)=\frac{1}{\mathcal{Z}} \, e^{-\frac{N\mathcal{D}}{4} \sum_{i=1}^{N} E_i^2} \, \prod_{j<k} | E_j - E_k |^{\mathcal{D}} \,,
\end{equation}
where $\mathcal{Z}$ is the normalization constant, and $\mathcal{D} = 1,\,2,\,4$ is referred to as the Dyson index, corresponding to GOE, GUE, and GSE, respectively. Notably, $\rho$ here does not represent the density matrix.

Three remarks are in order. Firstly, \eqref{MKC} is equivalent to the arithmetic mean, i.e.,
\begin{equation}\label{CKMET44}
\begin{aligned}
\bar{C}_K = \frac{1}{\mathcal{N}_{\text{it}}} \sum_{i=1}^{\mathcal{N}_{\text{it}}} C_K^{(i)} \,,
\end{aligned}
\end{equation}
where $C_K^{(i)}$ denotes the Krylov complexity at the $i$-th result in the total $\mathcal{N}_{\text{it}}$ iterations. Secondly, for a sufficiently large dimension of the Hilbert space (or size of the matrix), $\bar{C}_K \approx C_K$, implying no need for averaging.\footnote{We have confirmed this by the numerical calculation.} Thirdly, the average Spread complexity can also be defined similarly: $\bar{C}_S := \int_{-\infty}^{\infty} \dd E_i \,\, \big[ \rho\left(E_i\right) \, C_S\left(E_i\right) \big]$.\\

The primary objective of this section is then to explore the relationship between $\bar{C}_K$ and $\bar{C}_S$ specifically in the context of the TFD when $N\geq2$. It is important to note that, for the TFD case, the autocorrelation function \eqref{autocorr} becomes determinable once the energy eigenvalue $E_n$ is acquired. Consequently, by employing the moment method outlined in \eqref{OCF}-\eqref{OCF2}, the coefficients $b_n$ can be determined. Subsequently, one can solve the recursion equation \eqref{dse} to calculate $C_K$. In simpler terms, the ``sole" input required to compute $C_K$ in this section is to ascertain the energy eigenvalue $E_n$ for the random matrix \eqref{CLASSRMT}.\footnote{The energy eigenvalues for each class can be easily evaluated using functions within \textit{Mathematica}.}

\subsection{Krylov complexity in random matrix theories}

\subsubsection{Analytic results ($N=2$)}
In general, numerical methods are typically employed to compute \eqref{MKC}-\eqref{MKC2}. However, for the specific case of $N=2$, analytical computation becomes feasible.

For the RMT with $N=2$, applicable to our two-dimensional Hilbert space results, both $C_K$ in \eqref{TFDCKTD1} and $C_S$ in \eqref{SCRMTN2} can be determined analytically:
\begin{align}\label{N2RESULT1}
\begin{split}
C_K &= \frac{1}{2}\sech^2 \left[\frac{\beta\Delta E}{2}\right] \, \sin^2 \left[\Delta E \,t\right]  \,+\,  2\cosh[\beta \Delta E] \sech^4 \left[\frac{\beta\Delta E}{2}\right] \, \sin^4 \left[\frac{\Delta E}{2}t\right] \,, \\
C_{S} &= \sech^2 \left[\frac{\beta\Delta E}{2}\right] \, \sin^2 \left[\frac{\Delta E}{2}t\right] \,,
\end{split}
\end{align}
where $\Delta E := E_2-E_1$. Recall that $C_K = 2 C_S$ holds for the maximally entangled state \eqref{MCR} throughout all time range, i.e., also does for the RMT when $N=2$.

Furthermore, when $N=2$, \eqref{MKC} can be further expressed as follows:
\begin{align}\label{N2RESULT2}
\begin{split}
\bar{C}_K &:= \int_{-\infty}^{\infty} \dd E_1 \dd E_2 \,\, \big[ \rho\left(E_1, E_2\right) \, C_K\left(E_1, E_2\right) \big]  \\
&= \int_{-\infty}^{\infty} \dd E_1 \dd E_2 \,\, \left[ 
\int_{0}^{\infty} \dd \Delta E \,\, \delta\left(\Delta E-\abs{E_{2}-E_{1}}\right)\right]  \big[\rho\left(E_1, E_2\right) \, C_K\left(E_1, E_2\right) \big] \\
&= \int_{0}^{\infty} \dd \Delta E \,\, \left[ 
\int_{-\infty}^{\infty} \dd E_1 \dd E_2 \,\, \delta\left(\Delta E-\abs{E_{2}-E_{1}}\right) \, \rho\left(E_1, E_2\right) \right] C_K\left(E_1, E_2\right)  \\
&= \int_{0}^{\infty} \dd \Delta E \,\, P(\Delta E) \,\, C_K\left(\Delta E \right) \,,
\end{split}
\end{align}
where $P(\Delta E)$ is the probability distribution function of the level spacing $\Delta E$. This distribution function plays a crucial role in the subsequent subsection. 

For $N=2$, $P(\Delta E)$ can be computed analytically for each class
\begin{align}\label{N2RESULT3}
\begin{split}
P(\Delta E) =  
\begin{cases}
      \,\sqrt{\frac{2}{\pi}}e^{-\frac{(\Delta E)^2}{2}} (\Delta E)^2  & \text{for GUE \,,}  \\
      \, \frac{\Delta E}{2} e^{-\frac{(\Delta E)^2}{4}}  & \text{for GOE \,,} \\
      \, \frac{8}{3\sqrt{\pi}}(\Delta E)^{4} e^{-(\Delta E)^2}  & \text{for GSE \,.} 
\end{cases} 
\end{split}
\end{align}
By plugging \eqref{N2RESULT3} and \eqref{N2RESULT1} into \eqref{N2RESULT2}, we obtain the analytical expression for $\bar{C}_K$ when $\beta=0$ (i.e., maximally entangled state):
\begin{align}\label{N2RESULT4}
\begin{split}
\bar{C}_K =  
\begin{cases}
      \, 1+e^{-\frac{t^2}{2}}(t^2-1)  & \text{for GUE \,,}  \\
      \, 2t \, D(t)  & \text{for GOE \,,} \\
      \, 1+e^{-\frac{t^2}{4}}\left(t^2-\frac{t^4}{12}-1\right)  & \text{for GSE \,.} 
\end{cases} 
\end{split}
\end{align}
where $D(t)=e^{-t^2} \int_0^t e^{x^2} d x$ is the Dawson function. 

\begin{figure}[]
 \centering
     \subfigure[$\bar{C}_K$ vs. $t$ at $\beta=0$]
     {\includegraphics[width=7.1cm]{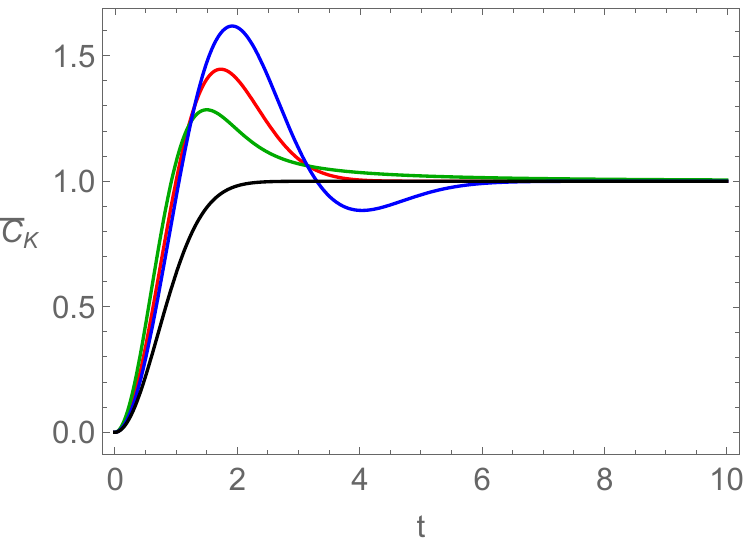} \label{}}
     \subfigure[$\bar{C}_K$ vs. $t$ at $\beta\neq0$]
     {\includegraphics[width=7.1cm]{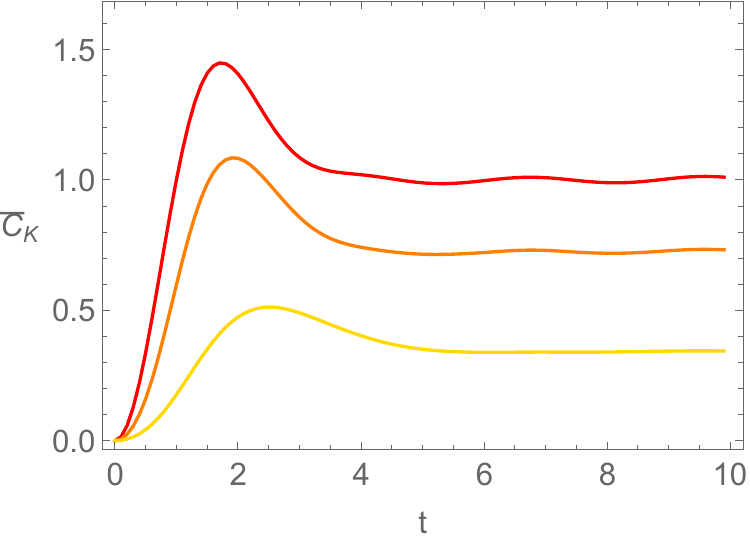} \label{}}
\caption{The average Krylov complexity $\bar{C}_K$ is illustrated in the \textit{left} panel, showcasing its values for various classes: GUE (red), GOE (green), GSE (blue), and IID (black), specifically when $\beta=0$, as per \eqref{N2RESULT4}. On the \textit{right} panel, the $\beta$ dependence of $\bar{C}_K$ is displayed for the case of GUE, with $\beta=0,1,2$ (red, orange, yellow). This $\beta$-dependent data is computed numerically with $\mathcal{N}_{\text{it}}=10^4$. Notably, the consistency between the red (analytical) data on the left and the red (numerical) data on the right panel is evident.}\label{RMTN2fig}
\end{figure}

To contrast with the non-chaotic model, we also examined Independent and Identically Distributed random variables (IID) from the standard normal distribution, where the probability distribution of spacing is given by
\begin{align}\label{eIID}
\begin{split}
P(\Delta E) = \frac{e^{-\frac{(\Delta E)^2}{4}}}{\sqrt{\pi}} \,, \qquad \bar{C}_K = 1-e^{-t^2} \quad\,\,\, \text{for IID} \,.
\end{split}
\end{align}
In Fig. \ref{RMTN2fig}, we display the average Krylov complexity when $N=2$.
In the left panel, we observe that the Krylov complexity of the RMT aligns with the Spread complexity of the RMT, as presented in \cite{Balasubramanian:2022tpr}. In other words, both complexities exhibit a characteristic peak for quantum chaotic cases (GUE, GOE, GSE), in contrast to the integrable cases (IID). This aligns with our analytical proof establishing that $C_K = 2 C_S$. Furthermore, in the right panel, the $\beta$-dependence is also consistently observed: complexities are suppressed by the finite $\beta$.

\subsubsection{Numerical results ($N>2$)}
We proceed with the exploration of the average Krylov complexity and Spread complexity for the case when $N>2$. In this scenario, numerical methods are employed, as outlined below \eqref{CKMET44}. Essentially, once we compute the energy eigenvalues, $\bar{C}_{K}$ can be evaluated.

\begin{figure}[]
 \centering
     \subfigure[$\bar{C}_K$ at $N=3$]
     {\includegraphics[width=5.4cm]{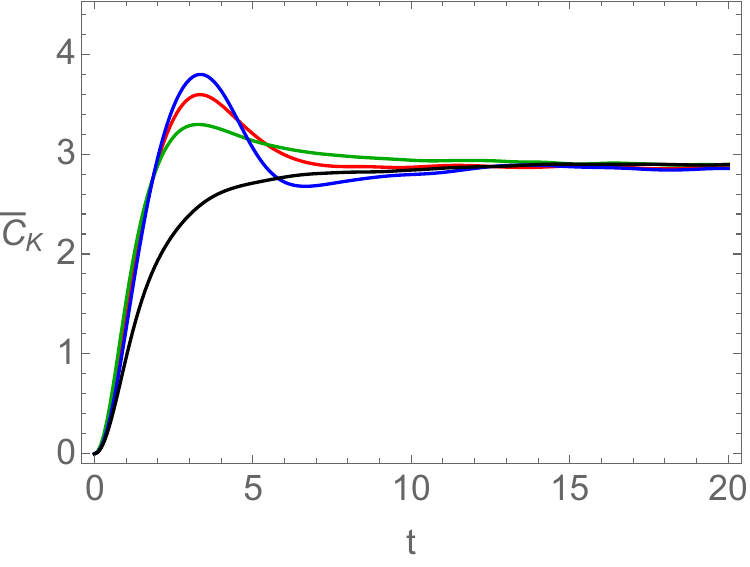} \label{RMTN348figa}}
\qquad
     \subfigure[$3\bar{C}_S$ at $N=3$]
     {\includegraphics[width=5.4cm]{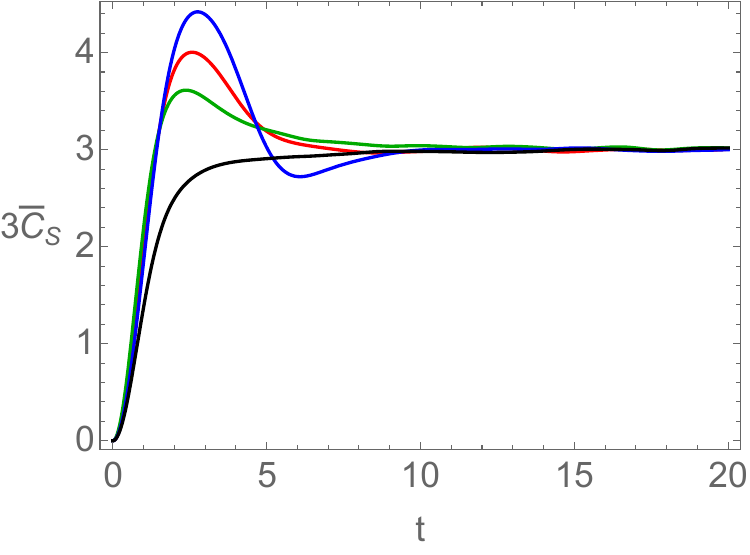} \label{RMTN348figd}}
     
     \subfigure[$\bar{C}_K$ at $N=4$]
     {\includegraphics[width=5.4cm]{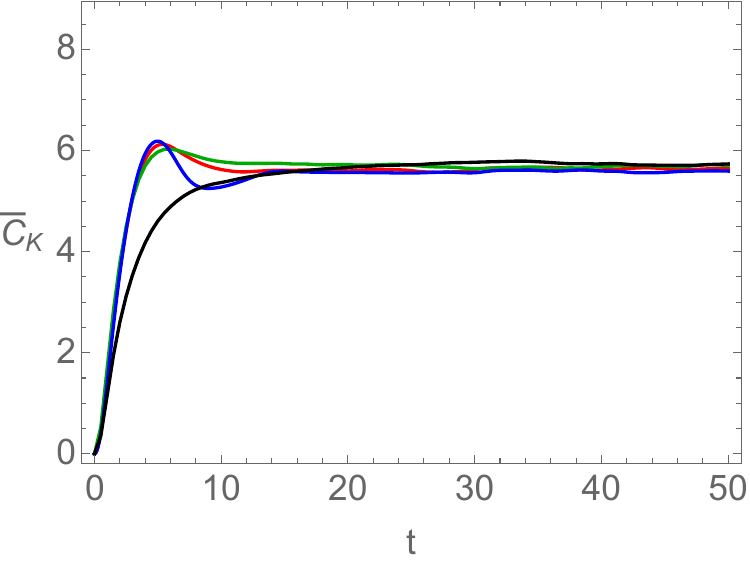} \label{RMTN348figb}}
\qquad
     \subfigure[$4\bar{C}_S$ at $N=4$]
     {\includegraphics[width=5.4cm]{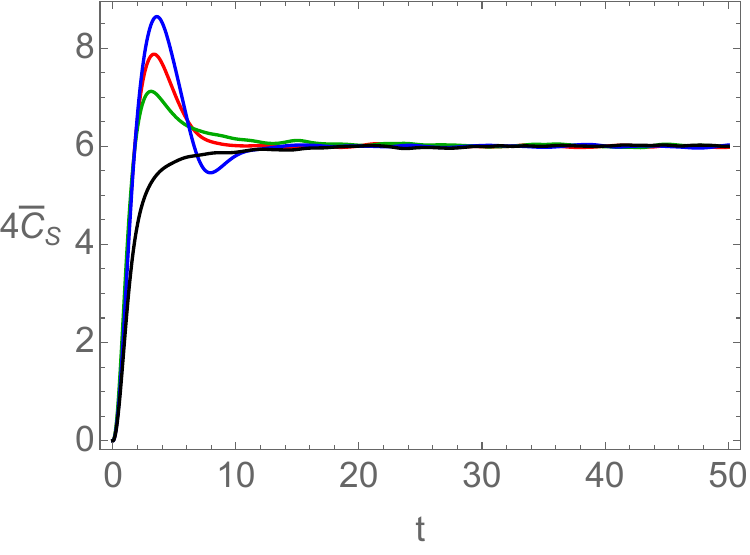} \label{RMTN348fige}}
     
     \subfigure[$\bar{C}_K$ at $N=8$]
     {\includegraphics[width=5.4cm]{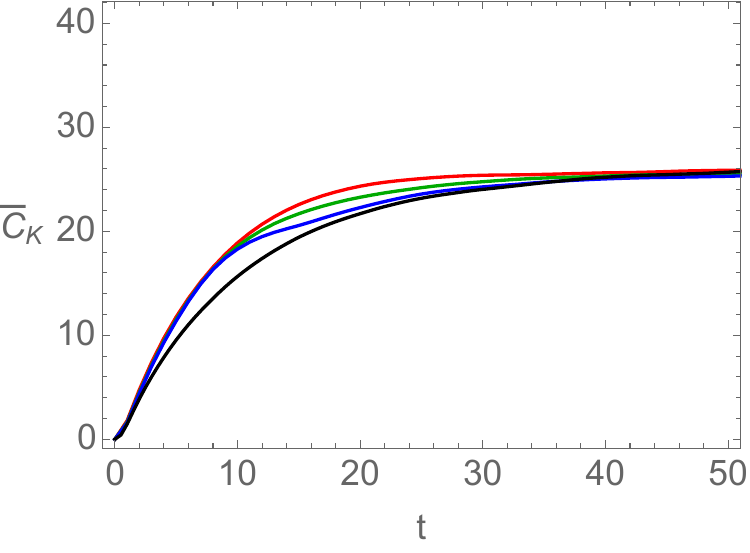} \label{RMTN348figc}}
\qquad
     \subfigure[$8\bar{C}_S$ at $N=8$]
     {\includegraphics[width=5.4cm]{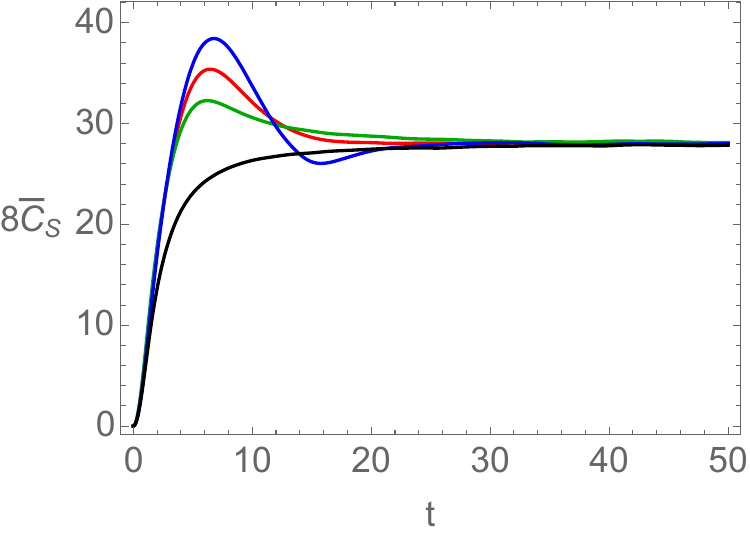} \label{RMTN348figf}}

     \subfigure[$\bar{C}_K$ at $N=10$]
     {\includegraphics[width=5.4cm]{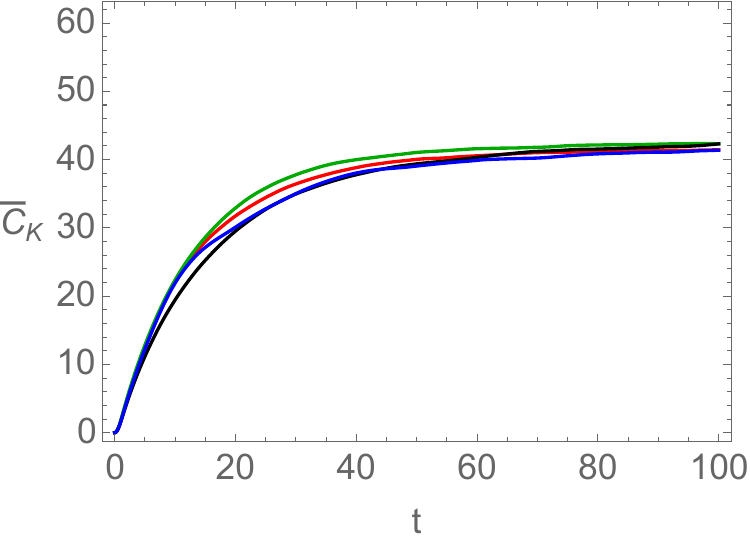} \label{RMTN10figa}}
\qquad
     \subfigure[$10\bar{C}_S$ at $N=10$]
     {\includegraphics[width=5.4cm]{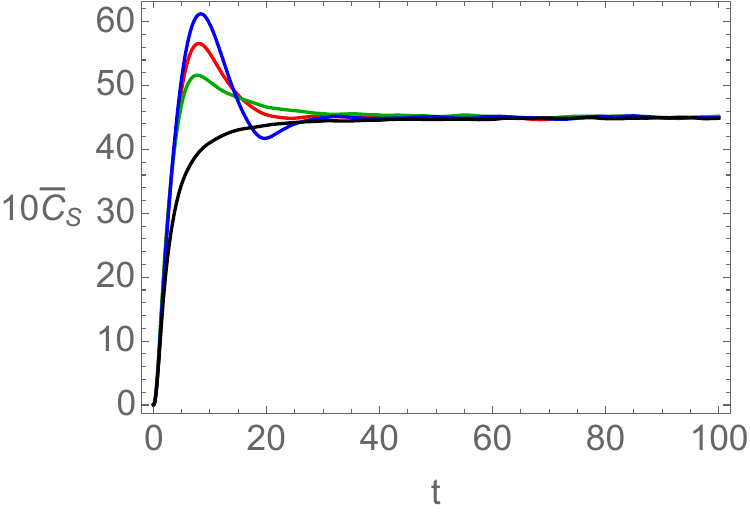} \label{RMTN10figb}}
\caption{Average Krylov and Spread complexities at $\beta=0$. In the \textit{left} column, $\bar{C}_K$ is presented for different dimensions: $N=3$ (a), $N=4$ (c), $N=8$ (e), and $N=10$ (g). The \textit{right} column depict $N \bar{C}_S$ when: $N=3$ (b), $N=4$ (d), $N=8$ (f), and $N=10$ (h). In all figures, various colors denote the different classes: GUE (red), GOE (green), GSE (blue), and IID (black).}\label{RMTN348fig}
\end{figure}

In Fig. \ref{RMTN348fig}, we present both $\bar{C}_{K}$ and $\bar{C}_{S}$ for various values of $N$, with a specific focus on $N=3,\,4,\,8$ and $10$ when $\beta=0$ in this paper.\footnote{$\bar{C}_S$ is numerically computed with $\mathcal{N}_{\text{it}}=10^4$ for $N=3,\,4,\,8$ and $\mathcal{N}_{\text{it}}=5\times10^3$ for $N=10$. On the other hand, we compute $\bar{C}_K$ with $\mathcal{N}_{\text{it}}=10^4,\,5\times10^3,\,10^3,\,3\times10^2$ for $N=3,\,4,\,8,\,10$, respectively.}
We observe two main features for higher-dimensional Hilbert space.
\begin{itemize}
\item{The peak in $\bar{C}_K$ diminishes, unlike $\bar{C}_S$ (e.g., see the left column in Fig. \ref{RMTN348fig}).}
\item{In the intermediate time regime, $\bar{C}_K$ appears to deviate from its similarity with $\bar{C}_S$ (e.g., see Fig. \ref{RMTN10figa} vs. Fig. \ref{RMTN10figb}).}
\end{itemize}
In the subsequent subsection, we provide a potential origin of the difference (observed in the intermediate time regime) between $\bar{C}_K$ and $\bar{C}_S$ through an analysis of the energy spectrum.

Two remarks are in order. Firstly, across all figures, it is evident that the initial growth rate of $\bar{C}_K$ (and $\bar{C}_S$) in the chaotic cases (denoted by red, green, blue) surpasses that of the integrable case (denoted in black). Secondly, the value of the late-time plateau is exactly the same for integrable and RMT cases. 

Additionally, as a by-product, we discover a potential connection between the two complexities, at least in late times ($t\gg1$): $\bar{C}_K\approx N \bar{C}_S$. Combining the results from both two (and infinite)-dimensional Hilbert spaces in the preceding sections with the finite-dimensional outcomes here, we also observe a relationship for the Krylov complexity of the maximally entangled state as:
\begin{align}\label{}
\begin{split}
(\text{when} \,\, t\gg1): \quad 
C_K \cong  
\begin{cases}
      \, N C_S  & \text{for $N$-dimensional Hilbert space \,,}  \\
      \, C_S^{2}  & \text{for infinite-dimensional Hilbert space \,.}
\end{cases} 
\end{split}
\end{align}
Therefore, for the maximally entangled pure state in a (finite) $N$-Hilbert dimension, we conclude that
\begin{itemize}
\centering
\item{$C_K \,\cong\, 2 C_S$ \,\,($t\ll1$) \quad\,$\xrightarrow{\makebox[1.2cm]{$t$}}$\quad\,  $C_K \,\cong\, N C_S$ \,\,($t\gg1$),}
\end{itemize}
where $C_K \,=\, 2 C_S$ holds within all time window when $N=2$. It is also important to recall that $C_K \,=\, 2 C_S$ in early times is universally valid for general pure state (see below \eqref{NER66}).

\subsection{Energy spectrum analysis}
In this section, we consider one possible scenario for understanding the difference between the Liouvillian and the Hamiltonian dynamics in Krylov space during intermediate times. Our analysis closely draws from the conjecture concerning the dimension of the Krylov space ($D_K$), as introduced in \cite{Rabinovici:2020ryf}, which we elaborate on below.\footnote{See \cite{Balasubramanian:2023kwd} for more discussion on its validity and potential subtleties.}

\paragraph{The dimension of the Krylov space ($D_K$) and quantum chaos.}

We begin by considering the eigen-equation of the Liouvillian superoperator $\mathcal{L}$ \eqref{KBEOM1},
\begin{equation}\label{LOU11}
\mathcal{L} | \omega_{a b}) \,=\, \omega_{a b} | \omega_{a b}) \,,
\end{equation}
where
\begin{equation}\label{BSIS11}
\omega_{a b} := E_a - E_b \,, \qquad   | \omega_{a b}) := \left|E_a\right\rangle \left\langle E_b\right| \,.
\end{equation}
Here, $E_a$ and $\left|E_a\right\rangle$ denote the energy eigenvalues and eigenstates of the given Hamiltonian.
Let us consider an arbitrary operator $O$, expressed in the basis \eqref{BSIS11} as
\begin{equation}\label{nested0}
| {O}) = \sum_{a=1}^{D_H} O_{a a} \, | \omega_{a a})+\sum_{\substack{a, b=1 \\ a \neq b}}^{D_H} O_{a b} \,  | \omega_{a b}) \,,
\end{equation}
where $D_H$ is the dimension of the Hilbert space. Acting with $\mathcal{L}^n$ on $| {O})$ yields
\begin{align}\label{nested}
\begin{split}
\mathcal{L}^n | {O}) &= \delta_{n 0} \sum_{a=1}^{D_H} O_{a a} \, | \omega_{a a})+\sum_{\substack{a, b=1 \\ a \neq b}}^{D_H} O_{a b} \, \omega_{a b}^n  \, | \omega_{a b}) \\
&:= P_{\text{diagonal}} \,+\, P_{\text{off-diagonal}} \,,
\end{split}
\end{align}
which generates the Krylov subspace (analogous to \eqref{KBEOM2} for the case of the density matrix). The dimension of this subspace is denoted as $D_K$, and it satisfies: 
\begin{equation}\label{bound}
1 \leq D_K \leq D_H^2 - D_H + 1\,.
\end{equation}
The terms $D_H^2 - D_H$ can be obtained from the off-diagonal term in \eqref{nested}. They represent the number of the off-diagonal components of a $D_H \times D_H$ matrix. Additionally, we have $+1$ from the diagonal term, which appear both in the left and right side of \eqref{bound}.

In \cite{Rabinovici:2020ryf}, it was conjectured that quantum chaos implies large values of $D_K$. In particular,
\begin{align}\label{MRDDS}
\begin{split}
(\text{Chaotic system}): \quad D_K &\cong  D_H^2 - D_H + 1 \,, \\
(\text{Non-chaotic system}): \quad D_K &\ll  D_H^2 - D_H + 1 \,.
\end{split}
\end{align}
Thus, if true, it is crucial to determine when and how $D_K$ can attain its upper bound.

\paragraph{Reduced $D_K$ by type-II degeneracies.}

Following \cite{Rabinovici:2020ryf} we now suppose that for some set of $m$ pairs of indices $I$, the eigenvalue of $\mathcal{L}$ is degenerate, i.e.,
\begin{equation}
\mathcal{L} | \omega_{a b}) \,=\, \omega | \omega_{a b}) \,,\qquad \forall(a, b) \in I\,.
\end{equation}
In this case, we can express the off-diagonal part $P_{\text{off-diagonal}}$ in \eqref{nested} as follows:
\begin{equation}\label{DEGEQ1}
P_{\text{off-diagonal}} = \omega^n \sum_{(a, b) \in I} O_{a b} \,| \omega_{a b}) + \sum_{(a, b) \notin I} O_{a b} \, \omega_{a b}^n \, | \omega_{a b}) \,,
\end{equation}
where 
\begin{equation}\label{degener}
\omega_{ab}=\omega\,, \qquad  \forall(a, b) \in I \,.
\end{equation} 
In terms of the spectrum, \eqref{degener} implies that there are $m$ pairs of energies with the same gap, e.g., $E_a - E_b = E_c - E_d = \omega$, for some $(a, b)$ and $(c, d)$. We refer to this property as type-II degeneracy.\footnote{In \cite{Rabinovici:2020ryf}, it is referred to simply as ``degeneracy.'' We append ``type-II'' to avoid any confusion with the conventional definition of energy degeneracy, where $E_a=E_b$.} Importantly, when \eqref{degener} occurs, \cite{Rabinovici:2020ryf} showed that the upper bound of $D_K$ is reduced by the number of type-II degeneracies $m$, such that
\begin{equation}\label{bound2}
1 \leq D_K \leq D_H^2 - D_H+1-m \,.
\end{equation}
The reduction in $D_K$ may play an important role in understanding the emergence of chaotic characteristics within Krylov complexity. Notably, even amidst chaotic behavior in the system's Hamiltonian, the condition $D_K \cong D_H^2 - D_H + 1$ in \eqref{MRDDS} may not be fulfilled due to a large amount of type-II degeneracy. In such cases, the chaotic attributes stemming from Krylov complexity might not be distinctly apparent. 

\paragraph{Generalized type-II degeneracy in RMT.} 
Given the preceding discussion, it is reasonable to speculate that within the framework of random matrix theory, type-II degeneracies might serve as a potential source of distinction between $C_K$ ($N > 2$) and $C_S$ within the intermediate time regime.

\begin{figure}[]
 \centering
     {\includegraphics[width=7.1cm]{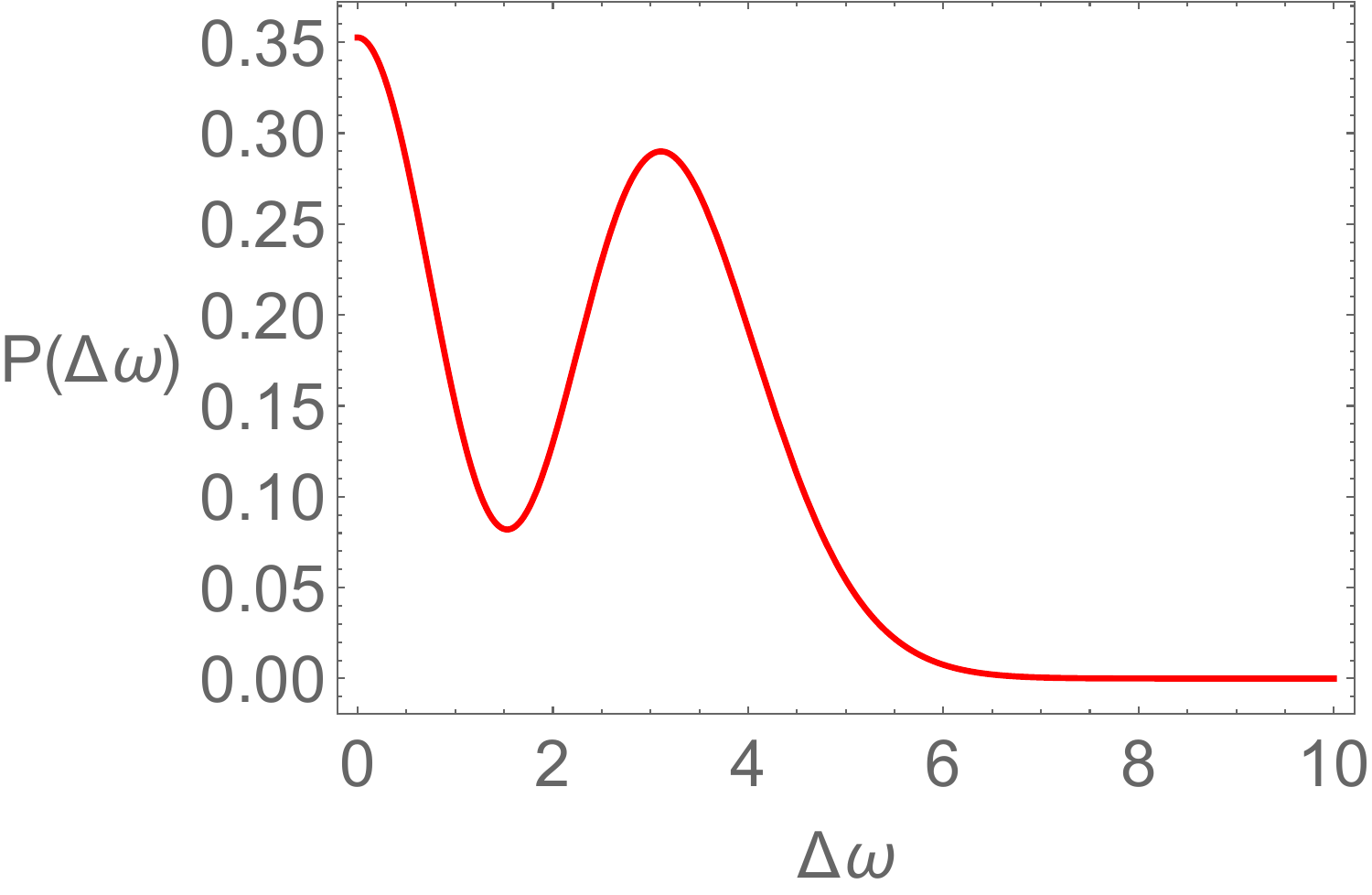} \label{}}
      {\includegraphics[width=7.1cm]{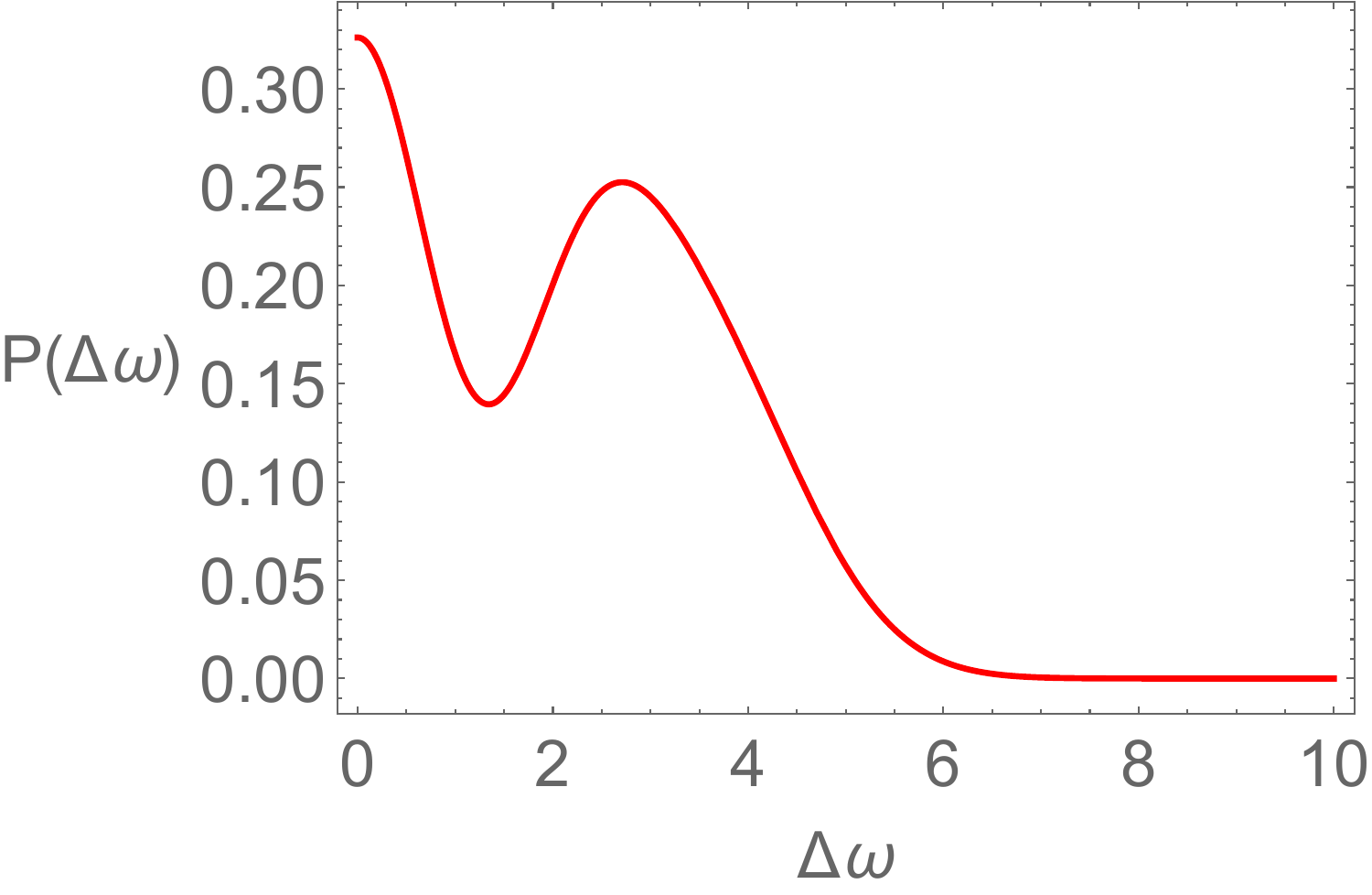} \label{FPDTOMG}}
\caption{$P(\Delta \omega)$ of RMT (GUE) for $N=3,\,4$ (left, right).}\label{TIIDEGPLOT}
\end{figure}

An important subtlety arises: since we are dealing with continuous probability distribution functions, it becomes evident that the probability of any specific value for $\Delta \omega := \omega_{a b} - \omega_{c d}$ must be zero. Instead, attention is directed towards the probability associated with intervals, where the probability density function encompasses non-zero probabilities. In this context, we can define probability distribution function $P(\Delta \omega)$ as
\begin{equation}
\begin{aligned}
P(\Delta \omega) = \int_{-\infty}^{\infty} \dd E_i \, \rho\left(E_i\right) \delta\left(\Delta \omega-\abs{\omega_{jk}-\omega_{lm}}\right) \,,
\end{aligned}
\end{equation}
where $\rho\left(E_i\right)$ is in \eqref{MKC2}. Note that, by definition, $P(\Delta \omega=0)=0$ for $N=2$, as there are only two eigenvalues. However, when $N>2$, we find $P(\Delta \omega=0)\neq0$ and is in fact the maximum value of the distribution. For instance, for the case of GUE and $N=3,\,4$, the probability distribution function is given by:
\begin{align}\label{pdfps}
\begin{split} 
P(\Delta \omega) =  
\begin{cases}
      \, \frac{e^{-\frac{(\Delta \omega)^2}{4}}\left(20-12 (\Delta \omega)^2+3 (\Delta \omega)^4\right)}{32 \sqrt{\pi}}  & \text{($N=3$) \,,}  \\
      \, \frac{e^{-\frac{(\Delta \omega)^2}{3}}\left(210195-108135 (\Delta \omega)^2+47088 (\Delta \omega)^4-696 (\Delta \omega)^6-48(\Delta \omega)^8+16(\Delta \omega)^{10}\right)}{209952 \sqrt{ 3 \pi}} &  \text{($N=4$) \,,}
\end{cases}\nonumber
\end{split}
\end{align}
where $\Delta \omega=\abs{\omega_{12}-\omega_{23}}$. These probability distributions are illustrated in Fig. \ref{TIIDEGPLOT}. 

Hence, we can deduce that in RMT, the probability of having $\Delta \omega=0$ is precisely zero, while there exists a non-zero probability of having $\Delta \omega\approx0$ within a finite interval. This prompts the question: are the complexities of nearly degenerate and exactly degenerate systems nearly identical? The arguments below suggest us that the answer is affirmative.

To begin, consider the scenario with exact type-II degeneracy. Let us suppose that \( b_\alpha \) represents the final Lanczos coefficient, so \( b_{\alpha+1} = 0 \). Due to type-II degeneracy, $\alpha$ will be smaller than \( D_H^2 - D_H +1 \). If we slightly alter the spectrum such that none of the \( \Delta \omega\) are strictly zero, the previous Lanczos coefficients (\( b_1, b_2, \cdots, b_\alpha \)) will only experience a minor shift. However, as \( b_{\alpha+1} \) becomes slightly larger than zero, \( b_{\alpha+2} \) will take a considerably large value, which can be inferred from \eqref{OCF2}. Consequently, the subsequent Lanczos coefficients \( b_{\alpha+3}, \cdots, b_{D_H^2-D_H +1} \) are not expected to vanish in most cases. It is crucial to emphasize that the progression of the transition amplitude occurs from \( \varphi_{0} \) to \( \varphi_{D_H^2-D_H +1} \), indicating that the early deviations are primarily derived from \( \varphi_{\alpha} \). However, given the small value of \( b_{\alpha+1} \), and the fact that the initial increment of the transition amplitude \( \varphi_{\alpha+1}(t) \) relies entirely on \( b_{\alpha+1} \varphi_{\alpha}(t) \), \( \varphi_{\alpha+1}(t) \) can be neglected for some period of time. Even though \( b_{\alpha+2} \) assumes a considerable magnitude, the negligible value of \( \varphi_{\alpha+1} \) during such period implies that \( \varphi_{\alpha+2} \) can also be disregarded in that same period. Consequently, the transition amplitude essentially halts its propagation at \( \varphi_{\alpha} \). Hence, the significance of the subsequent transition amplitudes (\( \varphi_{\alpha+1}(t), \varphi_{\alpha+2}(t), \ldots \)) can be neglected within eq.~\eqref{Kcom}. Given that the changes in the previous Lanczos coefficients (\(b_1, b_2, \ldots, b_\alpha\)) are minimal, the preceding transition amplitudes (\(\varphi_{0}(t), \varphi_{1}(t), \ldots, \varphi_{\alpha-1}(t)\)) similarly experience negligible changes. Consequently, the overall change in complexity should remain minimal.

Given the similarity between the complexities of systems exhibiting nearly type-II degeneracy and those with exact type-II degeneracy, it becomes natural to extend the concept of type-II degeneracy to encompass scenarios involving continuous probability distributions. Specifically, we define the generalized type-II degeneracy so that 
\begin{equation}\label{TIIDEG}
\begin{aligned}
    &(\exists \, \,\text{Generalized Type-II degeneracy}): \quad P \left( \Delta \omega=0 \right)\neq0 \,.
\end{aligned}
\end{equation}
Crucially, even though the complexity of systems satysfying \eqref{TIIDEG} resemble those with exact degeneracy, generalized degeneracies do \emph{not} lower the dimension of the Krlylov space. 

One important remark about RMT with $N>2$ is in order, namely, the fact that it is more likely to encounter generalized type-II degeneracies as $N$ increases. This can be understood as follows. First, consider all possible $\Delta \omega$ that are non-degenerate. These include
\begin{align}\label{}
\begin{split} 
\Delta \omega_{\text{non-deg}} =  
\begin{cases}
      \, \omega_{ij}-\omega_{kj}\,=\,E_i-E_k \,\neq\, 0\,,  &  \\
      \, \omega_{ij}-\omega_{ji}\,=\,2(E_i-E_j) \,\neq\, 0 \,, & 
\end{cases} 
\longrightarrow\quad P \left( \Delta \omega=0 \right) \,=\, 0  \,,
\end{split}
\end{align}
where we employed \eqref{BSIS11} and considered the fact that RMT exhibits the level repulsion in energy eigenvalues, i.e., $E_i\neq E_j$. Subsequently, we can determine the number of $\Delta \omega_{\text{non-deg}}$ as $2N(N-1)(N-2)+N(N-1)=N(2N^2-5N+3)$ together with the total number $\Delta \omega_{\text{total}}$ as $N(N-1)(N(N-1)-1)=N^4-2N^3+N$.
Consequently, the number of $\Delta \omega$ that \emph{can} be (nearly) degenerate is given by
\begin{align}
\begin{split}
N^4-2N^3+N-N(2N^2-5N+3) = N(N-2)(N-1)^2 \,, 
\end{split}
\end{align}
\color{black}indicating that as we increase the size of the Hilbert space, it is likely to find a higher number of the generalized type-II degeneracies. Our numerical results support this observation: as we increased $N$, the peak observed in $\bar{C}_K$ diminished. This loss of sensitivity to quantum chaos could then be attributed to the growing prevalence of type-II degeneracies.

\paragraph{Spread complexity and the degeneracy.} A natural question one may ask is if Spread complexity
is also sensitive to (nearly) degeneracies in the spectrum. To analyze this question, we can explore the dimension of the subspace concerning the Spread complexity $C_S$, following the same logic outlined for Krylov complexity.
For the Spread complexity \cite{Balasubramanian:2022tpr}, analogous equations to \eqref{nested0}-\eqref{nested} can be identified from
\begin{equation}\label{FEQOM1}
|\psi\rangle = \sum_{a=1}^{D_H} \psi_{a} |E_a\rangle \,, \qquad H^n|\psi\rangle=\sum_{a=1}^{D_H} \psi_{a} E_a^n |E_a\rangle \,,
\end{equation}
where $\mathcal{L}=H$. The dimension of the Spread subspace, denoted as $D_S$, is then given by
\begin{equation}\label{bound3}
1 \leq D_S \leq D_H \,.
\end{equation}
Furthermore, similar to \eqref{DEGEQ1}, the second equation in \eqref{FEQOM1} can be rewritten as
\begin{equation}
H^n|\psi\rangle=E^n \sum_{a  \in I} \psi_{a} |E_a\rangle+\sum_{a \notin I} \psi_{a} E_a^n |E_a\rangle \,,
\end{equation}
where we used the fact that the spectrum may contain (standard, or type-I) degeneracies,
\begin{equation}
H|E_a\rangle=E|E_a\rangle \quad \text{for \,} \forall a \in I \,.
\end{equation}
Assuming that the number of degeneracies is $m$, this leads to the reduction of \eqref{bound3} by
\begin{equation}\label{bound4}
1 \leq D_S \leq D_H - m \,.
\end{equation}
Similarly, one can also define generalized (type-I) degeneracy in the presence of  continuous probability distributions, such that
\begin{equation}
(\exists \, \,\text{Generalized degeneracy}): \quad P \left( \Delta E=0 \right)\neq0 \,.
\end{equation}
A couple of comments are in order. First, note that, by definition, $P(\Delta E=0)\neq0$ for the IID ensemble, shown in Fig. \ref{FPDTOMG}, so there exists generalized degeneracy in this case. We suspect this is the reason why there is no peak in the Spread complexity of IID, which goes hand in hand with the conjecture relating the peak to quantum chaos. Second, note that if there exists generalized degeneracy ($\Delta E \approx 0$), there exists generalized type-II degeneracy as well ($\Delta\omega\approx 0$), which explains why there is no peak in the Krylov complexity of IID. 
\begin{figure}[]
 \centering
     {\includegraphics[width=7.1cm]{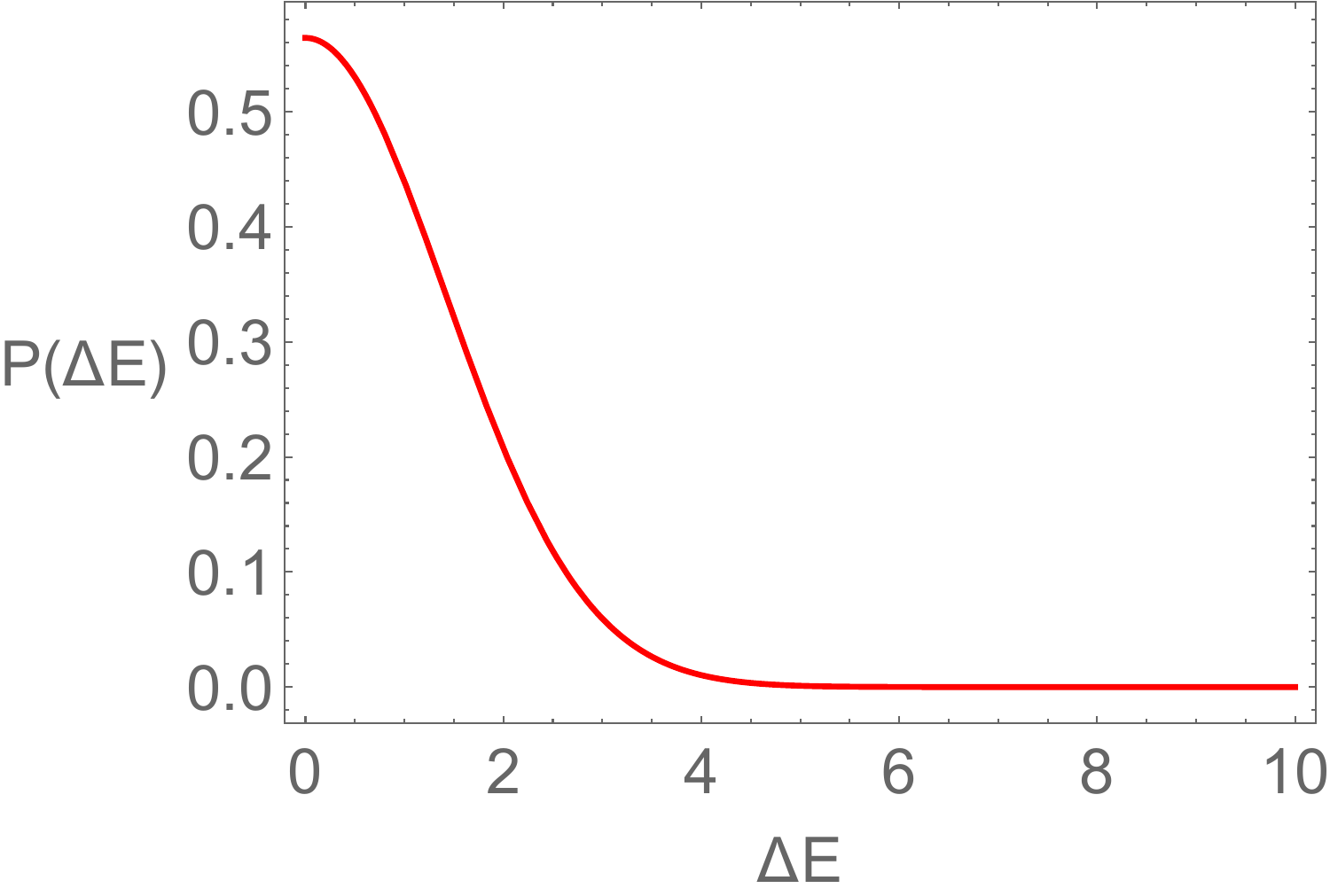} \label{}}
      {\includegraphics[width=7.1cm]{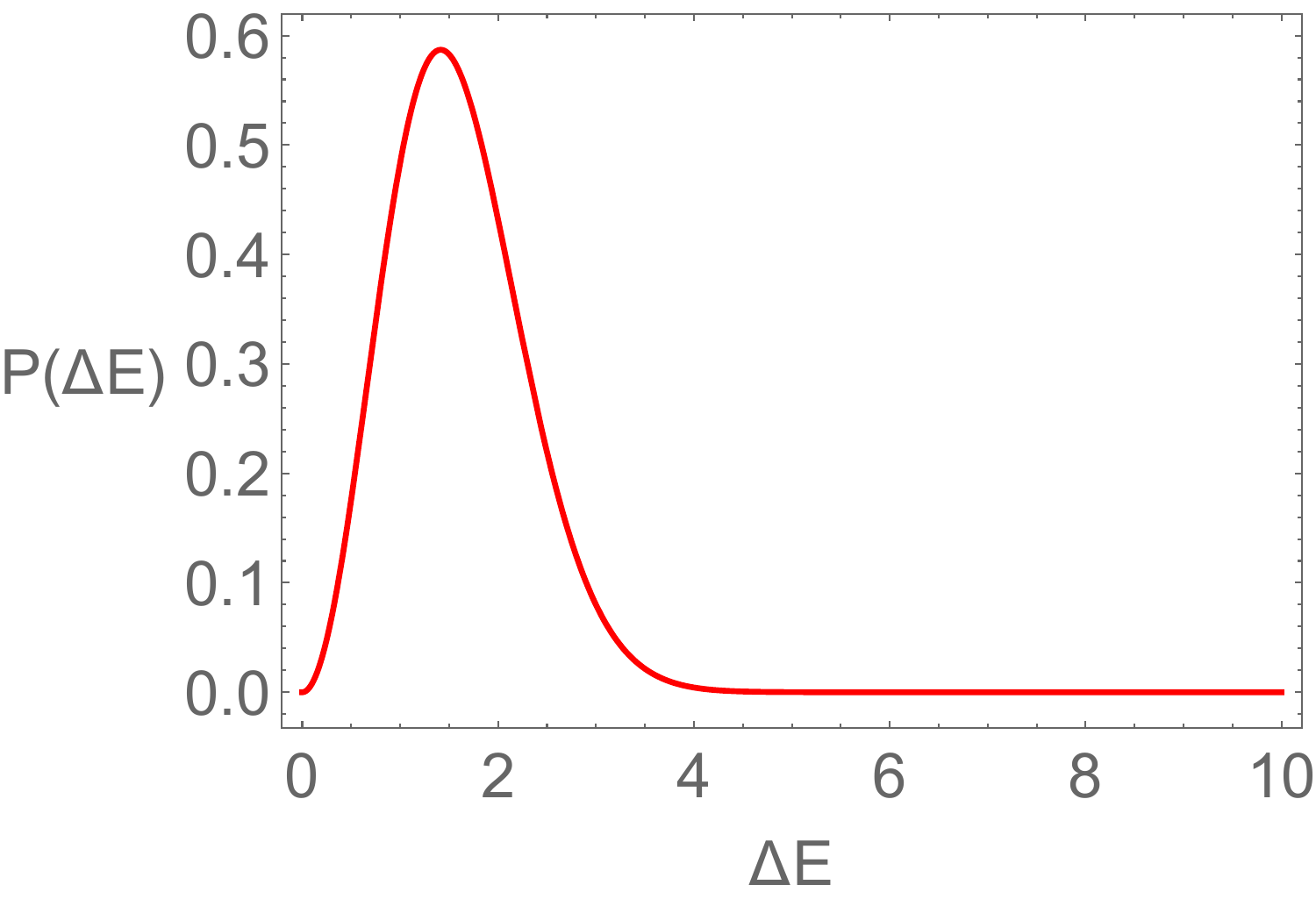} \label{}}
\caption{$P(\Delta E)$ in \eqref{eIID} of IID (left), and $P(\Delta E)$ in \eqref{N2RESULT3} of GUE (right). }\label{FPDTOMG}
\end{figure}
Notably, random matrix theories (GUE, GOE and GSE) do not exhibit generalized (type-I) degeneracies due to the level repulsion between energy eigenvalues, which explains the presence of the peak in Spread complexity for chaotic RMT with $N\geq2$.  

Summarising, we argued that the vanishing of the peak in Krylov complexity may be attributed to the existence on approximate type-II degeneracies in RMT. In contrast, the persistence of the peak in Spread complexity may be attributed to the absence of standard degeneracies in RMT. These observations help elucidate the qualitative distinctions between $C_K$ and $C_S$ for $N>2$ within the intermediate time regime. However, it will be interesting to better understand different scenarios for explaining the decreasing peak in the evolution of Krylov complexity and their compatibility with the analysis of this section. For instance, \cite{Balasubramanian:2023kwd} noticed that the diminishing of the peak is also correlated with differences in variances for different Dyson index $\mathcal{D}$ of random matrix models.\footnote{We thank Javier Magan for pointing this potential connection.} Intuitively, this may be related to introducing extra noise in the Lanczos spectrum when focusing on the Liouvillian for $\rho(t)=\ket{\psi(t)}\bra{\psi(t)}$ instead of just the Hamiltonian for a single state $\ket{\psi(t)}$.

\subsection{Comments on averaging}\label{SECTION6}
\begin{figure}[]
\centering
\includegraphics[width=3.0in]{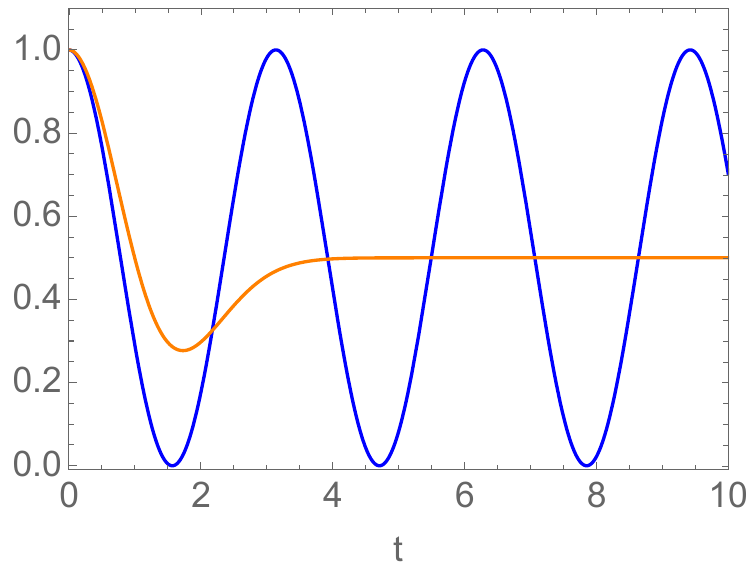}
\caption{Comparison of the SFF (blue, for $\Delta E=2$) in our toy model and its GUE average (orange) for $\beta=0$. The dip occurs at $t_d=\sqrt{3}$ and value of the plateau is 1/2.}\label{FigSFFAv}
\end{figure}

\begin{figure}[t!]
 \centering
     \subfigure[]
     {\includegraphics[width=6.6cm]{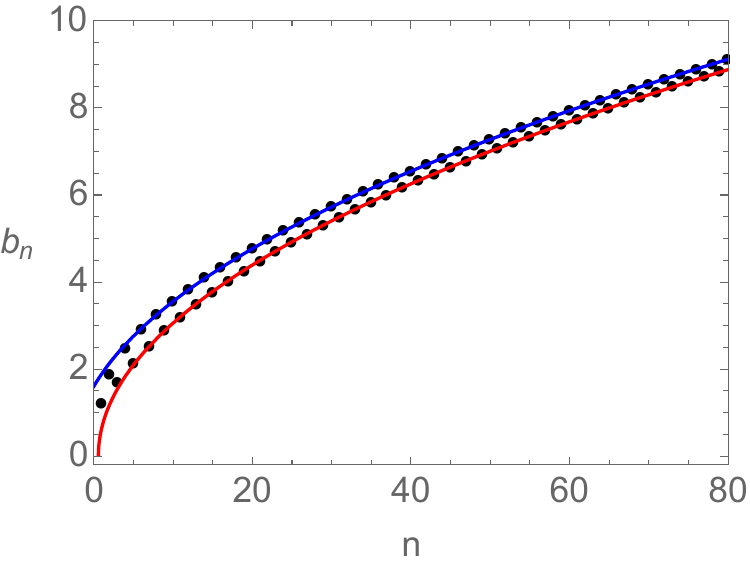} }
\quad
     \subfigure[]
     {\includegraphics[width=6.8cm]{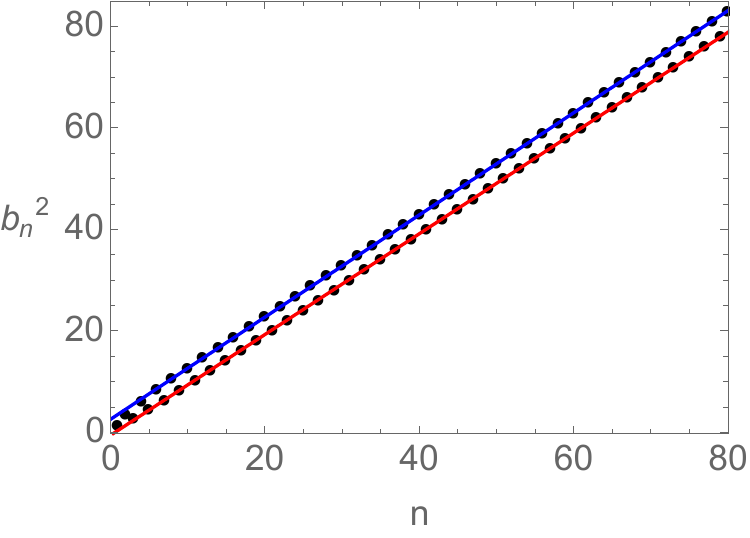}}
\caption{Numerical plot of $b_n$ and $b^2_n$ Lanczos coefficients from the averaged return amplitude.}\label{BnsAv}
\end{figure}

In this section we discuss some subtleties in taking the average at different stages of computation for the Krylov complexity of the pure density matrix. This is clearly relevant when we are interested in understanding wormhole contributions to the spectral form factor from the perspective of complexity.\\
First, we will start a simple toy model with two energies $E_1$ and $E_2$ and partition function
\be
Z(\beta)=e^{-\beta E_1}+e^{-\beta E_2}.
\ee
As we argued before, when computing the Krylov complexity of $\rho(t)$ for the evolution of the TFD state the return amplitude is given by the spectral from factor itself. Let us first compute everything for some general choice of $E_1$ and $E_2$. The return amplitude becomes
\be
S(t)=(\rho_\beta(t)|\rho_\beta)=\frac{Z(\beta-it)}{Z(\beta)}\frac{Z(\beta+it)}{Z(\beta)}=1-\frac{\sin^2\left(\frac{\Delta E t}{2}\right)}{\cosh^2\left(\frac{\beta\Delta E}{2}\right)},\label{SFF2En}
\ee
where $\Delta E=E_1-E_2$. The return amplitude is clearly periodic with period $2\pi/\Delta E$.\\ From this expression we only get two non-trivial Lanczos coefficients
\be
b_1=\frac{\Delta E}{\sqrt{1+\cosh(\beta\Delta E)}},\qquad b_2=\frac{\Delta E\sqrt{\cosh(\beta\Delta E)}}{\sqrt{1+\cosh(\beta\Delta E)}},
\ee
and all the others vanish. Expanding the state
\be
|\rho_\beta(t))=\sum_n i^n\varphi_n(t)|\rho_n),
\ee
we find three solutions of \eqref{dse} that read
\be
\varphi_0(t)=S(t),\qquad \varphi_1(t)=-\frac{\sin(\Delta Et)}{\sqrt{1+\cosh(\beta\Delta E)}},\quad \varphi_2(t)=\cosh^{1/2}(\beta\Delta E)\frac{\sin^2\left(\frac{\Delta E t}{2}\right)}{\cosh^2\left(\frac{\beta\Delta E}{2}\right)}.
\ee 
The Krylov complexity becomes
\be
{C_K}(t)=\sum^2_{n=0}n\varphi_n(t)=\left(1+\frac{\cosh(\beta\Delta E)\tan^2\left(\frac{\Delta E\,t}{2}\right)}{\cosh^2\left(\frac{\beta \Delta E}{2}\right)}\right)\frac{\sin^2(\Delta E\, t)}{2\cosh^2\left(\frac{\beta \Delta E}{2}\right)},
\ee
so has the same period as the return amplitude. \\
To keep our formulas simple, let us consider $\beta=0$ case where, $b_1=b_2=\frac{\Delta E}{\sqrt{2}}$ and 
\be
S^{\beta=0}(t)=\cos^2\left(\frac{\Delta E\, t}{2}\right),\qquad {C_K}^{\beta=0}(t)=2\sin^2\left(\frac{\Delta E\,t}{2}\right),
\ee
such that we simply have the relation between Krylov complexity and spectral form factor 
\be
{C_K}^{\beta=0}(t)=2\left(1-S^{\beta=0}(t)\right).\label{AvB0CS}
\ee
Let us now consider averaged return amplitude and averaged Krylov complexity by simply taking the GUE average over $\Delta E$ \eqref{N2RESULT3}
\be
\langle A\rangle =\int^\infty_0 \sqrt{\frac{2}{\pi}}e^{-\frac{(\Delta E)^2}{2}}(\Delta E)^2 \,\,A\,\, \dd(\Delta E).
\ee
This yields our results for the $N=2$ RMT
\be
\langle S^{\beta=0}(t)\rangle=\frac{1}{2}\left(1-e^{-t^2/2}(t^2-1)\right),\qquad \langle{C_K}^{\beta=0}(t)\rangle=1+e^{-t^2/2}(t^2-1),
\ee
and of course the averaged relation \eqref{AvB0CS} holds. Interestingly, the averaging of the completely periodic spectral form factor turns it into the one of the RMT with the famous slope-dip-ramp-plateau structure. Of course the times of these features are very short here e.g. the dip occurs at $t_d=\sqrt{3}$  (i.e. $t_d=\sqrt{\mathcal{K}}$ where $\mathcal{K}=D_H^2 - D_H + 1$) and SFF becomes constant. Moreover, the ramp (with value $1/2$) already starts around twice this time (see Fig. \ref{FigSFFAv}).
Next we may consider an alternative procedure when we first take the average spectral form factor and then compute from it the moments and Lanczos coefficients to finally evaluate the Krylov complexity. It may be natural to expect that this is a reasonable approach\footnote{It would be interesting to check it numerically in large-$N$ matrix models.} but we can see that this is not always the case. Indeed if we can write our averaged return amplitude in terms of even moments $\mu_{2k}$ only as
\be
\langle S^{\beta=0}(t)\rangle=\sum^\infty_{k=0}\mu_{2k}\frac{t^{2k}}{(2k)!},\qquad \mu_{2k}=\frac{1}{2}\delta_{k,0}+(-1)^{k}2^{k-1}\left(\frac{3}{2}\right)_k,
\ee
and this gives rise to the infinite number of ``staggered" Lanczos coefficients where even and odd $b_n$ grow differently (similarly to coefficients found in \cite{Avdoshkin:2022xuw,Camargo:2022rnt}). More explicitly, we can show that even $b_{2n}$ and odd $b_{2n+1}$ satisfy 
\be
b^2_1=\frac{3}{2},\qquad b^{2}_{2n}+b^2_{2n-1}=4n+1,\qquad b^2_{2n}b^2_{2n+1}=2n(2n+3).
\ee
We can rewrite these relations as $b^2_1=3/2$ as well as two equations separately for even and odd coefficients for $n\ge 1$ 
\be
b^2_{2n+2}=4n+5-\frac{2n(2n+3)}{b^2_{2n}},\qquad b^2_{2n-1}=4n+1-\frac{2n(2n+3)}{b^2_{2n+1}}.
\ee
Even though we were not able to find a closed form solution for these coefficients, we can generate a large number and check their scaling with $n$. This is shown on Fig. \ref{BnsAv}. Both, even and odd coefficients grow as $\sim \sqrt{n}$ and they can be fit into 
\be
b^2_{n-even}=1.007 n+2.565,\qquad b^2_{n-odd}=0.993 n-0.580.
\ee
Already at this point we see that taking the naive average on the level of the return amplitude would change the dimension of the Krylov subspace to infinite.

\begin{figure}[t!]
 \centering
     \subfigure[]
     {\includegraphics[width=6.9cm]{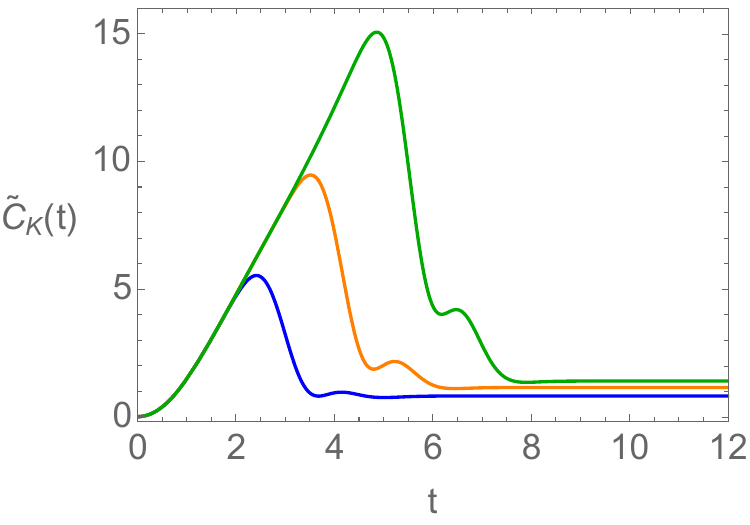} }
\quad
     \subfigure[]
     {\includegraphics[width=6.7cm]{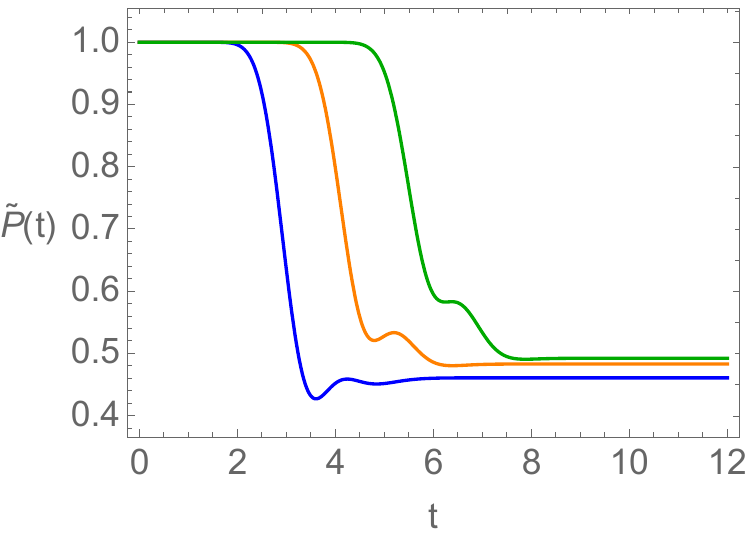}}
\caption{Numerical plots of regulated Krylov complexity (a) and regulated probability (b) for cut-off $\Lambda=10$ (blue), $\Lambda=20$ (orange) and  $\Lambda=35$ (green).}\label{CPt}
\end{figure}

Moreover, with these $b_n$, we can now in principle find solutions of the Schrödinger equation. In fact, for odd $n=2k+1$ we have a simple expression
\be
\varphi_n(t)=e^{-t^2/2}t^n\frac{(t^2-n-2)}{2\prod^n_{l=1}b_l},\qquad \varphi'_n(t)=e^{-t^2/2}t^{n-1}\frac{(2t^2-n)(n+2)-t^4}{2\prod^n_{l=1}b_l}.
\ee
However, the even wave functions with $n=2k$ appear much more complicated. Nevertheless, using the Schrödinger equation, we can express them via derivatives of the odd ones as
\be
\varphi_n(t)=\left(\prod^{\frac{n}{2}-1}_{l=0}\frac{b_{2l+1}}{b_{2l+2}}\right)\varphi_0(t)+\frac{\varphi'_{n-1}(t)}{b_{n}}+\sum^{\frac{n}{2}-2}_{m=0}\left(\prod^{\frac{n}{2}-1}_{l=m+1}\frac{b_{2l+1}}{b_{2l}}\right)\frac{\varphi'_{2m+1}(t)}{b_n}.
\ee
Unfortunately, we were also not able to evaluate the Krylov complexity analytically from these expressions (i.e. perform the sum to $n=\infty$ and we leave it as an interesting future problem). Nevertheless, we can use these wave functions to compute the answer for arbitrarily large $n$ cut-off $\Lambda$. We then introduce
\be
\tilde{C}_K(t)=\sum^\Lambda_{n=0}n|\varphi_n(t)|^2,\qquad \tilde{P}(t)=\sum^\Lambda_{n=0}|\varphi_n(t)|^2,
\ee
and plot them on Fig. \ref{CPt}. Observe that after the initial quadratic growth, complexity starts growing linearly. Since the dimension of the Krylov basis is not bounded, we do not expect saturation and this growth is likely to persist when the cut-off is removed. On the other hand, we can see that peaks of the complexity are cut-off effects and they take place exactly at times when the probability sum deviates from $1$. The second peak and plateau for later times are more mysterious and it would be interesting to investigate them in e.g. random matrix models after performing the average on the level of the return amplitude.

Concluding this section, we want to stress again that naive average at the level of the return amplitude, however tempting and interesting from the gravity and wormholes perspective, may be quite subtle and can lead to wrong conclusions. Nevertheless, it will be interesting to explore it further in JT gravity or large-$N$ matrix model with the aim to better understand how/if wormhole contributions to the return amplitudes (spectral form factor) can be deduced from the scaling of Lanczos coefficients of the growth of Krylov complexity. We leave this as another important future problem. In addition, we have mostly focused on the $\beta=0$ example and understanding finite-temperature effects on Lanczos coefficients (even in our toy model) would also be interesting.

\section{Formal approach}\label{SECTION4}
In this last section we finally take a slightly more abstract approach and ask how the Krylov basis for the pure density matrix operator $\rho(t)=|\psi(t)\rangle \langle \psi(t)|$, is related to the one computed for the state $|\psi(t)\rangle$. We will then illustrate our arguments for the Harmonic oscillator example.\\
Firstly, recall that expanding the quantum state in the Krylov basis
\be
|\psi(t)\rangle=e^{-iHt}\ket{\psi_0}=\sum_n\psi_n(t)\ket{K_n}\,.
\ee
The Krylov basis $\ket{K_n}$ is constructed via Lanczos algorithm and the sets of  Lanczos coefficients $\{a_n\}$ and $\{b_n\}$ can be extracted from the moments $\mu_n$, return amplitude 
\be
S(t)=\psi^*_0(t)= \sum_{n=0}^{\infty} \frac{(-it)^n}{n!} \mu_n\,.
\ee 
Moreover, the coefficients of the expansion satisfy the Schrödinger equation
\begin{equation}
i \partial_t \psi_n(t)=a_n \psi_n(t)+b_{n+1} \psi_{n+1}(t)+b_n \psi_{n-1}(t)\,.
\end{equation}
Let us assume that we have solved this problem and know all the ingredients above.\\
Then, in this work, we are interested in the density matrix operator $\rho(t)=|\psi(t)\rangle \langle \psi(t)|$  that evolves according the the Heisenberg equation and, after choosing our inner product \eqref{inner}, we want to turn it into a state in the Hilbert space $H\otimes \bar{H}$
\begin{equation}
\rho (t) = e^{-i H t} \rho (0) e^{i H t} \quad\Longleftrightarrow\quad | \rho (t) ) = e^{-i \mathcal{L} t} | \rho (0) )\,.
\label{evorho}
\end{equation}
In addition, we want to expand this state in the Krylov basis for the operators
\be
| \rho (t) ) =\sum_ni^n \varphi_n(t)|\rho_n),
\ee
where the Schrödinger equation is now
\bea
\partial_t \varphi_n(t) = B_{n+1} \, \varphi_{n+1}(t) - B_n \, \varphi_{n-1}(t) \,,\label{SE2}
\eea
and Lanczos coefficients $B_n$ (we use capital letter to distinguish them from $b_n$'s) that enter the Lanczos algorithm for $|\rho_n)$ are encoded in the return amplitude
\be
G(t)=(\rho(t)|\rho(0))=\varphi^*_0(t)=|\psi_0(t)|^2=|S(t)|^2\,.\label{RetAmplits}
\ee
Before proceeding, we have three general useful facts for our problem. Firstly, the form of the super-operator Liouvillian in $\eqref{evorho}$ acting on vectorized $\rho(0)$ (on the Kronecker product of $\bar{H}\otimes H$) is known in terms of the original Hamiltonian and its copy, and reads 
 \begin{equation}
\mathcal{L}= \mathbb{I} \otimes H - H^T \otimes \mathbb{I}\,. 
\label{liouhami}
\end{equation}
 Secondly, since the return amplitudes in the two approaches are simply related \eqref{RetAmplits}, the moments of $G(t)$ defined via
 \be
 G(t) = \sum_{n=0}^{\infty} \frac{(-it)^n}{n!} \mathcal{M}_n\,,
 \ee
 are non-zero only for $n$-even and can be expressed in terms of $\mu_n$'s as
\begin{equation}
\mathcal{M}_n =\sum_{k=0}^n \binom{n}{k} (-1)^k \mu_k \mu_{n-k}\,.
\label{momentrel}
\end{equation}
Still the relation between $B_n$'s and $\{a_n,b_n\}$'s of $\psi^*_0(t)$ is quite complicated. For example, we have the first three
\be
B^2_1=2b^2_1,\qquad B^2_2=b^2_2+2b^2_1+(a_1-a_0)^2,\qquad B^2_3=b^2_2\frac{b^2_3+9b^2_1+(a_2+a_1-2a_0)^2}{b^2_2+2b^2_1+(a_1-a_0)^2}\,,\label{Bnbn}
\ee
where the first result corresponds to \eqref{NER55}.
Last but not the least, we can also use the state-channel map to rewrite the density matrix in the Krylov basis $\ket{K_n}$ as\footnote{The second vector should be a CPT conjugated.}
\be
|\rho(t))=\sum_{n,m}\psi_n(t)\bar{\psi}_m(t)|K_n,K_m)\,,
\ee
this allows us to read off the formal relation between the amplitudes in two approaches
\be
i^k \varphi_k(t)\equiv (\rho_k|\rho(t))=\sum_{n,m}\psi_n(t)\bar{\psi}_m(t) (\rho_k|K_n,K_m).\label{RelWF}
\ee
In general, these hints may not simplify the task of evaluating the Krylov complexity of $\rho(t)$ too significantly (and we simply have to compute it directly) but we will illustrate below these steps for the Hamiltonian constructed from the Heisenberg-Weyl algebra generators. We expect an elegant generalisation of these steps for general semi-simple Lie groups and their discrete series representations (and leave it for the future work). 
\subsection{Toy model: harmonic oscillator}
As an example, we consider a state that evolves according to the Hamiltonian built from the Heisenberg-Weyl generators
\begin{equation}
\ket{\psi(t)}=e^{-iHt}\ket{0},\qquad H = \lambda ( a^{\dagger} + a) + \omega \hat{n},
\label{hamiHO}
\end{equation}
where without loss of generality, we let $\lambda \in \mathbb{R}$ and $\hat{n}= a^{\dagger} a$ is the number operator, $a$ the annihilation operator and $a^\dagger$ is the creation operator that satisfy 
\begin{equation}
[ a, a^{\dagger} ] =1, \qquad [ \hat{n}, a] = -a, \qquad [ \hat{n}, a^{\dagger}] = a^{\dagger}.
\end{equation}
The Krylov basis for this case is simply the algebra basis (see \cite{Caputa:2021sib})  
\begin{equation}
| K_n \rangle \equiv |n \rangle=\frac{(a^\dagger)^n}{\sqrt{n!}}\ket{0}\,,
\end{equation}
and from the action of $H$ 
\begin{equation}
H | n \rangle =  \omega n |n \rangle + \lambda \sqrt{n} | n - 1\rangle+ \lambda \sqrt{n+1} |n+1 \rangle\,,
\end{equation}
we simply read off Lanczos coefficients
\be
a_n=\omega n,\qquad b_n=\lambda \sqrt{n}\,.
\ee
They are encoded in the return amplitude
\be
\psi^*_0(t)=e^{-\frac{\lambda^2}{\omega^2}\left(i\omega t+1-e^{i\omega t}\right)}\,,
\ee
and the solutions of the Schrödinger equations read
\be
\psi_n(t)=e^{\frac{\lambda^2}{\omega^2}\left(i\omega t-1+e^{-i\omega t}\right)}\frac{\left(\frac{\lambda}{\omega}(e^{-i\omega t}-1)\right)^n}{\sqrt{n!}}\,.
\ee

Now we turn to our main problem. According to \eqref{liouhami}, the Liouvillian super-operator becomes
\begin{equation}
\mathcal{L} = \left[-\lambda ( a^{\dagger} + a)^T - \omega \hat{n}^T \right]  \oplus \left[ \lambda ( a^{\dagger} + a) + \omega \hat{n} \right] =\lambda  (a^{\dagger}-\bar{a}^{\dagger})+ \lambda  (a -\bar{a}  ) + \omega (\hat{n} - \hat{\bar{n}} ),
\label{liouHO}
\end{equation}
where we omit $\otimes 1$ (in $a^\dagger\equiv 1\otimes a^\dagger$ and $\bar{a}^\dagger\equiv \bar{a}^\dagger\otimes 1 $)
and we defined the second copy as
\begin{equation}
a^T \equiv \bar{a}^{\dagger}, \qquad \left(a^{\dagger}\right)^T \equiv \bar{a}, \qquad \hat{n}^T =\bar{a}^{\dagger}\bar{a} \equiv \hat{\bar{n}}.
\end{equation}
Next, we compute the return amplitude
\be
G(t)=\varphi_0(t)=|\psi_0(t)|^2=e^{-\frac{4\lambda^2}{\omega^2}\sin^2\left(\frac{\omega t}{2}\right)}\,,
\ee
that allows us to extract $B_n$'s (consistent with \eqref{Bnbn})
\be
B_1=\sqrt{2}\lambda,\qquad B_2=\sqrt{4\lambda^2+\omega^2},\qquad B_3=\frac{\sqrt{6}\lambda\sqrt{4\lambda^2+3\omega^2}}{\sqrt{4\lambda^2+\omega^2}},\qquad \cdots
\label{bn}
\ee
From the Schrödinger equation \eqref{SE2} we  find a couple of  amplitudes
\bea
\varphi_1(t)&=&-\frac{\sqrt{2}\lambda}{\omega}\sin(\omega t)e^{-\frac{4\lambda^2}{\omega^2}\sin^2\left(\frac{\omega t}{2}\right)}\,,\nn\\
\varphi_2(t)&=&2\sqrt{2}\lambda\frac{\sin^2\left(\frac{\omega t}{2}\right)+\frac{\lambda^2}{\omega^2}\sin^2(\omega t)}{\sqrt{4\lambda^2+\omega^2}}e^{-\frac{4\lambda^2}{\omega^2}\sin^2\left(\frac{\omega t}{2}\right)}\,,
\eea
and so on. We will get back to them momentarily but first let us elaborate more on the Krylov basis $|\rho_n)$. From the direct application of the Gram-Schmidt procedure we have
\be
|\rho_0)=\ket{\bar{0}0},\quad |\rho_1)=\frac{1}{b_1}\mathcal{L}|\rho_0)=\frac{\ket{\bar{0}1}-\ket{\bar{1}0}}{\sqrt{2}}\,,
\ee
and similarly
\be
|\rho_2)=b^{-1}_2\left(\mathcal{L}|\rho_1)-b_1|\rho_0)\right)= \frac{\lambda   \left(  |\bar{2}0\rangle -\sqrt{2}  |\bar{1}1\rangle +  |\bar{0} 2 \rangle \right) +  \frac{\omega}{\sqrt{2}} \left(  |\bar{1} 0 \rangle +| \bar{0} 1 \rangle \right)}{\sqrt{4 \lambda^2 +\omega^2}}.
\ee
Note that since  $\ket{K_n,K_m}=\ket{n,m}$, we have e.g. the overlap
\be
(\rho_1|K_n,K_m)=\frac{\delta_{n0}\delta_{m1}-\delta_{n1}\delta_{m0}}{\sqrt{2}},
\ee
and indeed $\varphi_n(t)$ satisfy the relations between the wave functions \eqref{RelWF}. The procedure, even though straightforward, gets of course quite complicated with every iteration.

Finally, we explore another avenue that gives more mileage in deriving the Krylov basis. Namely, following the intuition from \cite{Caputa:2021sib}, we know that action of the Liouvillian in the Krylov basis
\begin{equation}
\mathcal{L} | \rho_n ) = B_{n+1} | \rho_{n+1} ) + B_n  | \rho_{n-1} ),\,
\label{liourec}
\end{equation}
allows us to formally represent the Liouvillian as a sum of ``raising" and ``lowering" ladder operators. It is clear that in our case they should consist of $a^{\dagger}, \bar{a}^{\dagger}$ and $a, \bar{a}$ but the presence of $n, \bar{n}$ is a bit confusing, so let us check how this works in detail. For that, we introduce two sets of ladder operators
\begin{equation}
\alpha_{\mp} \equiv \frac{1}{\sqrt{2}}(  a \mp \bar{a})~, 
\end{equation}
that satisfy the commutation relations
\begin{equation}
[ \alpha_{\mp}, \alpha^{\dagger}_{\mp} ] =  1~, \qquad
~[ \alpha_{\mp}, \alpha_{\pm}^{\dagger}] =0~, \qquad 
~[ \alpha_{\mp}, \alpha_{\mp}] = [ \alpha_{\pm}, \alpha_{\mp}] = 0~.
\end{equation}
This way, the number operator term $(\hat{n}-\hat{\bar{n}})$ becomes
\begin{equation}
\hat{n}-\hat{\bar{n}} = \alpha^{\dagger}_- \alpha_+ + \alpha^{\dagger}_+ \alpha_- ~.
\label{Nin2}
\end{equation}
The RHS of \eqref{Nin2}, together with their commutator $[ \alpha^{\dagger}_+ \alpha_-,  \alpha^{\dagger}_- \alpha_+]= \alpha^{\dagger}_+\alpha_+ - \alpha_-^{\dagger}\alpha_- $ forms the $\operatorname{SU}(2)$ algebra. This implies that as $a^{\dagger}- \bar{a}^{\dagger}$ and $a- \bar{a}$, the number operator term can also be written as a sum of ``raising" and ``lowering" ladder operators, although they do not change total number of oscillators
\begin{equation}
[\hat{n} + \hat{\bar{n}}, \alpha^{\dagger}_- \alpha_+ ] = [ \alpha^{\dagger}_+\alpha_+ + \alpha^{\dagger}_-\alpha_-,  \alpha^{\dagger}_- \alpha_+ ] =0~.
\end{equation}
Thus our Liouvillian can be rewritten as
\begin{equation}
\mathcal{L} = \sqrt{2} \lambda (\alpha^{\dagger}_- + \alpha_-) + \omega \left( \alpha^{\dagger}_- \alpha_+ + \alpha_+^{\dagger} \alpha_- \right) \equiv 
\omega (\ell_{1}+\ell_{-1})\,,
\end{equation}
where we  defined
\begin{equation}
\ell_{-1} \equiv \sqrt{2} \frac{\lambda}{\omega} \alpha_- + \alpha^{\dagger}_- \alpha_+ = \frac{\lambda}{\omega}(-\bar{a}+a) + \frac{1}{2}(-\hat{\bar{n}}+\hat{n} -\bar{a}^{\dagger}a + a^{\dagger} \bar{a})=\ell_{+1}^{\dagger}~.
\end{equation}
The effective number operator $\ell_0$ should be a superposition of the total number operator $\hat{\bar{n}} + \hat{n}= \alpha_+^{\dagger}\alpha_+ + \alpha_-^{\dagger}\alpha_-$, and number operator of the $\operatorname{SU}(2)$ algebra,  $\alpha_+^{\dagger}\alpha_+ - \alpha_-^{\dagger}\alpha_-$.  Utilising the commutator
\begin{equation}
[ \ell_0 , \ell_{\pm 1} ] = \pm \ell_{\pm 1 },
\end{equation}
we find
\begin{equation}
\ell_0 \equiv \frac{1}{2} \left[ 3 \left( \alpha_+^{\dagger}\alpha_+ + \alpha_-^{\dagger}\alpha_- \right) + \left( \alpha_+^{\dagger}\alpha_+ - \alpha_-^{\dagger}\alpha_- \right)\right]= 2\alpha_+^{\dagger}\alpha_+ + \alpha_-^{\dagger}\alpha_-~.
\end{equation}
It is not hard to see that $|\rho_0)=\ket{\bar{0}0}$
is indeed the lowest eigenstate of $\ell_0$.\\
Higher eigenstates of $\ell_0$ can be constructed as
\begin{equation}
| \rho_l ) = \mathcal{N}_l \ell_{+1}^l |\rho_0 ) ~,
\end{equation}
where $\mathcal{N}_l$ is the normalisation factor satisfying
\begin{equation}
    \mathcal{N}_l \equiv \left[(\rho_0 | \ell_{-1}^l \ell_{+1}^l |\rho_0 )\right]^{-1/2} = \mathcal{N}_{l-1} \left[(\rho_{l-1} | \ell_{-1} \ell_{+1} |\rho_{l-1} )\right]^{-1/2} ~,
\end{equation}
and $\mathcal{N}_0 =1$. This implies the recursion relation
\begin{equation}
    \ell_{+1}  |\rho_l) = \frac{\mathcal{N}_{l}}{\mathcal{N}_{l+1}} |\rho_{l+1})~, \qquad
    \ell_{-1}  |\rho_l) = \frac{\mathcal{N}_{l-1}}{\mathcal{N}_{l}} |\rho_{l-1})~,
\end{equation}
which, compared with \eqref{liourec}, allows us to derive the Lanczos coefficients formally as
\begin{equation}
B_n = \omega \frac{\mathcal{N}_{n-1}}{\mathcal{N}_{n}} = \omega \left[\frac{(\rho_0 | \ell_{-1}^l \ell_{+1}^l |\rho_0 )}{(\rho_0 | \ell_{-1}^{l-1} \ell_{+1}^{l-1} |\rho_0 )} \right]^{1/2}, \qquad  \forall~ n>0\,.
\label{genbn}
\end{equation}

In order to derive more explicit results, we express the ladder operators in terms of the creation and annihilation operators of the harmonic oscillator. Firstly, we have the Krylov basis 
\begin{equation}
| \rho_l ) =  \mathcal{N}_l  \left(\sqrt{2}\frac{\lambda}{\omega} \alpha_-^{\dagger} + \alpha_+^{\dagger} \alpha_- \right)^{l}  | \rho_0 )\,,  
\end{equation}
and we can series expand it such that every term contains a product of $k$ $\alpha_-^{\dagger}$'s and $(l-k)$ $\alpha_+^{\dagger}\alpha_-$'s acting on $|\rho_0)$. However, we observe that due to the commutator
\begin{equation}
  \left[ \frac{1}{2} \left(\sqrt{2}\frac{\lambda}{\omega} \right)^{-1} \alpha_+^{\dagger} \alpha_-^2, \sqrt{2}\frac{\lambda}{\omega} \alpha_-^{\dagger} \right] = \alpha_+^{\dagger} \alpha_-\,,
\end{equation}
all terms with $(l-k)$ $\alpha_+^{\dagger}\alpha_-$'s can be equivalently generated by acting $(l-k)$ times with the operator $\frac{1}{2} \left(\sqrt{2}\frac{\lambda}{\omega} \right)^{-1} \alpha_+^{\dagger} \alpha_-^2$ on the left of $\left( \alpha_-^{\dagger} \right)^l$. Denoting $\kappa \equiv \sqrt{2} \frac{\lambda}{\omega}$ and $j_{-1} = \frac{1}{2}  \alpha_+^{\dagger} \alpha_-^2 $, we then have the $\ell_0 = l$ eigenstate
\bea
| \rho_l ) &= & \mathcal{N}_l \cdot \left[\kappa^l \left(  \alpha_-^{\dagger} \right)^l | \rho_0 ) + \kappa^{l-1} j_{-1}  \left( \alpha_-^{\dagger} \right)^l | \rho_0 ) + \cdots \right]=\mathcal{N}  \sum_{k=0}^{\lfloor \frac{l}{2} \rfloor} \kappa^{l-k}\left( j_{-1} \right)^k \left( \alpha^{\dagger}_- \right)^l | \rho_0 )\nonumber  \\
&=& \mathcal{N} \sum_{k=0}^{\lfloor \frac{l}{2} \rfloor}2^{-k} \kappa^{l-k}\left( \alpha_+^{\dagger} \right)^k \times l(l-1)\cdots(l-2k+1) \left( \alpha_-^{\dagger} \right)^{l-2k} | \rho_0 )\nonumber  \\
&=&  \mathcal{N} \sum_{k=0}^{\lfloor \frac{l}{2} \rfloor} \frac{l !}{(l-2k)!}2^{-(l+k)/2}\left(\sqrt{2}\frac{\lambda}{\omega} \right)^{l-k}\left( \bar{a}^{\dagger} +a^{\dagger} \right)^k  \left( a^{\dagger}-\bar{a}^{\dagger}  \right)^{l-2k} \ket{\bar{0}0}~\,.
\label{krylovb}
\eea
The normalisation factor $\mathcal{N}$ can be then written as
\begin{equation}
    \mathcal{N}^{-2}_l = \sum_{k=0}^{\lfloor \frac{l}{2} \rfloor} \frac{(l !)^2}{[(l-2k)!]^2} 2^{-2k} \kappa^{2(l-k)} k! (l-2k)! =(l!)^2  \sum_{k=0}^{\lfloor \frac{l}{2} \rfloor}  \frac{k!}{(l-2k)!} 2^{-2k} \left(\sqrt{2}\frac{\lambda}{\omega}\right)^{2(l-k)}~.
\end{equation}
Using \eqref{genbn}, one can check that
\begin{eqnarray}
\mathcal{N}_1^{-2} &=& \left(\sqrt{2}\frac{\lambda}{\omega}\right)^{2}, \qquad \mathcal{N}_2^{-2} = 2 \left(\sqrt{2}\frac{\lambda}{\omega}\right)^{4} + \left(\sqrt{2}\frac{\lambda}{\omega}\right)^{2}~, \nonumber \\
\mathcal{N}_3^{-2} &=& 6 \left(\sqrt{2}\frac{\lambda}{\omega}\right)^{6} + 9 \left(\sqrt{2}\frac{\lambda}{\omega}\right)^{4}~, \cdots
\end{eqnarray}
indeed generates the $B_n$'s in \eqref{bn}.

\section{Conclusions}\label{SECTION7}

Quantifying complexity within quantum systems has emerged as a focal point, spanning investigations from quantum many-body systems to inquiries into quantum field theory and quantum gravity. This endeavor has led to the development of various physically-motivated measures of quantum complexity, serving as novel tools for probing the intricacies of the quantum domain. Among recent advancements, one of the most promising definitions is the concept of Krylov-based complexity, which encompasses Krylov complexity ($C_K$) \cite{Parker:2018yvk} and Spread complexity ($C_S$) \cite{Balasubramanian:2022tpr}.

In this paper, we sought to explore an intriguing question in the domain of Krylov-based complexities, namely, to elucidate the potential relationship between these two complexity notions and their ability to encode information about system dynamics, including phenomena like quantum chaos. For this purpose, we mainly investigated $C_K$ associated with the density matrix operator of a pure state, alongside $C_S$ calculated for the same state. Consequently, our work was dedicated to exploring the similarities and discrepancies between these two approaches, namely Krylov and Spread complexity, concerning pure states.

Our main findings can be summarized as follows:\vspace{1mm}\\
\textbf{(Result I) General pure state.}
Considering an scenario involving a \textit{general pure state}, we validated analytically that
\begin{itemize}
\item{The moment-generating function $G(t)$, pivotal for assessing the Lanczos coefficients, can be represented in terms of the survival (or return) amplitude $S(t)$ as follows:
\begin{align}\label{}
\begin{split}
\text{for $C_K$:} \quad G(t) = |S(t)|^2 \,,  \quad\quad  \text{for $C_S$:} \quad G(t) =  S(t) \,.
\end{split}
\end{align}
}
\item{In the \textit{early} time regime ($t\ll1$), we find $C_K \,=\, 2 C_S$.}
\end{itemize}
These two findings illustrate the explicit relationship between $C_K$ and $C_S$.\vspace{1mm}\\
\textbf{(Result II) Maximally entangled pure state.}
In the scenario involving a \textit{maximally entangled pure state} (i.e., the thermofield double state with $\beta=0$), we also observe that, within finite $N$-dimensional Hilbert spaces,
\begin{itemize}
\item{$G(t)$ for $C_K$ becomes the Spectral Form Factor (SFF), since $|S(t)| = \sqrt{\text{SFF}}$.}
\item{$C_K \,=\, 2 C_S$ remains valid for all times when $N=2$.}
\item{In the \textit{late} time regime ($t\gg1$), we find $C_K \,\cong\, N C_S$ when $N\geq2$.}
\end{itemize}
Note that the Thermofield Double (TFD) state can be particularly intriguing. The utilization of this state in Ref. \cite{Balasubramanian:2022tpr} led the authors to unveil new features in $C_S$, which were conjectured to represent universal characteristics of chaotic systems.

The first finding listed in Result II is noteworthy for two reasons. Firstly, it is analytically verified and holds true across any $N$-dimensional Hilbert space.\footnote{It also remains valid for all values of $\beta$ encompassed within the thermofield double state.} Secondly, as discussed in the introduction, averaging over the return amplitude in theories dual to gravity may include contributions from gravitational wormholes in the bulk. This connection offers relevance to investigations concerning wormhole contributions to the Spectral Form Factor (SFF) from a complexity perspective. 

The second finding listed in Result II, concerning the $N=2$ case, is also derived from scenarios where analytical examination is feasible: the general two-dimensional Hilbert space and qubit states undergoing modular Hamiltonian evolution. In these models, we also investigated $C_K$ associated with the density matrix of a general \emph{mixed} state. Interestingly, we consistently observe that the $C_K$ for a pure state surpasses that of a generic mixed state, aligning with the findings of circuit complexity studies~\cite{Caceres:2019pgf}.

The analysis for $N=2$ was also extended to higher-dimensional Hilbert spaces ($N>2$), leading to our third finding listed in Result II. For this extension we focused on random matrix theories (RMT), where the Spread complexity ($C_S$) of the Thermofield Double (TFD) state has emerged as a reliable measure for quantum chaos~\cite{Balasubramanian:2022tpr}.

Given the inherent randomness in random matrix theories (RMT), we introduced the averaged $C_K$ (and $C_S$) using the probability distribution \eqref{MKC}, which is equivalent to the arithmetic mean \eqref{CKMET44} when the total iterations (drawn from specific ensembles) are sufficiently high. Through numerical computations up to $N=10$, we discerned that $C_K$ approximately equaled $N C_S$ at late times. Additionally, we observed that in the intermediate time regime, $C_K$ could deviate from its resemblance to $C_S$, with $C_K$ lacking a discernible peak unlike $C_S$. We speculated that the absence of this peak in $C_K$ could be related to the manifestation of type-II degeneracy \eqref{TIIDEG}.

Regarding the averaging process, we also discussed some subtleties associated with averaging at different computational stages for $C_K$. We explicitly demonstrated that a naive averaging approach at the level of the return amplitude could be subtle and potentially lead to erroneous conclusions.

Last but not least, in the context of (inverted) harmonic oscillators, we conducted analytical computations for both $C_K$ and $C_S$ as a toy example of an infinite dimensional Hilbert space ($N\rightarrow \infty$). We observed a curious relation at late times, $C_K =C_S^2$, suggesting the need for additional analysis to understand their correlation.

From a bigger picture, it is worth recalling that the exploration of complexity and chaos has played a pivotal role in recent developments of the AdS/CFT correspondence and the study of black holes. Various conjectures suggest that the complexity of holographic quantum systems may offer insights into the black hole interior~\cite{Stanford:2014jda,Brown:2015bva,Brown:2015lvg,Couch:2016exn,Belin:2021bga}, and potentially shed light on the emergence of spacetime~\cite{Czech:2017ryf,Susskind:2019ddc,Pedraza:2021mkh,Pedraza:2021fgp,Pedraza:2022dqi,Carrasco:2023fcj}. In this context, a pressing concern arises from the ambiguity resulting from various proposals defining complexity. In contrast, Krylov (and Spread) complexity provides a clear and unambiguous definition. This leads to a critical question: can Krylov complexity effectively capture the growth of a black hole interior? Relevant investigations in this direction can be found in~\cite{Kar:2021nbm}. From this standpoint, comprehending the relationship between complexities (such as conventional computational complexity, Krylov, and Spread complexity) emerges as a significant consideration. While proposals exist for the gravity dual interpretation of conventional circuit complexity, recent inquiries have also explored a dual interpretation of Krylov complexity~\cite{Rabinovici:2023yex}. We anticipate that our examination of the relationship (using the density matrix operator) between Krylov complexity, Spread complexity, and the Spectral Form Factor could not only contribute to refining definitions of quantum chaos and complexity but also offer insights into the mystery of black hole interiors and the emergence of spacetime.

\acknowledgments
We would like to thank Jos\'e Barb\'on, Hugo A. Camargo, Bowen Chen, Kyoung-Bum Huh, Viktor Jahnke, Hong-Yue Jiang, Keun-Young Kim, Yu-Xiao Liu, Javier Mag\'an, Mitsuhiro Nishida and Shan-Ming Ruan for valuable discussions and correspondence. PC and SL are supported by NAWA “Polish Returns 2019” PPN/PPO/2019/1/00010/U/0001 and NCN Sonata Bis 9 2019/34/E/ST2/00123 grants. PC would also like to thank the Isaac Newton Institute for Mathematical Sciences, Cambridge, for support and hospitality during the programme \emph{Black Holes: bridges between number theory and holographic quantum information}, where work on this paper was undertaken. HSJ and JFP are supported by the Spanish MINECO ‘Centro de Excelencia Severo Ochoa' program under grant SEV-2012-0249, the Comunidad de Madrid ‘Atracci\'on de Talento’ program (ATCAM) grant 2020-T1/TIC-20495, the Spanish Research Agency via grants CEX2020-001007-S and PID2021-123017NB-I00, funded by MCIN/AEI/10.13039/501100011033, and ERDF A way of making Europe. LCQ acknowledges the financial support provided by the scholarship granted by the Chinese Scholarship Council (CSC). LCQ would also like to thank the Galileo Galilei Institute for Theoretical Physics for the hospitality and the INFN for partial support during the completion of this work. All authors contributed equally to this paper and should be considered as co-first authors. LCQ is the corresponding author.

\appendix
\section{Spread complexity: a quick review}\label{appena}
In the appendix, we review the definition of Spread complexity \cite{Balasubramanian:2022tpr}. Consider a quantum system with a time-independent Hamiltonian $H$. The time evolution of a state $ \ket{\psi(t)}$ is governed by the Schrödinger equation
\begin{equation}
\label{Schr}
|\psi(t)\rangle=e^{-i H t}|\psi(0)\rangle\, ,
\end{equation}
which is nothing but a linear combination of $\left\{|\psi(0)\rangle, H|\psi(0)\rangle, H^2|\psi(0)\rangle, \cdots\right\}$\,.
We may orthonormalize \eqref{Schr} via the Lanczos method by utilizing the Gram-Schmidt process in conjunction with the natural inner product, 
\begin{equation}
\left|A_{n+1}\right\rangle=\left(H-a_n\right)\left|K_n\right\rangle-b_n\left|K_{n-1}\right\rangle\,, \quad\left|K_n\right\rangle=b_n^{-1}\left|A_n\right\rangle\,,
\end{equation}
where the Lanczos coefficients $a_n$ and $b_n$ are defined as
\begin{equation}
a_n=\left\langle K_n|H| K_n\right\rangle, \quad b_n=\left\langle A_n \mid A_n\right\rangle^{1 / 2}\,,
\end{equation}
with the initial condition
\begin{equation}
b_0:=0, \quad\left|K_0\right\rangle:=|\psi(0)\rangle \,.
\end{equation}
By employing the acquired Krylov basis, the \eqref{Schr} may be extended as
\begin{equation}
|\psi(t)\rangle=\sum_{n=0} \psi_n(t)\left|K_n\right\rangle\,,
\end{equation}
which results in the Schrödinger equation
\begin{equation}
i \partial_t \psi_n(t)=a_n \psi_n(t)+b_{n+1} \psi_{n+1}(t)+b_n \psi_{n-1}(t)\,,
\end{equation}
where, by definition, we have the initial condition $\psi_n(0)=\delta_{n 0}$. Finally, the Spread complexity
of the state $ \ket{\psi(t)}$ is defined as
\begin{equation}
C_S(t):=\sum_{n=0} n\left|\psi_n(t)\right|^2, \quad \sum_{n=0}\left|\psi_n(t)\right|^2=1\,.
\end{equation}
We also introduce a method for computing the Lanczos coefficients, known as the moment method. To begin with, the moments $\mu_{n}$ are defined through the moment-generating function $G(t)$ as
\begin{equation}
\mu_{n} := \left.\frac{d^{n}}{d t^{n}} {G}(t)\right|_{t=0}, \qquad G(t) = S(t) :=\langle\psi_{}(t)|\psi_{}(0)\rangle \,,
\end{equation}
where $G(t)$ is chosen as the survival amplitude $S(t)$ for the Spread complexity. Finally, we can calculate the Lanczos coefficients by using the Markov chain described in Ref.~\cite{Balasubramanian:2022tpr}.

\bibliography{Refs}

\providecommand{\href}[2]{#2}\begingroup\raggedright\begin{thebibliography}{10}

\bibitem{shannon1949synthesis}
C.~E. Shannon, \emph{The synthesis of two-terminal switching circuits},
  {\emph{The Bell System Technical Journal} {\bf 28} (1949) 59--98}.

\bibitem{Dowling:aa}
M.~R. Dowling and M.~A. Nielsen, \emph{The geometry of quantum computation},
  \href{http://arxiv.org/abs/quant-ph/0701004}{{\tt quant-ph/0701004}}.

\bibitem{Jefferson:2017sdb}
R.~Jefferson and R.~C. Myers, \emph{{Circuit complexity in quantum field
  theory}}, \href{http://dx.doi.org/10.1007/JHEP10(2017)107}{\emph{JHEP} {\bf
  10} (2017) 107}, [\href{http://arxiv.org/abs/1707.08570}{{\tt 1707.08570}}].

\bibitem{Caputa:2017yrh}
P.~Caputa, N.~Kundu, M.~Miyaji, T.~Takayanagi and K.~Watanabe, \emph{{Liouville
  Action as Path-Integral Complexity: From Continuous Tensor Networks to
  AdS/CFT}}, \href{http://dx.doi.org/10.1007/JHEP11(2017)097}{\emph{JHEP} {\bf
  11} (2017) 097}, [\href{http://arxiv.org/abs/1706.07056}{{\tt 1706.07056}}].

\bibitem{Chapman:2017rqy}
S.~Chapman, M.~P. Heller, H.~Marrochio and F.~Pastawski, \emph{{Toward a
  Definition of Complexity for Quantum Field Theory States}},
  \href{http://dx.doi.org/10.1103/PhysRevLett.120.121602}{\emph{Phys. Rev.
  Lett.} {\bf 120} (2018) 121602}, [\href{http://arxiv.org/abs/1707.08582}{{\tt
  1707.08582}}].

\bibitem{Magan:2018nmu}
J.~M. Mag{\'a}n, \emph{{Black holes, complexity and quantum chaos}},
  \href{http://dx.doi.org/10.1007/JHEP09(2018)043}{\emph{JHEP} {\bf 09} (2018)
  043}, [\href{http://arxiv.org/abs/1805.05839}{{\tt 1805.05839}}].

\bibitem{Caputa:2018kdj}
P.~Caputa and J.~M. Magan, \emph{{Quantum Computation as Gravity}},
  \href{http://arxiv.org/abs/1807.04422}{{\tt 1807.04422}}.

\bibitem{Stanford:2014jda}
D.~Stanford and L.~Susskind, \emph{{Complexity and Shock Wave Geometries}},
  \href{http://dx.doi.org/10.1103/PhysRevD.90.126007}{\emph{Phys. Rev.} {\bf
  D90} (2014) 126007}, [\href{http://arxiv.org/abs/1406.2678}{{\tt
  1406.2678}}].

\bibitem{Brown:2015bva}
A.~R. Brown, D.~A. Roberts, L.~Susskind, B.~Swingle and Y.~Zhao,
  \emph{{Holographic Complexity Equals Bulk Action?}},
  \href{http://dx.doi.org/10.1103/PhysRevLett.116.191301}{\emph{Phys. Rev.
  Lett.} {\bf 116} (2016) 191301}, [\href{http://arxiv.org/abs/1509.07876}{{\tt
  1509.07876}}].

\bibitem{Brown:2015lvg}
A.~R. Brown, D.~A. Roberts, L.~Susskind, B.~Swingle and Y.~Zhao,
  \emph{{Complexity, action, and black holes}},
  \href{http://dx.doi.org/10.1103/PhysRevD.93.086006}{\emph{Phys. Rev.} {\bf
  D93} (2016) 086006}, [\href{http://arxiv.org/abs/1512.04993}{{\tt
  1512.04993}}].

\bibitem{Couch:2016exn}
J.~Couch, W.~Fischler and P.~H. Nguyen, \emph{{Noether charge, black hole
  volume, and complexity}},
  \href{http://dx.doi.org/10.1007/JHEP03(2017)119}{\emph{JHEP} {\bf 03} (2017)
  119}, [\href{http://arxiv.org/abs/1610.02038}{{\tt 1610.02038}}].

\bibitem{Belin:2021bga}
A.~Belin, R.~C. Myers, S.-M. Ruan, G.~S\'arosi and A.~J. Speranza, \emph{{Does
  Complexity Equal Anything?}},
  \href{http://dx.doi.org/10.1103/PhysRevLett.128.081602}{\emph{Phys. Rev.
  Lett.} {\bf 128} (2022) 081602}, [\href{http://arxiv.org/abs/2111.02429}{{\tt
  2111.02429}}].

\bibitem{Parker:2018yvk}
D.~E. Parker, X.~Cao, A.~Avdoshkin, T.~Scaffidi and E.~Altman, \emph{{A
  Universal Operator Growth Hypothesis}},
  \href{http://dx.doi.org/10.1103/PhysRevX.9.041017}{\emph{Phys. Rev. X} {\bf
  9} (2019) 041017}, [\href{http://arxiv.org/abs/1812.08657}{{\tt
  1812.08657}}].

\bibitem{Roberts:2018mnp}
D.~A. Roberts, D.~Stanford and A.~Streicher, \emph{{Operator growth in the SYK
  model}}, \href{http://dx.doi.org/10.1007/JHEP06(2018)122}{\emph{JHEP} {\bf
  06} (2018) 122}, [\href{http://arxiv.org/abs/1802.02633}{{\tt 1802.02633}}].

\bibitem{Balasubramanian:2022tpr}
V.~Balasubramanian, P.~Caputa, J.~M. Magan and Q.~Wu, \emph{{Quantum chaos and
  the complexity of spread of states}},
  \href{http://dx.doi.org/10.1103/PhysRevD.106.046007}{\emph{Phys. Rev. D} {\bf
  106} (2022) 046007}, [\href{http://arxiv.org/abs/2202.06957}{{\tt
  2202.06957}}].

\bibitem{Rabinovici:2022beu}
E.~Rabinovici, A.~S\'anchez-Garrido, R.~Shir and J.~Sonner, \emph{{Krylov
  complexity from integrability to chaos}},
  \href{http://dx.doi.org/10.1007/JHEP07(2022)151}{\emph{JHEP} {\bf 07} (2022)
  151}, [\href{http://arxiv.org/abs/2207.07701}{{\tt 2207.07701}}].

\bibitem{Balasubramanian:2023kwd}
V.~Balasubramanian, J.~M. Magan and Q.~Wu, \emph{{Quantum chaos, integrability,
  and late times in the Krylov basis}},
  \href{http://arxiv.org/abs/2312.03848}{{\tt 2312.03848}}.

\bibitem{Caputa:2022eye}
P.~Caputa and S.~Liu, \emph{{Quantum complexity and topological phases of
  matter}}, \href{http://dx.doi.org/10.1103/PhysRevB.106.195125}{\emph{Phys.
  Rev. B} {\bf 106} (2022) 195125},
  [\href{http://arxiv.org/abs/2205.05688}{{\tt 2205.05688}}].

\bibitem{Caputa:2022yju}
P.~Caputa, N.~Gupta, S.~S. Haque, S.~Liu, J.~Murugan and H.~J.~R. Van~Zyl,
  \emph{{Spread complexity and topological transitions in the Kitaev chain}},
  \href{http://dx.doi.org/10.1007/JHEP01(2023)120}{\emph{JHEP} {\bf 01} (2023)
  120}, [\href{http://arxiv.org/abs/2208.06311}{{\tt 2208.06311}}].

\bibitem{Balasubramanian:2022dnj}
V.~Balasubramanian, J.~M. Magan and Q.~Wu, \emph{{Tridiagonalizing random
  matrices}}, \href{http://dx.doi.org/10.1103/PhysRevD.107.126001}{\emph{Phys.
  Rev. D} {\bf 107} (2023) 126001},
  [\href{http://arxiv.org/abs/2208.08452}{{\tt 2208.08452}}].

\bibitem{Lin:2022rbf}
H.~W. Lin, \emph{{The bulk Hilbert space of double scaled SYK}},
  \href{http://dx.doi.org/10.1007/JHEP11(2022)060}{\emph{JHEP} {\bf 11} (2022)
  060}, [\href{http://arxiv.org/abs/2208.07032}{{\tt 2208.07032}}].

\bibitem{Rabinovici:2023yex}
E.~Rabinovici, A.~S\'anchez-Garrido, R.~Shir and J.~Sonner, \emph{{A bulk
  manifestation of Krylov complexity}},
  \href{http://dx.doi.org/10.1007/JHEP08(2023)213}{\emph{JHEP} {\bf 08} (2023)
  213}, [\href{http://arxiv.org/abs/2305.04355}{{\tt 2305.04355}}].

\bibitem{Barbon:2019wsy}
J.~L.~F. Barb\'on, E.~Rabinovici, R.~Shir and R.~Sinha, \emph{{On The Evolution
  Of Operator Complexity Beyond Scrambling}},
  \href{http://dx.doi.org/10.1007/JHEP10(2019)264}{\emph{JHEP} {\bf 10} (2019)
  264}, [\href{http://arxiv.org/abs/1907.05393}{{\tt 1907.05393}}].

\bibitem{Avdoshkin:2019trj}
A.~Avdoshkin and A.~Dymarsky, \emph{{Euclidean operator growth and quantum
  chaos}},
  \href{http://dx.doi.org/10.1103/PhysRevResearch.2.043234}{\emph{Phys. Rev.
  Res.} {\bf 2} (2020) 043234}, [\href{http://arxiv.org/abs/1911.09672}{{\tt
  1911.09672}}].

\bibitem{Dymarsky:2019elm}
A.~Dymarsky and A.~Gorsky, \emph{{Quantum chaos as delocalization in Krylov
  space}}, \href{http://dx.doi.org/10.1103/PhysRevB.102.085137}{\emph{Phys.
  Rev. B} {\bf 102} (2020) 085137},
  [\href{http://arxiv.org/abs/1912.12227}{{\tt 1912.12227}}].

\bibitem{Magan:2020iac}
J.~M. Mag\'an and J.~Sim\'on, \emph{{On operator growth and emergent Poincar\'e
  symmetries}}, \href{http://dx.doi.org/10.1007/JHEP05(2020)071}{\emph{JHEP}
  {\bf 05} (2020) 071}, [\href{http://arxiv.org/abs/2002.03865}{{\tt
  2002.03865}}].

\bibitem{Rabinovici:2020ryf}
E.~Rabinovici, A.~S\'anchez-Garrido, R.~Shir and J.~Sonner, \emph{{Operator
  complexity: a journey to the edge of Krylov space}},
  \href{http://dx.doi.org/10.1007/JHEP06(2021)062}{\emph{JHEP} {\bf 06} (2021)
  062}, [\href{http://arxiv.org/abs/2009.01862}{{\tt 2009.01862}}].

\bibitem{Cao:2020zls}
X.~Cao, \emph{{A statistical mechanism for operator growth}},
  \href{http://dx.doi.org/10.1088/1751-8121/abe77c}{\emph{J. Phys. A} {\bf 54}
  (2021) 144001}, [\href{http://arxiv.org/abs/2012.06544}{{\tt 2012.06544}}].

\bibitem{Jian:2020qpp}
S.-K. Jian, B.~Swingle and Z.-Y. Xian, \emph{{Complexity growth of operators in
  the SYK model and in JT gravity}},
  \href{http://dx.doi.org/10.1007/JHEP03(2021)014}{\emph{JHEP} {\bf 03} (2021)
  014}, [\href{http://arxiv.org/abs/2008.12274}{{\tt 2008.12274}}].

\bibitem{Kim:2021okd}
J.~Kim, J.~Murugan, J.~Olle and D.~Rosa, \emph{{Operator delocalization in
  quantum networks}},
  \href{http://dx.doi.org/10.1103/PhysRevA.105.L010201}{\emph{Phys. Rev. A}
  {\bf 105} (2022) L010201}, [\href{http://arxiv.org/abs/2109.05301}{{\tt
  2109.05301}}].

\bibitem{Rabinovici:2021qqt}
E.~Rabinovici, A.~S\'anchez-Garrido, R.~Shir and J.~Sonner, \emph{{Krylov
  localization and suppression of complexity}},
  \href{http://dx.doi.org/10.1007/JHEP03(2022)211}{\emph{JHEP} {\bf 03} (2022)
  211}, [\href{http://arxiv.org/abs/2112.12128}{{\tt 2112.12128}}].

\bibitem{Dymarsky:2021bjq}
A.~Dymarsky and M.~Smolkin, \emph{{Krylov complexity in conformal field
  theory}}, \href{http://dx.doi.org/10.1103/PhysRevD.104.L081702}{\emph{Phys.
  Rev. D} {\bf 104} (2021) L081702},
  [\href{http://arxiv.org/abs/2104.09514}{{\tt 2104.09514}}].

\bibitem{Caputa:2021sib}
P.~Caputa, J.~M. Magan and D.~Patramanis, \emph{{Geometry of Krylov
  complexity}},
  \href{http://dx.doi.org/10.1103/PhysRevResearch.4.013041}{\emph{Phys. Rev.
  Res.} {\bf 4} (2022) 013041}, [\href{http://arxiv.org/abs/2109.03824}{{\tt
  2109.03824}}].

\bibitem{Patramanis:2021lkx}
D.~Patramanis, \emph{{Probing the entanglement of operator growth}},
  \href{http://dx.doi.org/10.1093/ptep/ptac081}{\emph{PTEP} {\bf 2022} (2022)
  063A01}, [\href{http://arxiv.org/abs/2111.03424}{{\tt 2111.03424}}].

\bibitem{Caputa:2021ori}
P.~Caputa and S.~Datta, \emph{{Operator growth in 2d CFT}},
  \href{http://dx.doi.org/10.1007/JHEP12(2021)188}{\emph{JHEP} {\bf 12} (2021)
  188}, [\href{http://arxiv.org/abs/2110.10519}{{\tt 2110.10519}}].

\bibitem{Trigueros:2021rwj}
F.~B. Trigueros and C.-J. Lin, \emph{{Krylov complexity of many-body
  localization: Operator localization in Krylov basis}},
  \href{http://arxiv.org/abs/2112.04722}{{\tt 2112.04722}}.

\bibitem{Fan:2022xaa}
Z.-Y. Fan, \emph{{Universal relation for operator complexity}},
  \href{http://dx.doi.org/10.1103/PhysRevA.105.062210}{\emph{Phys. Rev. A} {\bf
  105} (2022) 062210}, [\href{http://arxiv.org/abs/2202.07220}{{\tt
  2202.07220}}].

\bibitem{Heveling:2022hth}
R.~Heveling, J.~Wang and J.~Gemmer, \emph{{Numerically Probing the Universal
  Operator Growth Hypothesis}},  \href{http://arxiv.org/abs/2203.00533}{{\tt
  2203.00533}}.

\bibitem{Bhattacharjee:2022vlt}
B.~Bhattacharjee, X.~Cao, P.~Nandy and T.~Pathak, \emph{{Krylov complexity in
  saddle-dominated scrambling}},
  \href{http://dx.doi.org/10.1007/JHEP05(2022)174}{\emph{JHEP} {\bf 05} (2022)
  174}, [\href{http://arxiv.org/abs/2203.03534}{{\tt 2203.03534}}].

\bibitem{Muck:2022xfc}
W.~M\"uck and Y.~Yang, \emph{{Krylov complexity and orthogonal polynomials}},
  \href{http://dx.doi.org/10.1016/j.nuclphysb.2022.115948}{\emph{Nucl. Phys. B}
  {\bf 984} (2022) 115948}, [\href{http://arxiv.org/abs/2205.12815}{{\tt
  2205.12815}}].

\bibitem{He:2022ryk}
S.~He, P.~H.~C. Lau, Z.-Y. Xian and L.~Zhao, \emph{{Quantum chaos, scrambling
  and operator growth in $ T\overline{T} $ deformed SYK models}},
  \href{http://dx.doi.org/10.1007/JHEP12(2022)070}{\emph{JHEP} {\bf 12} (2022)
  070}, [\href{http://arxiv.org/abs/2209.14936}{{\tt 2209.14936}}].

\bibitem{Hornedal:2022pkc}
N.~H\"ornedal, N.~Carabba, A.~S. Matsoukas-Roubeas and A.~del Campo,
  \emph{{Ultimate Physical Limits to the Growth of Operator Complexity}},
  \href{http://arxiv.org/abs/2202.05006}{{\tt 2202.05006}}.

\bibitem{Alishahiha:2022anw}
M.~Alishahiha and S.~Banerjee, \emph{{A universal approach to Krylov State and
  Operator complexities}},  \href{http://arxiv.org/abs/2212.10583}{{\tt
  2212.10583}}.

\bibitem{Camargo:2022rnt}
H.~A. Camargo, V.~Jahnke, K.-Y. Kim and M.~Nishida, \emph{{Krylov complexity in
  free and interacting scalar field theories with bounded power spectrum}},
  \href{http://dx.doi.org/10.1007/JHEP05(2023)226}{\emph{JHEP} {\bf 05} (2023)
  226}, [\href{http://arxiv.org/abs/2212.14702}{{\tt 2212.14702}}].

\bibitem{Baek:2022pkt}
S.~Baek, \emph{{Krylov complexity in inverted harmonic oscillator}},
  \href{http://arxiv.org/abs/2210.06815}{{\tt 2210.06815}}.

\bibitem{Bhattacharya:2022gbz}
A.~Bhattacharya, P.~Nandy, P.~P. Nath and H.~Sahu, \emph{{Operator growth and
  Krylov construction in dissipative open quantum systems}},
  \href{http://dx.doi.org/10.1007/JHEP12(2022)081}{\emph{JHEP} {\bf 12} (2022)
  081}, [\href{http://arxiv.org/abs/2207.05347}{{\tt 2207.05347}}].

\bibitem{Liu:2022god}
C.~Liu, H.~Tang and H.~Zhai, \emph{{Krylov complexity in open quantum
  systems}},
  \href{http://dx.doi.org/10.1103/PhysRevResearch.5.033085}{\emph{Phys. Rev.
  Res.} {\bf 5} (2023) 033085}, [\href{http://arxiv.org/abs/2207.13603}{{\tt
  2207.13603}}].

\bibitem{Bhattacharjee:2022lzy}
B.~Bhattacharjee, X.~Cao, P.~Nandy and T.~Pathak, \emph{{Operator growth in
  open quantum systems: lessons from the dissipative SYK}},
  \href{http://dx.doi.org/10.1007/JHEP03(2023)054}{\emph{JHEP} {\bf 03} (2023)
  054}, [\href{http://arxiv.org/abs/2212.06180}{{\tt 2212.06180}}].

\bibitem{Afrasiar:2022efk}
M.~Afrasiar, J.~Kumar~Basak, B.~Dey, K.~Pal and K.~Pal, \emph{{Time evolution
  of spread complexity in quenched Lipkin-Meshkov-Glick model}},
  \href{http://arxiv.org/abs/2208.10520}{{\tt 2208.10520}}.

\bibitem{Avdoshkin:2022xuw}
A.~Avdoshkin, A.~Dymarsky and M.~Smolkin, \emph{{Krylov complexity in quantum
  field theory, and beyond}},  \href{http://arxiv.org/abs/2212.14429}{{\tt
  2212.14429}}.

\bibitem{Erdmenger:2023wjg}
J.~Erdmenger, S.-K. Jian and Z.-Y. Xian, \emph{{Universal chaotic dynamics from
  Krylov space}}, \href{http://dx.doi.org/10.1007/JHEP08(2023)176}{\emph{JHEP}
  {\bf 08} (2023) 176}, [\href{http://arxiv.org/abs/2303.12151}{{\tt
  2303.12151}}].

\bibitem{Hashimoto:2023swv}
K.~Hashimoto, K.~Murata, N.~Tanahashi and R.~Watanabe, \emph{{Krylov complexity
  and chaos in quantum mechanics}},
  \href{http://dx.doi.org/10.1007/JHEP11(2023)040}{\emph{JHEP} {\bf 11} (2023)
  040}, [\href{http://arxiv.org/abs/2305.16669}{{\tt 2305.16669}}].

\bibitem{Camargo:2023eev}
H.~A. Camargo, V.~Jahnke, H.-S. Jeong, K.-Y. Kim and M.~Nishida,
  \emph{{Spectral and Krylov Complexity in Billiard Systems}},
  \href{http://arxiv.org/abs/2306.11632}{{\tt 2306.11632}}.

\bibitem{Bhattacharya:2023yec}
A.~Bhattacharya, R.~N. Das, B.~Dey and J.~Erdmenger, \emph{{Spread complexity
  for measurement-induced non-unitary dynamics and Zeno effect}},
  \href{http://arxiv.org/abs/2312.11635}{{\tt 2312.11635}}.

\bibitem{Vasli:2023syq}
M.~J. Vasli, K.~Babaei~Velni, M.~R. Mohammadi~Mozaffar, A.~Mollabashi and
  M.~Alishahiha, \emph{{Krylov Complexity in Lifshitz-type Scalar Field
  Theories}},  \href{http://arxiv.org/abs/2307.08307}{{\tt 2307.08307}}.

\bibitem{Patramanis:2023cwz}
D.~Patramanis and W.~Sybesma, \emph{{Krylov complexity in a natural basis for
  the Schr\"odinger algebra}},  \href{http://arxiv.org/abs/2306.03133}{{\tt
  2306.03133}}.

\bibitem{Chattopadhyay:2023fob}
A.~Chattopadhyay, A.~Mitra and H.~J.~R. van Zyl, \emph{{Spread complexity as
  classical dilaton solutions}},
  \href{http://dx.doi.org/10.1103/PhysRevD.108.025013}{\emph{Phys. Rev. D} {\bf
  108} (2023) 025013}, [\href{http://arxiv.org/abs/2302.10489}{{\tt
  2302.10489}}].

\bibitem{Iizuka:2023pov}
N.~Iizuka and M.~Nishida, \emph{{Krylov complexity in the IP matrix model}},
  \href{http://dx.doi.org/10.1007/JHEP11(2023)065}{\emph{JHEP} {\bf 11} (2023)
  065}, [\href{http://arxiv.org/abs/2306.04805}{{\tt 2306.04805}}].

\bibitem{Huh:2023jxt}
K.-B. Huh, H.-S. Jeong and J.~F. Pedraza, \emph{{Spread complexity in
  saddle-dominated scrambling}},  \href{http://arxiv.org/abs/2312.12593}{{\tt
  2312.12593}}.

\bibitem{Pal:2023yik}
K.~Pal, K.~Pal, A.~Gill and T.~Sarkar, \emph{{Time evolution of spread
  complexity and statistics of work done in quantum quenches}},
  \href{http://arxiv.org/abs/2304.09636}{{\tt 2304.09636}}.

\bibitem{Bhattacharya:2023zqt}
A.~Bhattacharya, P.~Nandy, P.~P. Nath and H.~Sahu, \emph{{On Krylov complexity
  in open systems: an approach via bi-Lanczos algorithm}},
  \href{http://arxiv.org/abs/2303.04175}{{\tt 2303.04175}}.

\bibitem{Li:2024kfm}
T.~Li and L.-H. Liu, \emph{{Inflationary Krylov complexity}},
  \href{http://arxiv.org/abs/2401.09307}{{\tt 2401.09307}}.

\bibitem{Anegawa:2024wov}
T.~Anegawa, N.~Iizuka and M.~Nishida, \emph{{Krylov complexity as an order
  parameter for deconfinement phase transitions at large $N$}},
  \href{http://arxiv.org/abs/2401.04383}{{\tt 2401.04383}}.

\bibitem{Beetar:2023mfn}
C.~Beetar, N.~Gupta, S.~S. Haque, J.~Murugan and H.~J.~R. Van~Zyl,
  \emph{{Complexity and Operator Growth for Quantum Systems in Dynamic
  Equilibrium}},  \href{http://arxiv.org/abs/2312.15790}{{\tt 2312.15790}}.

\bibitem{Loc:2024oen}
T.~Q. Loc, \emph{{Lanczos spectrum for random operator growth}},
  \href{http://arxiv.org/abs/2402.07980}{{\tt 2402.07980}}.

\bibitem{adhikari2023krylov}
K.~Adhikari, S.~Choudhury and A.~Roy, \emph{Krylov complexity in quantum field
  theory}, {\emph{Nuclear Physics B} (2023) 116263}.

\bibitem{adhikari2022cosmological}
K.~Adhikari and S.~Choudhury, \emph{Cosmological krylov complexity},
  {\emph{arXiv preprint arXiv:2203.14330} (2022) }.

\bibitem{Papadodimas:2015xma}
K.~Papadodimas and S.~Raju, \emph{{Local Operators in the Eternal Black Hole}},
  \href{http://dx.doi.org/10.1103/PhysRevLett.115.211601}{\emph{Phys. Rev.
  Lett.} {\bf 115} (2015) 211601}, [\href{http://arxiv.org/abs/1502.06692}{{\tt
  1502.06692}}].

\bibitem{delCampo:2017bzr}
A.~del Campo, J.~Molina-Vilaplana and J.~Sonner, \emph{{Scrambling the spectral
  form factor: unitarity constraints and exact results}},
  \href{http://dx.doi.org/10.1103/PhysRevD.95.126008}{\emph{Phys. Rev. D} {\bf
  95} (2017) 126008}, [\href{http://arxiv.org/abs/1702.04350}{{\tt
  1702.04350}}].

\bibitem{Saad:2018bqo}
P.~Saad, S.~H. Shenker and D.~Stanford, \emph{{A semiclassical ramp in SYK and
  in gravity}},  \href{http://arxiv.org/abs/1806.06840}{{\tt 1806.06840}}.

\bibitem{Saad:2019lba}
P.~Saad, S.~H. Shenker and D.~Stanford, \emph{{JT gravity as a matrix
  integral}},  \href{http://arxiv.org/abs/1903.11115}{{\tt 1903.11115}}.

\bibitem{Chen:2023hra}
Y.~Chen, V.~Ivo and J.~Maldacena, \emph{{Comments on the double cone
  wormhole}},  \href{http://arxiv.org/abs/2310.11617}{{\tt 2310.11617}}.

\bibitem{cotler2017black}
J.~S. Cotler, G.~Gur-Ari, M.~Hanada, J.~Polchinski, P.~Saad, S.~H. Shenker
  et~al., \emph{Black holes and random matrices}, {\emph{Journal of High Energy
  Physics} {\bf 2017} (2017) 1--54}.

\bibitem{saad2018semiclassical}
P.~Saad, S.~H. Shenker and D.~Stanford, \emph{A semiclassical ramp in syk and
  in gravity}, {\emph{arXiv preprint arXiv:1806.06840} (2018) }.

\bibitem{rabinovici2022krylov}
E.~Rabinovici, A.~S{\'a}nchez-Garrido, R.~Shir and J.~Sonner, \emph{Krylov
  complexity from integrability to chaos}, {\emph{Journal of High Energy
  Physics} {\bf 2022} (2022) 1--29}.

\bibitem{Caputa:2023vyr}
P.~Caputa, J.~M. Magan, D.~Patramanis and E.~Tonni, \emph{{Krylov complexity of
  modular Hamiltonian evolution}},  \href{http://arxiv.org/abs/2306.14732}{{\tt
  2306.14732}}.

\bibitem{Caceres:2019pgf}
E.~Caceres, S.~Chapman, J.~D. Couch, J.~P. Hernandez, R.~C. Myers and S.-M.
  Ruan, \emph{{Complexity of Mixed States in QFT and Holography}},
  \href{http://dx.doi.org/10.1007/JHEP03(2020)012}{\emph{JHEP} {\bf 03} (2020)
  012}, [\href{http://arxiv.org/abs/1909.10557}{{\tt 1909.10557}}].

\bibitem{Haag_1996}
R.~Haag, \emph{Local Quantum Physics}.
\newblock Springer Berlin Heidelberg, 1996,
  \href{http://dx.doi.org/10.1007/978-3-642-61458-3}{10.1007/978-3-642-61458-3}.

\bibitem{ali2020chaos}
T.~Ali, A.~Bhattacharyya, S.~S. Haque, E.~H. Kim, N.~Moynihan and J.~Murugan,
  \emph{Chaos and complexity in quantum mechanics},
  \href{http://dx.doi.org/10.1103/PhysRevD.101.026021}{\emph{Physical Review D}
  {\bf 101} (2020) 026021}.

\bibitem{qu2022chaos}
L.-C. Qu, J.~Chen and Y.-X. Liu, \emph{Chaos and complexity for inverted
  harmonic oscillators},
  \href{http://dx.doi.org/10.1103/PhysRevD.105.126015}{\emph{Physical Review D}
  {\bf 105} (2022) 126015}.

\bibitem{qu2022chaos1}
L.-C. Qu, H.-Y. Jiang and Y.-X. Liu, \emph{Chaos and multifold complexity for
  an inverted harmonic oscillator},
  \href{http://dx.doi.org/10.1007/JHEP12(2022)065}{\emph{Journal of High Energy
  Physics} {\bf 2022} (2022) 1--24}.

\bibitem{Tang:2023ocr}
H.~Tang, \emph{{Operator Krylov complexity in random matrix theory}},
  \href{http://arxiv.org/abs/2312.17416}{{\tt 2312.17416}}.

\bibitem{Dyson_1962}
F.~J. Dyson, \emph{Statistical theory of the energy levels of complex systems.
  i}, \href{http://dx.doi.org/10.1063/1.1703773}{\emph{Journal of Mathematical
  Physics} {\bf 3} (Jan., 1962) 140--156}.

\bibitem{Bohigas:1984aa}
O.~Bohigas, \emph{Characterization of chaotic quantum spectra and universality
  of level fluctuation laws},
  \href{http://dx.doi.org/10.1103/PhysRevLett.52.1}{\emph{Physical Review
  Letters} {\bf 52} (1984) 1--4}.

\bibitem{Guhr:1997ve}
T.~Guhr, A.~Muller-Groeling and H.~A. Weidenmuller, \emph{{Random matrix
  theories in quantum physics: Common concepts}},
  \href{http://dx.doi.org/10.1016/S0370-1573(97)00088-4}{\emph{Phys. Rept.}
  {\bf 299} (1998) 189--425},
  [\href{http://arxiv.org/abs/cond-mat/9707301}{{\tt cond-mat/9707301}}].

\bibitem{MEHTA1960395}
M.~Mehta, \emph{On the statistical properties of the level-spacings in nuclear
  spectra},
  \href{http://dx.doi.org/https://doi.org/10.1016/0029-5582(60)90413-2}{\emph{Nuclear
  Physics} {\bf 18} (1960) 395--419}.

\bibitem{GAUDIN1961447}
M.~Gaudin, \emph{Sur la loi limite de l'espacement des valeurs propres d'une
  matrice ale´atoire},
  \href{http://dx.doi.org/https://doi.org/10.1016/0029-5582(61)90176-6}{\emph{Nuclear
  Physics} {\bf 25} (1961) 447--458}.

\bibitem{Dyson_1962III}
F.~J. Dyson, \emph{Statistical theory of the energy levels of complex systems.
  iii}, \href{http://dx.doi.org/10.1063/1.1703775}{\emph{Journal of
  Mathematical Physics} {\bf 3} (Jan., 1962) 166--175}.

\bibitem{Czech:2017ryf}
B.~Czech, \emph{{Einstein Equations from Varying Complexity}},
  \href{http://dx.doi.org/10.1103/PhysRevLett.120.031601}{\emph{Phys. Rev.
  Lett.} {\bf 120} (2018) 031601}, [\href{http://arxiv.org/abs/1706.00965}{{\tt
  1706.00965}}].

\bibitem{Susskind:2019ddc}
L.~Susskind, \emph{{Complexity and Newton's Laws}},
  \href{http://dx.doi.org/10.3389/fphy.2020.00262}{\emph{Front. in Phys.} {\bf
  8} (2020) 262}, [\href{http://arxiv.org/abs/1904.12819}{{\tt 1904.12819}}].

\bibitem{Pedraza:2021mkh}
J.~F. Pedraza, A.~Russo, A.~Svesko and Z.~Weller-Davies, \emph{{Lorentzian
  Threads as Gatelines and Holographic Complexity}},
  \href{http://dx.doi.org/10.1103/PhysRevLett.127.271602}{\emph{Phys. Rev.
  Lett.} {\bf 127} (2021) 271602}, [\href{http://arxiv.org/abs/2105.12735}{{\tt
  2105.12735}}].

\bibitem{Pedraza:2021fgp}
J.~F. Pedraza, A.~Russo, A.~Svesko and Z.~Weller-Davies, \emph{{Sewing
  spacetime with Lorentzian threads: complexity and the emergence of time in
  quantum gravity}},
  \href{http://dx.doi.org/10.1007/JHEP02(2022)093}{\emph{JHEP} {\bf 02} (2022)
  093}, [\href{http://arxiv.org/abs/2106.12585}{{\tt 2106.12585}}].

\bibitem{Pedraza:2022dqi}
J.~F. Pedraza, A.~Russo, A.~Svesko and Z.~Weller-Davies, \emph{{Computing
  spacetime}}, \href{http://dx.doi.org/10.1142/S021827182242010X}{\emph{Int. J.
  Mod. Phys. D} {\bf 31} (2022) 2242010},
  [\href{http://arxiv.org/abs/2205.05705}{{\tt 2205.05705}}].

\bibitem{Carrasco:2023fcj}
R.~Carrasco, J.~F. Pedraza, A.~Svesko and Z.~Weller-Davies, \emph{{Gravitation
  from optimized computation: Einstein and beyond}},
  \href{http://dx.doi.org/10.1007/JHEP09(2023)167}{\emph{JHEP} {\bf 09} (2023)
  167}, [\href{http://arxiv.org/abs/2306.08503}{{\tt 2306.08503}}].

\bibitem{Kar:2021nbm}
A.~Kar, L.~Lamprou, M.~Rozali and J.~Sully, \emph{{Random matrix theory for
  complexity growth and black hole interiors}},
  \href{http://dx.doi.org/10.1007/JHEP01(2022)016}{\emph{JHEP} {\bf 01} (2022)
  016}, [\href{http://arxiv.org/abs/2106.02046}{{\tt 2106.02046}}].

\end{thebibliography}\endgroup
\bibliographystyle{JHEP}

\end{document}